\documentclass[english,
]{aa}
\usepackage{epsf,amsfonts,amssymb,graphicx,fancyheadings}
\usepackage{caption}
\usepackage{longtable}
\usepackage{multirow}
\usepackage{supertabular}
\usepackage{natbib}
\usepackage{babel}
\usepackage{txfonts}
\usepackage{rotating}
\usepackage{enumitem}

\bibpunct{(}{)}{;}{a}{}{,} 

\begin{document}

\title{A plethora of new R Coronae Borealis stars discovered from a dedicated spectroscopic follow-up survey
\thanks{Full Table~\ref{tab.short_version} and the spectra presented in Figs.~\ref{fig_SpectroCoolRCB},~\ref{fig_SpectroCoolRCBblue}, ~\ref{fig_KDM6546spectrum},  ~\ref{fig_SpectroWarmRCB} and~\ref{fig_SpectroHotRCB} are only available at the CDS via anonymous ftp to cdsarc.u-strasbg.fr (130.79.128.5) or via http://cdsarc. u-strasbg.fr/viz-bin/cat/J/A+A/vol/page}}

\author{
P.~Tisserand\inst{1,2},
G.C.~Clayton\inst{3},
M.S.~Bessell\inst{2},
D.L.~Welch\inst{4},
D.~Kamath\inst{2,6},
P.R.~Wood\inst{2},
P.~Wils\inst{5},
\L{}.~Wyrzykowski\inst{7},
P.~Mr\'oz\inst{7},
A.~Udalski\inst{7}
}



\institute{
Sorbonne Universit\'es, UPMC Univ Paris 6 et CNRS, UMR 7095, Institut d'Astrophysique de Paris, IAP, F-75014 Paris, France \and
Research School of Astronomy and Astrophysics, Australian National University, Cotter Rd, Weston Creek ACT 2611, Australia \and
Department of Physics \& Astronomy, Louisiana State University, Baton Rouge, LA 70803, USA \and
Department of Physics \& Astronomy, McMaster University, Hamilton, Ontario, L8S 4M1, Canada \and
Vereniging Voor Sterrenkunde (VVS), Brugge, Belgium\and
Department of Physics and Astronomy, Macquarie University, Sydney, NSW 2109, Australia \and
Astronomical Observatory, University of Warsaw, Al. Ujazdowskie 4, 00-478 Warszawa, Poland
}

\offprints{Patrick Tisserand; \email{tisserand@iap.fr}}

\date{}


\abstract {It is more and more suspected that R Coronae Borealis (RCB) stars - rare hydrogen-deficient and carbon-rich supergiant stars - are the products of mergers of CO/He white-dwarf binary systems in the intermediate mass regime ($0.6<M_{Tot}<1.2 M_{\odot}$). Following the merger, a short-lived cool supergiant phase starts. 
RCB stars are extremely rare as only 77 have hitherto been known in the Galaxy, while up to 1000 have been predicted from population synthesis models.}
{The goal is to significantly increase the number of known RCB stars in order to better understand their evolutionary paths, their spatial distribution, and their formation rate in the context of population synthesis results. A list of 2356 RCB star candidates was selected using infrared colours from the all-sky 2MASS and WISE surveys. The objective is to follow them up spectroscopically to classify the candidates and, thus, to distinguish RCB stars from other dust-producing stars.}
{A series of brightness and colour-colour cuts that were used as selection criteria were then tested using the sample of known Galactic and Magellanic RCB stars. 
RCB spectral energy distribution models were also used to understand the effects of each selection criterion in terms of circumstellar shell temperature. Optical, low-resolution spectra were obtained for nearly 500 of the candidate stars. These spectra were compared to synthetic spectra from a new grid of MARCs hydrogen-deficient atmospheric models. This allowed us to define a spectroscopic classification system for RCB stars depending on their effective temperature and photometric status.}
{This programme has found 45 new RCB stars, including 30 Cold ($4000<T_\mathrm{eff}<6800$ K), 14 Warm ($6800<T_\mathrm{eff}<8500$ K), and one Hot ($T_\mathrm{eff}>15000$ K). Forty of these belong to the Milky Way and five are located in the Magellanic Clouds. We also confirmed that the candidate KDM 5651 is indeed a new RCB star, increasing the total number of Magellanic RCB stars to 30.}
{We increased the total number of RCB stars known by $\sim$50\%, bringing it up to 147. In addition, we compiled a list of 14 strong RCB candidates, most certainly observed during a dust obscuration phase. From the detection efficiency and success rate so far, we estimate that there should be no more than 500 RCB stars existing in the Milky Way, all HdC stars included.}


\authorrunning{Tisserand, P. }
\titlerunning{A plethora of new R Coronae Borealis stars}

\maketitle

\section{Introduction \label{sec_intro}}

R Coronae Borealis (RCB) stars are rare hydrogen-deficient, carbon-rich, supergiant stars that are increasingly suspected of having resulted from the merger of one CO- + one He- white dwarfs \citep{2012JAVSO..40..539C}. 
Therefore, they may be low-mass analogues of Type Ia supernova progenitors.
The double-degenerate scenario has been strongly supported by the observations of abundance anomalies in RCB stars including a large $^{18}$O over-abundance in their atmospheres \citep{2007ApJ...662.1220C,2010ApJ...714..144G} and of surface abundance anomalies for a few elements, fluorine in particular \citep{2008ApJ...674.1068P,2011MNRAS.414.3599J}. Furthermore, the abundances computed by simulations of such merging events agree well with the peculiar and disparate atmosphere abundances observed in RCB stars \citep{2011MNRAS.414.3599J,2012ApJ...757...76S,2013ApJ...772...59M,2014MNRAS.445..660Z,2019MNRAS.488..438L}.


Interestingly, a second evolutionary scenario, the final helium shell flash, has also been proposed to explain the origin of RCB stars \citep{1996ApJ...456..750I,1990ASPC...11..549R}. The final-flash in a star on the verge of becoming a WD, causes it to expand into a cool supergiant star similar to RCB stars. Therefore, a fraction of RCB stars may result from the final-flash scenario. Such objects could be identified from the detection of hydrogen-rich nebulae around them \citep{2011ApJ...743...44C}. However, recent studies of some RCB's immediate circumstellar environment do not favour that scenario \citep{Montiel_2015,2018AJ....156..148M}.


Our goal is to test these two scenarios by increasing the numbers of known RCB stars and consequently studying their sky distribution and formation rate. In the double-degenerate scenario, it is estimated that the He-CO WDs merger birthrate should be between $\sim10^{-3}$ and $\sim5\times10^{-3}$ per year \citep{2001A&A...365..491N,2009ApJ...699.2026R,2015ApJ...809..184K} and that an RCB phase lifetime should last about $10^5$ years, as predicted by theoretical evolution models \citep{2002MNRAS.333..121S}. Therefore, we can expect between 100 and 500 RCB stars to exist in our Galaxy. 

RCB stars possess a large range of photospheric temperatures, mostly between $\sim$4000 and 8000 K \citep{2009A&A...501..985T,2012A&A...539A..51T}, but some are also known at hotter temperature ($>$12000 K) \citep{2002AJ....123.3387D}. 
This wide range of effective temperatures supports the scenario that after the cataclysmic event that creates an RCB, it goes through a supergiant phase which then evolves from a cold to a warm state while the helium-rich atmosphere contracts \citep{2011MNRAS.414.3599J,2019MNRAS.488..438L}. Fortunately, RCB stars are also known to be very bright, $-5\leqslant M_V\leqslant-3.5$ \citep[Fig. 3]{2009A&A...501..985T}. All RCB stars have an IR excess due to the presence of a warm circumstellar dust shell with $300<T_{eff,shell}<1000$ K \citep[Fig. 2]{2012A&A...539A..51T}. These two characteristics facilitate the search reported here using the ensemble of photometric datasets available. Finally, we note that RCB stars are members of a larger class of stars called the Hydrogen deficient Carbon stars (HdC stars) that share similar spectroscopic characteristics, but RCB stars have the particularity of being surrounded by dust and of undergoing unpredictable fast and large photometric declines due to clouds of dust newly produced. 

We search for new RCB stars located in the Milky Way and the Magellanic Clouds in two steps: firstly, by selecting a short-list of targets of interest (ToI) among the 500 million objects catalogued within the two all-sky near- and mid- infrared (IR) surveys, 2MASS
\citep{2006AJ....131.1163S} and WISE
\citep{2010AJ....140.1868W}; secondly, by following them up spectroscopically, possibly with added support from photometric monitoring surveys.

Section~\ref{sec_IRcuts} presents the WISE survey and describes the broad-band, colour selection criteria applied to the 2MASS and WISE ALL-Sky catalogues. The resulting list of RCB star candidates, their characteristics and the subsequent classification into priority groups for the spectroscopic follow-up are discussed in subsections~\ref{subsec_priority} and~\ref{subsec_cat}. Then in Section~\ref{sec_data}, we detail the new spectra obtained, the light curves, and the stellar atmosphere models used. The spectral analysis and the classification system developed to identify new RCB stars are discussed in Section~\ref{sec_ana}, while the status of previously discovered RCB candidates is reviewed in Section~\ref{sec_cand}. In Section~\ref{sec_result}, we explore the RCB star spatial distribution, and estimate the total number of RCB stars in the Milky Way. Finally, we summarise our results in Section~\ref{sec_summary}.

\begin{figure*}
\centering
\includegraphics[scale=0.55]{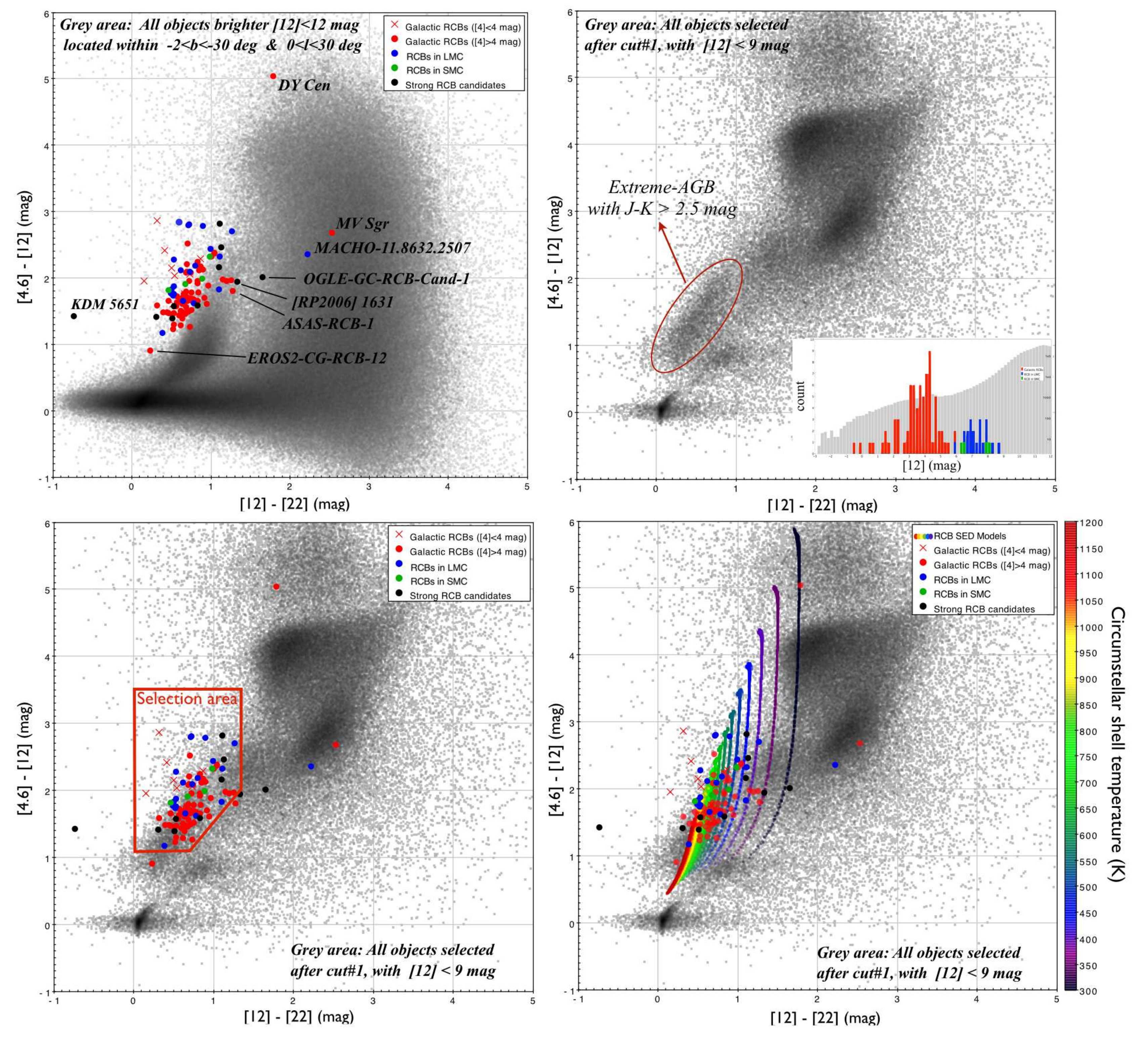}
\caption{Four panels representing the same colour-colour $[4.6]-[12]$ versus $[12]-[22]$ diagram for different WISE source subsamples. See the explanations in the text and within the figures. The [12] magnitude distribution represented inside the top-right panel corresponds to the known RCB stars (linear scale, coloured bins: red for Galactic, blue for LMC, green for SMC) and to the WISE objects selected after cut \#0 (grey bins, logarithmic scale). The names and positions of all known and candidate RCB stars that were rejected by the criterion cut \#2 are indicated on the top-left panel.}
\label{figcut_A}
\end{figure*}

\section{Infra-red broad bands selection of targets of interest \label{sec_IRcuts}}

\subsection{WISE All-Sky catalogue and Known RCB stars\label{sec_wise}}

The WISE Explorer mapped in 2010, during its full cryogenic phase, the entire sky in 3.4, 4.6, 12 and 22 $\mu$m with, respectively, angular resolutions of 6.1, 6.4, 6.5 and 12.0 arcsec and 0.08, 0.11, 1.0 and 6.0 mJy in point source sensitivities at 5 sigma \citep{2010AJ....140.1868W}. The four WISE photometric bands are hereafter named [3.4], [4.6], [12] and [22].
There was a Preliminary Data release in 2011
(WISE-PDR1) covering 57\% of the full sky area \citep{2011wise.rept....1C}, and the All-Sky Data Release
(WISE All-Sky) containing more than 563 million objects in 2012 \citep{2012wise.rept....1C}.

We chose to use the WISE All-Sky catalogue instead of the latest release, the ALLWISE catalogue \citep{2011ApJ...731...53M}, which was made after combining the data obtained during the cryogenic and post-cryogenic survey (NEOWISE) phases, for the following reasons. Firstly, it is preferable when selecting objects as highly variable as RCB stars to compare magnitudes taken at the same epoch (photometric variations as large as $\sim$0.5 mag can been observed at 3 microns). 
Secondly, the ALLWISE combined dataset observations were made at different temperature states of the spacecraft and the [3.4] and [4.6] photometric sensitivities changed following the depletion of the solid hydrogen cryogen and subsequent warm-up of the detectors (see the Data Release Explanatory Supplement for more information). 
Therefore the photometric bias observed for saturated sources has also changed and is therefore not uniform for all measurements listed in ALLWISE. This is important as the majority of known Galactic RCB stars are saturated in the [3.4] and [4.6] photometric bands. A correction of these biases would be complicated to apply on the entire ALLWISE dataset, instead we found a pragmatic solution with the entire WISE All-Sky catalogue as its dataset can be considered as homogeneous. 
Finally, RCB stars are bright objects (all known RCB stars are brighter than 9 mag in [12] - see histogram in Fig.~\ref{figcut_A}, top-right) and, with a magnitude limit of $\sim$13 mag in [12], the WISE All-Sky catalogue is sufficient for our search. A deeper dataset like ALLWISE would not improve it. The magnitude limit offered by the WISE All-Sky catalogue allow us to detect Galactic and Magellanic RCB stars within about 50 kpc in all four mid-infrared bands.

From the WISE-PDR1 catalogue, a preliminary list of candidate RCB stars was created \citep{2012A&A...539A..51T} using selection criteria based on the peculiar near- and mid-infrared colours of RCB stars and their circumstellar dust. This first list is now superseded by the sample selected here using the WISE All-Sky catalogue. 
The new catalogue is much improved in the sky coverage and the photometric sensitivity.
In addition, the Magellanic Clouds are now included in the catalogue. 
Furthermore, 
the WISE All-Sky detection algorithm, used in a highly crowded field, is more efficient. Consequently, some known RCB stars such as OGLE-GC-RCB-1 \& -2, that were not catalogued in the WISE-PDR1 despite being clearly bright, are now listed in the WISE All-Sky catalogue.

All 101 known RCB stars (77 of those are Galactic and 24 are Magellanic) have been catalogued in the WISE All-Sky release, as well as ten stars that are considered as strong RCB candidates \citep{2012JAVSO..40..539C,2013A&A...551A..77T}. The WISE magnitudes and associated 1$\sigma$ uncertainties are listed in Table~\ref{tab.WISEa}. This list of known RCB stars has been used as a benchmark in our search for new members of the class. An update on the status of the ten strong RCB star candidates is given in Section~\ref{sec_cand} after the analysis of their broad-band IR photometry and of the newly available spectra. 

We found small photometric zero-point shifts between the two WISE catalogues: All-Sky$ - $PDR1 $\sim-13$ mmags [3.4], $\sim-4$ mmags [4.6], $\sim29$ mmags [12], and $\sim-27$ mmags [22]. 
These differences are small but could have a significant impact on the selection thresholds already defined in \citet{2012A&A...539A..51T} if not taken into account (see Section~\ref{sec_ana}). 

Most known Galactic RCB stars are highly saturated in the [3.4] and [4.6] bands. Fortunately, photometry was nevertheless performed by fitting only the PSF wings. However, particularly in the [4.6] band, this resulted in a photometric bias for all objects brighter than about [4.6]$<6.5$ mag (see section III.3.c of WISE All-Sky Release Explanatory Supplement for more information). For the brightest objects, the [4.6] magnitude could be over-estimated by almost 1 magnitude. To correct for this effect, we used the same strategy as the one detailed in \citet[sect. 2.2.1]{2012A&A...539A..51T}. 
A corrected [4.6]$_{corr}$ magnitude was estimated from the spectral energy distributions (SEDs) of known RCB stars. We derived the following linear correction formula for the WISE All-Sky dataset: [4.6]$_{corr}$ = 0.83 x [4.6]$_{cat}$ + 1.09. We applied this correction to all objects brighter than [4.6]$_{cat}<6.35$ mag. For the brightest objects, [4.6]$_{cat}<4$ mag, we assume that we do not know the brightness better than 10\% and modified the respective WISE errors.


\begin{table*}[!htbp]
\caption{Number of selected Galactic (G) and Magellanic (M) objects remaining after each selection criterion.
\label{tab.Selection}}
\medskip
\centering
\begin{tabular}{lcccl}
\hline
\hline
Selection criterion & \multicolumn{2}{c}{Number of objects reported with} & Number of &  Known RCB stars  \\
 				                									&     		7 valid         & 		at least one upper    					&         known RCB stars             &  eliminated  \\
								  								   &     measurements    &   limit value 	in	    & 		       selected                               &	at each stage			 \\
								  								   &                                &  J, H, K, [3] or [22]  &  &  \\
\hline
0: [12]$<12$ mag  	  								  &  21532159 						&  33203403	     &  100  & V1157 Sgr  \\
1: Cut on ($J-H$ vs $H-K$)  					  & 1843558   						&  1582645	     &  95    & ASAS-RCB-8, XX Cam,\\
 														 			  &                                       &                        &          & Y Mus, UV Cas and \\ 
 														 			  &                                       &                        &          & EROS2-LMC-RCB-6 \\ 
2: Cut on ($[4.6]-[12]$						         &  14315       						&  243710	         &  90    & ASAS-RCB-1, DY Cen, \\ 
\hspace{3 mm} vs $[12]-[22]$) 					 &                                         &                      &          & EROS2-CG-RCB-12, MV Sgr \\
 														 			  &                                         &                      &          & and  MACHO-11.8632.2507 \\
3: Cut on ($J-K$ vs $J-[12]$) 	  		      & G: 11301, M: 482            & 232899        &  89    & MACHO-308.38099.66 \\
4: Cut on ([12] vs  $[12]-[22]$)   		      & G: 2568, M: 255             & 4279             &  & \\
5: Cut on (K vs $J-K$) 					      	  & G: 2024, M: 118             & 4204             &  & \\
6: Cut on ($[3.4]-[4.6]$) 			       		      & G: 1905, M: 112            & 2435              &  & \\
7: Cut on 2MASS-WISE                         & G: 1862, M: 109            & 1736               &  & \\
\hspace{3 mm}Association 										 &                                         &                      &    &  \\
  \multicolumn{5}{c}{Special supplementary cuts targeting objects with upper limit values in the J and/or [22] bands} \\
8: Outside $|b|<2$, $|l|<60$ deg &                                   & 757                & 88  &   EROS2-CG-RCB-8  \\
9: Strict cuts on blending   &                                        & 473                & 87  &   EROS2-CG-RCB-5  \\
\hline
\hspace{3 mm}Final, without known RCB & \multicolumn{2}{c}{G: 2194, M: 162}  &  & \\
\hspace{3 mm} stars and the known HdC      &                               &  & & \\
\hspace{3 mm} star HD 175893     				  &                               &  &  & \\
\hline
\end{tabular}
\end{table*}

\subsection{Selection criteria \label{sec_criteria}}

RCB stars are so rare and so diverse in terms of photospheric and circumstellar-shell luminosities and temperatures that to obtain a comprehensive view over the entire range of these parameters, we decided to cast a wide net over the entire 2MASS \citep{2006AJ....131.1163S} and WISE \citep{2010AJ....140.1868W} databases. The association between these two catalogues is already provided by the IR science archive (IRSA) from NASA. 
Firstly, the main selection criteria were applied to all catalogued objects that presented valid measurements in all seven (3 2MASS + 4 WISE) bands (Sect.~\ref{sec_mainana}). Subsequently, to catch RCB stars observed in a faint phases, new criteria are defined for a large number of detected objects that are listed with some upper limits values in up to two of these seven bands (Sect.~\ref{sec_upperlimits_ana}).

To simplify our search in these two cases, we required at the start of our analysis that each object should be detected in the WISE [12] band and be brighter than $[12]<12$ mag. That corresponds to $\sim$21.5 and $\sim$33.2 million objects respectively for the first and second group, which overall correspond to nearly 10\% of the 563 million objects detected in the WISE All-Sky survey. The [12] threshold was chosen to be conservative as it is fainter than the faintest known Magellanic RCB stars by three magnitudes
(see the [12] distribution in Fig.~\ref{figcut_A}, top-left). Using all known RCB stars, it is worth mentioning also that the median of the signal-to-noise ratio distribution in [12] is higher by a factor of two than those for the three remaining WISE bands. RCB stars are therefore most noticeable in the [12] bandpass.

For the definition of our selection criteria, we used all 102
known RCB stars as benchmarks, as well as comparing to RCB SED models with a range of photospheric (4000--8000 K) and circumstellar-shell temperatures (300--1200 K). The summary of the selection criteria applied, the resulting number of candidate stars selected, and the known RCB stars rejected are presented in Table~\ref{tab.Selection}.

\subsubsection{The main selection criteria \label{sec_mainana}}




\textbf{Cut 0}: We kept all objects that were detected with a valid measurement in each of the seven 2MASS+WISE bands, and were found to be brighter than $[12]<12$ mag. Only three known RCB stars did not pass these simple requirements. V1157 Sgr was not selected because it was not detected in both [12] and [22] as it is located in a part of the sky with no observational coverage in these two bandpasses.
MSX-SMC-014 and EROS2-CG-RCB-8 had two upper limit values in 2MASS J and H. However, these two RCB stars will still be considered in the second selection scenario applied to objects that possess upper limit values (see section~\ref{sec_upperlimits_ana}).


\textbf{Cut 1}: The first colour-colour selection criterion was applied on the $J-H$ versus $H-K$ diagram. It targets objects presenting a high near-infrared excess as RCB stars possess warm circumstellar shells. This selection criterion has already proved its high efficiency in previous studies \citep[Fig.6]{2012A&A...539A..51T} and remains the same. It was defined as a function of the Galactic latitude as the interstellar extinction affects these infrared magnitudes significantly. We have now rejected $\sim$91\% of all previously selected objects at this stage. However, four known RCB stars were also rejected: ASAS-RCB-8, XX Cam, Y Mus and UV Cas. This is because all of them are warm RCB stars ($T_\mathrm{eff}>7200$ K) with their respective shells not thick enough to impact on their near-IR magnitudes. Their SEDs are displayed in \citet{2012A&A...539A..51T} and \citet{2013A&A...551A..77T}.

It is important to mention here that the main locus of classical carbon stars (located at $J-H\sim0.6$ mag and $H-K\sim1.2$ mag) is entirely rejected at this stage. 


\begin{figure*}
\centering
\includegraphics[scale=0.52]{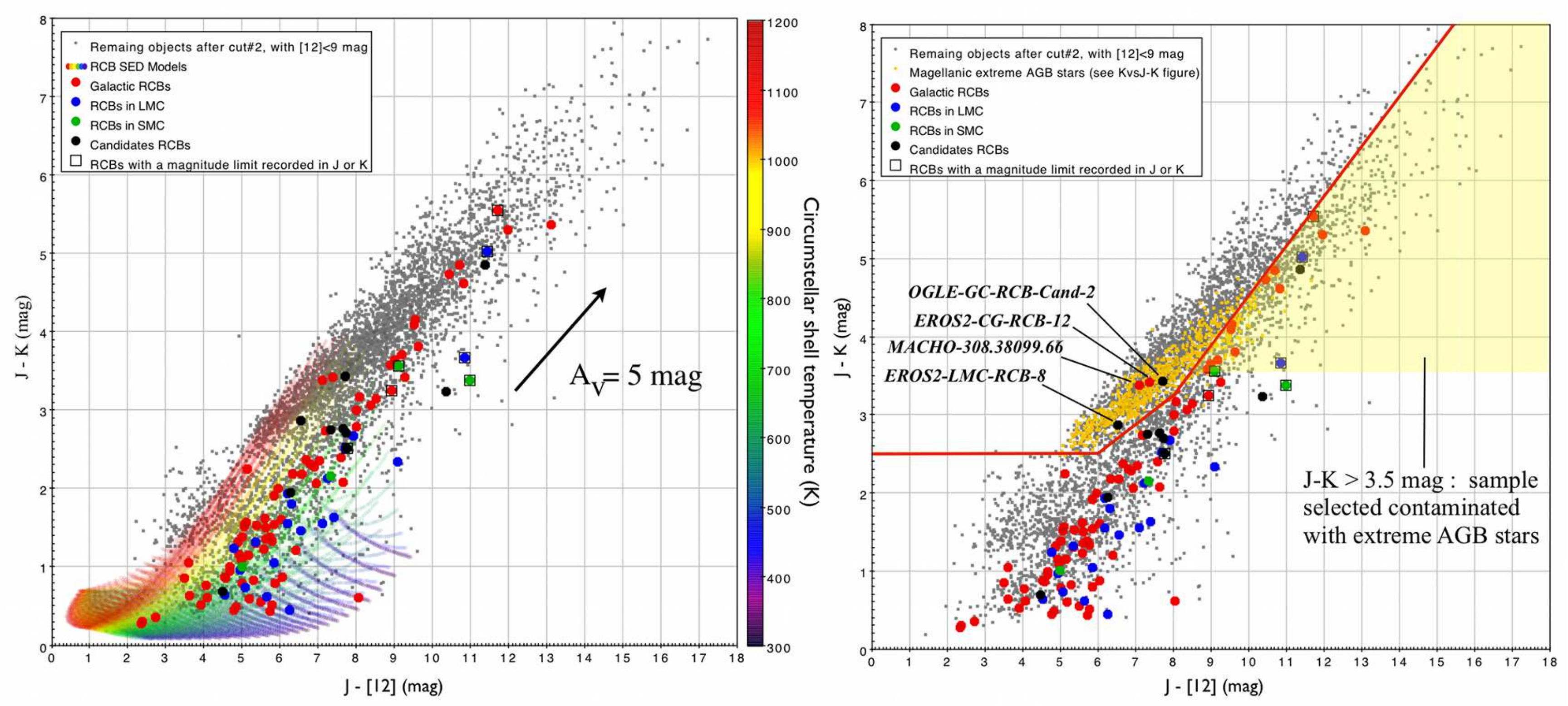}
\caption{$J-K$ versus $J-[12]$ colour-colour diagrams with all objects selected after cut \#2 and brighter than $[12]<9$ mag, and with all known and candidate RCB stars represented. On the left side, RCB SED models are overlayed with a colour scale related to the circumstellar shell temperature. The interstellar extinction arrow is also indicated, it shows that many RCB stars were indeed affected by extinction at the time of the 2MASS epochs. This extinction includes circumstellar and interstellar dust. RCB stars observed in a deep decline have their 2MASS J magnitudes recorded with a magnitude limit. On the right side, the Magellanic extreme-AGB stars are represented with orange dots (see Fig.~\ref{figcut_C}). The selection area of cut \#3 is delimited with red lines. Four known or candidate RCB stars that were not selected are marked with their names. This plot shows that all selected objects redder than $J-K>$3.5 mag will be contaminated by extreme AGB stars. RCB stars will also be even harder to find with such high extinction. }
\label{figcut_B}
\end{figure*}

\textbf{Cut 2}: The second colour-colour selection criterion was applied directly to the WISE photometry, and more specifically on the $[4.6]-[12]$ versus $[12]-[22]$ diagram. It therefore focuses on RCB dust shell temperatures. A similar criterion was used by \citet{2012A&A...539A..51T}, but we adjusted the selection thresholds here to better suit the new photometric datasets (i.e. due to the change in photometric zero-point between the preliminary and the All-Sky catalogues, see Section~\ref{sec_wise}) and the stellar distribution of the new known RCB stars. The selection area is illustrated in Fig.~\ref{figcut_A} (bottom, left) and the limits are the following:

\begin{eqnarray}
&& 1.1 < [4.6]-[12] < 3.5\hspace{1mm}and\hspace{1mm}0 < [12]-[22] < 1.35 \nonumber \\
&& \hspace{5 mm} and\hspace{1mm} [4.6]-[12] \geqslant 1.5\times ([12]-[22]) +0.05 
\label{eq.cut2}
\end{eqnarray}

Figure~\ref{figcut_A} shows four examples of the same $[4.6]-[12]$ versus $[12]-[22]$ colour-colour diagram using different object subsamples and with RCB SED models illustrated for different cicumstellar dust shell temperatures. First, top-left, the positions of the known RCB stars are indicated as well as a sample of catalogued objects selected after cut \#0. This panel shows that most RCB stars form a locus with mid-infrared colours that are uncharacteristic of ordinary catalogued objects. For the remaining three panels, the sample of objects plotted has passed the IR selection cut \#1 and, for illustrative purposes only, are brighter that $[12]<9$ mag, which is a magnitude corresponding more closely to our reference sample of known RCB stars (see, within the top-right graph, the [12] mag distribution for the known RCB stars and for the selected WISE objects after cut \#1). An elongated feature appears near the RCB-star locus, corresponding to highly enshrouded AGB stars with $J-K>2.5$ mag. Most of the known Galactic RCB stars are distributed over the redder part of that feature, while the Magellanic stars are more dispersed, with more than half lying above this AGB clump ($[4.6]-[12]>2$ mag). It could possibly indicate that Magellanic RCB stars have thicker circumstellar dust shells than their Galactic counterparts. More studies will be necessary to understand this difference. We display on the bottom-right panel the theoretical colours of RCB stars estimated from SED models. It shows that our selection criteria cut \#2 does not select RCB stars with very cold shell temperatures ($T_{shell}<400$ K) or with hot thin circumstellar shells.

A staggering 99\% of objects, that passed the selection cut \#1, were rejected by that particular criterion, while five known RCB stars were also not selected (i.e. ASAS-RCB-1, EROS2-CG-RCB-12, DY Cen, MV Sgr and MACHO-11.8632.2507), as well as three RCB candidates (i.e. KDM 5651, OGLE-GC-RCB-Cand-1 and [RP2006] 1631). Four of these eight objects, DY Cen, MV Sgr, MACHO-11.8632.2507 and OGLE-GC-RCB-Cand-1, lie far from the main RCB locus on the $[4.6]-[12]$ versus $[12]-[22]$ diagram. For the first three, this is explained by the presence of a second bright cold dust shell around them. This is not uncommon for an RCB star to have two dust shells around them - Radiative transfer modelling of seven RCB stars with Far-IR data found that all of them have two dust shells \citep{2018AJ....156..148M}. ASAS-RCB-1 and the candidate [RP2006] 1631 lie just outside the selection limit as they both have a cold thick shell ($T_{shell}\sim400$ K or colder). On the contrary, EROS2-CG-RCB-12 was eliminated because of its warm thin shell. Finally, the RCB candidate KDM 5651 was found with a negative $[12]-[22]$ colour, at odds with typical values found for known RCB stars. However, its [22] magnitude was reported with a very large error of 0.4 mag indicating a large uncertainty on its [22] brightness. The $[22]$ measurement reported in the subsequent WISE ALLWISE catalogue was corrected from this obvious photometric bias, and we found that KDM 5651 would then have passed selection cut \#2.


\begin{figure*}
\centering
\includegraphics[scale=0.36]{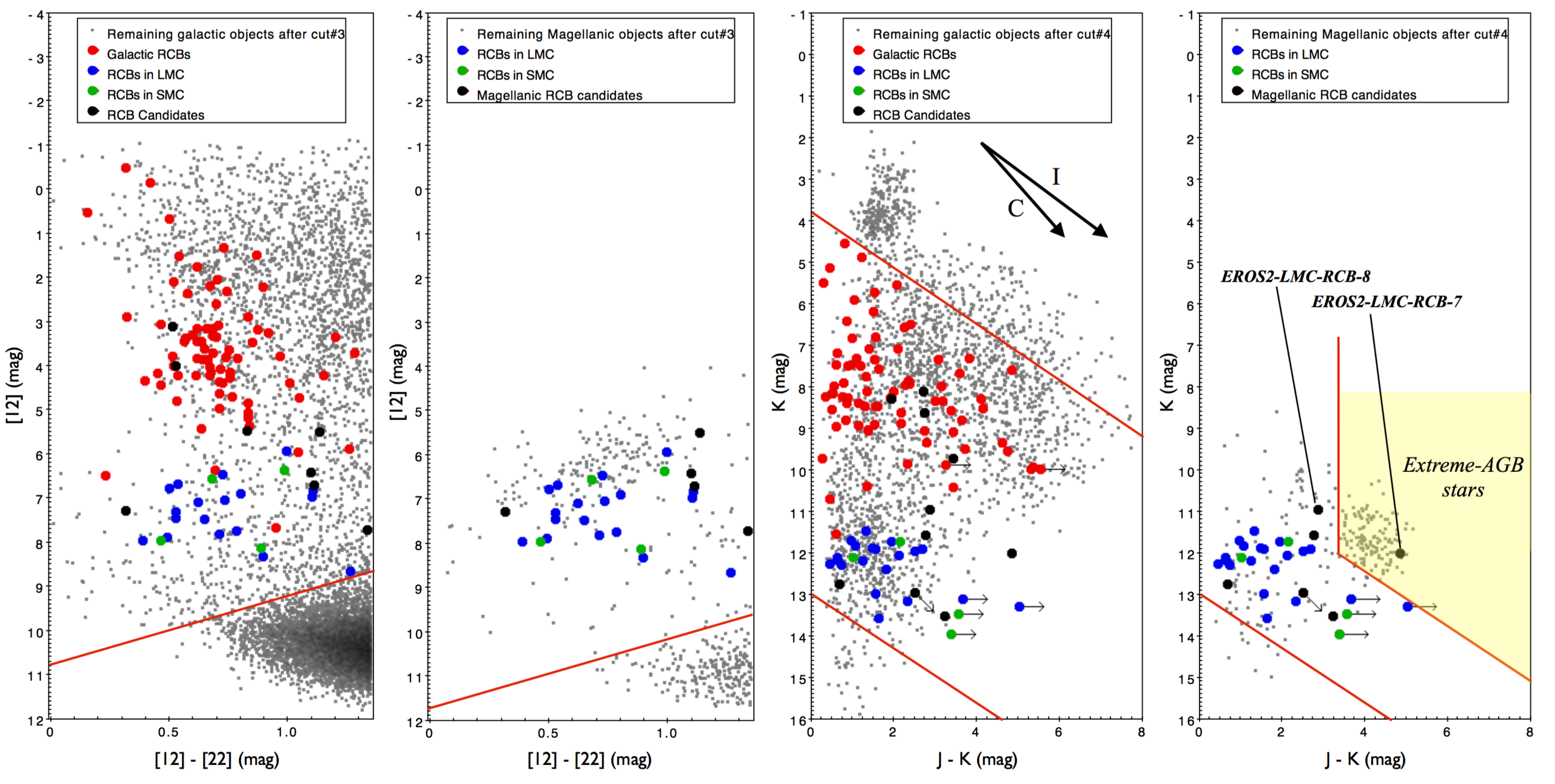}
\caption{Colour-Magnitude diagrams [12] versus $[12]-[22]$ and K versus $J-K$ for respectively the remaining Galactic and Magellanic objects sample selected after cut \#3. The selection areas of cut \#4 and cut \#5 are delimited with red lines. All known and candidate RCB stars are indicated also as a reference sample. The area where Magellanic extreme AGB stars were selected to be represented in Fig.~\ref{figcut_B} is indicated in orange. See discussion in Section~\ref{sec_mainana}, cut \#4 and cut \#5 for more details. The interstellar (I) and circumstellar (C) reddening effect are indicated with arrows.}
\label{figcut_C}
\end{figure*}

\textbf{Cut 3}: This criterion is less efficient than the previous ones, as it rejects only 20\% of the remaining sample (after cut \#2), but it is valuable as it targets mainly the more common AGB stars. We explain the reason of this selection with the two panels presented in Fig.~\ref{figcut_B}. There, we illustrate the distribution of RCB stars and of the remaining sample in the colour-colour $J-K$ versus $J-[12]$ diagram. For a better understanding, one can also look at figure 7 presented by \citet{2012A&A...539A..51T}, where the distributions of objects coloured coded with their respective SIMBAD classification are shown.

In the left-hand panel, the RCB SED models are colour-coded by circumstellar dust shell temperature. One can recognize that not all known RCB stars are located within the area covered by these models. This is the case for about a quarter of them, that have a $J-K$ colour index higher than about 3 mag. This is due to the combination of interstellar reddening and the characteristic high reddening that occurs when newly formed carbon dust clouds are on the line of sight and therefore obscure suddenly the photosphere. RCB stars that have been observed undergoing such rapid decline events, are found to have the highest $J-K$ colour index. This is why also the RCB stars with upper limit values in either the J or K 2MASS bands are all located outside the area covered by the RCB SED models. In the right-hand panel, we added the distribution of Magellanic extreme AGB stars (see cut \#5 and Fig.~\ref{figcut_C} for an explanation of the selection of these extreme AGB stars used here for illustration) as well as the limits of the selection criteria. This selection was designed to remove a high fraction of the extreme AGB stars. However, above a $J-K$ colour index of 3.5 mag, our sample will still be contaminated by such objects as we designed our cuts to keep also RCB stars reddened by decline events. On the other side of the diagram, our criteria have the negative effect of removing potential RCB stars that possess a warm dust shell ($T_{shell}>900$ K), in the following scenarios: RCB stars with thick shells and, RCB stars observed during a high extinction phase or located near the Galactic plane. This is the reason why two known RCB stars did not pass the selection, namely, MACHO-308.38099.66 and EROS2-CG-RCB-12 (which was already rejected by cut \#2). Two RCB candidates, OGLE-GC-RCB-Cand-2 and EROS2-LMC-RCB-8, were also rejected in a similar way. However, the situation here may be different as they are located in the diagram at the expected position of the AGB stars locus, and their respective light curves show unusually large amplitude oscillations for RCB stars. A more detailed discussion on their possible nature is given in Section~\ref{sec_cand}. Here are the limits of the selection applied:

\begin{equation}
$$ J-K\leqslant2.2 \hspace{2 mm} \mathrm{if}\hspace{2 mm} (J-[12])<6$$
\end{equation}

\begin{equation}
$$ J-K\leqslant(3.0\times (J-[12])-7.0)/5 \hspace{2 mm}\mathrm{if}\hspace{2 mm} (J-[12])\geqslant6$$
\label{eq.cut3}
\end{equation}

\textbf{Cut 4}: At this stage, we separate the selected sample into two groups: the objects detected towards the Magellanic Clouds and the Galactic sample. It was a necessary split as the following selection is on apparent magnitude.

The first selection focuses on the shell brightness using the [12] band as an indicator. The two panels on the left side of Fig.~\ref{figcut_C} represent the colour-magnitude diagrams [12] versus $[12] - [22]$ for both samples. The applied selection criteria are listed below in eq.~\ref{eq.cut4} and are represented on each panel. They correspond to faint limits and are used only to remove a group of faint red objects that are almost certainly all galaxies. For the Magellanic sample, we decided to be less restrictive by 1 mag as the number of objects added was small.
	
These criteria remove about 75\% of remaining objects selected after cut \#3. None of the known RCB stars were eliminated at that stage.
	
\begin{eqnarray}
&& Galactic : [12]\leqslant -1.53\times ([12]-[22])+10.76 \nonumber \\
&& Magellanic : [12]\leqslant -1.53\times ([12]-[22])+11.76
\label{eq.cut4}
\end{eqnarray}

\textbf{Cut 5}: Here we apply a selection cut on the photospheric brightness. We used the K band as an indicator as it is the photometric band that is the least affected by possible declines due to newly formed dust clouds that obscure the photosphere. The bright and faint limits are shown in the two colour-magnitude diagrams represented on the right side of Fig.~\ref{figcut_C} and are also listed in eq.~\ref{eq.cut5} below. For the Magellanic sample, the bright limit was truncated at $J-K >3.3$ mag to keep potential interesting Galactic RCB candidates located on the line of sight of the Magellanic Clouds but also to remove a group of extreme AGB stars concentrated in the bright red side of that diagram. About 28\% of the remaining objects were rejected at this stage, while keeping all known RCB stars. Only the Magellanic RCB candidate, EROS2-LMC-RCB-7, did not pass these selection criteria being located within the group of extreme AGB stars.

\begin{eqnarray}
&& Both\hspace{1 mm}samples: K\geqslant0.68\times (J-K)+4.0 \nonumber \\
&& Galactic: K\leqslant0.68\times (J-K)+13.0 \nonumber \\ 
&& Magellanic: K\leqslant0.68\times (J-K)+9.7 \hspace{2 mm}\mathrm{if}\hspace{2 mm} (J-K)\geqslant3.3 
\label{eq.cut5}
\end{eqnarray}

\textbf{Cut 6}: We apply here a selection using the last unused filter, the [3.4] WISE band. In a colour-colour $[3.4]-[4.6]$ versus $[12]-[22]$ diagram, we found that there was a possibility to clean up a bit more our sample by keeping all objects redder than $[3.4]-[4.6]>0.2$ mag. This criterion targets objects that one cannot justify to keep in the final catalogue because they are bluer that the bluer edge of all SED models calculated and of any known RCB stars (see Fig.~\ref{fig_W1W2W3W4}). Only about a hundred objects were eliminated with this criterion.

\textbf{Cut 7}: Finally, we studied the distribution of distances
between the 2MASS and WISE catalogues \citep{2012wise.rept....1C}. The RMS accuracy of the reconstructed WISE positions with respect to 2MASS for unsaturated sources with the signal-to-noise ratio $>50$ in unconfused regions of the sky is approximately 200 mas on each axis. Our targets are very bright stars located mostly in a crowded environment. We decided to apply no cuts if the stars were brighter than $[12]<4$ mag as we observed a wide range of variation for the association distance, however, for fainter objects we requested that the association distance to be lower than one arcsecond. Only 46 objects were rejected. 

\subsubsection{Description of the specific criteria targeting objects listed with upper limits \label{sec_upperlimits_ana}}

A majority of the WISE All-Sky objects detected are listed with an upper limit value in at least one of the seven bands (3 2MASS + 4 WISE). This could be due to multiple technical or physical reasons, but nevertheless some of these objects could be potentially interesting. Indeed, we found that 8 of the 101 known RCB stars are reported as such. Four of them are reported with only an upper limit value in the J band (EROS2-SMC-RCB-1, EROS2-LMC-RCB-6, EROS2-CG-RCB-5 and MACHO-6.6575.13); one with an upper limit in the K band (ASAS-RCB-20, with a surprising bright K upper limit value of $\sim$9.9 mag); two known RCB stars present upper limits values in the J and H bands (MSX-SMC-014 and EROS2-CG-RCB-8); and one known RCB star, V1157 Sgr, with no measurements reported in the WISE [12] and [22] bands (see discussion above in cut \#0). We are particularly familiar with the game of hide and seek played by RCB stars. The upper limits for 2MASS measurements
\footnote{The typical completeness limits of the 2MASS catalogue are 15.8, 15.1 and 14.3 mag, respectively for the J, H and Ks bands. It varies by one magnitude depending if observations were carried on towards the Galactic plane or at high Galactic latitude, because of the effects of confusion noise on the detection thresholds \citep{2003yCat.2246....0C}.} 
are in most cases due to observations made during a phase of a large photometric decline (which can go up to $\sim3.0$ mags in J and $\sim1.6$ mag in K during a maximum decline event of $\Delta\sim9$ mag observed in V). The light curves as well as the 2MASS epochs for each of the known RCB stars, listed above, can be found in \citet{2004A&A...424..245T,2008A&A...481..673T,2009A&A...501..985T}. In the particular case of ASAS-RCB-20, we found that its non-detection in the K band is due to a confusion between multiple sources that the 2MASS detector algorithm did not succeed in resolving it.

About 33.2 million objects, brighter than [12]$<12$ mag and detected in both [4.6] and [12] photometric bands, were reported with an upper limit in at least one of the remaining five bands : J, H, K, [3.4] or [22]. We concentrated our analysis on these objects. Understandably, for 29 million of them ($\sim$ 87\%), the [22] band was the only one affected. That is the least sensitive of the WISE photometric bands. We studied the 2MASS and WISE datasets more closely to analyse the reality of the situation and confirmed that when only the J or the [22] bands are reported with upper limit values, the objects are indeed predominantly not detected in these bands. However, in the following specific scenarios, when a measurement was only missing in the H or K bands, or only in both J and K bands, the related objects were in fact usually blended with one or more nearby objects. We have redesigned our selection criteria to suit each of these scenarios. The changes added in the main selection criteria for some specific cases are described below. 

\begin{itemize}[label={},leftmargin=*]		
\item \textbf{Scenario 1}: \textit{Objects listed with an upper limit in the H or K bands}. The H band measurements are used uniquely in selection cut \#1. There, if an object was to be selected at the brightness of the H upper limit value given, it would also remain selected for any fainter magnitude as the object would be shifted in the lower-right side of the IR colour-colour $J-H$ versus $H-K$ diagram used in cut \#1. Therefore, using the H magnitude limit value as a classical measurement when we applied the seven main selection criteria described in section~\ref{sec_mainana} adds no biases to our selection. Concerning the K band, the situation is not the same as for fainter magnitudes than the upper limit value given, an object would shift to the left in the $J-H$ versus $H-K$ diagram and therefore could potentially come out the selection area. As we saw for ASAS-RCB-20, if an object misses only the K band measurement, the issue is most certainly due to a confusion between multiple sources and not to a non-detection. Therefore we used the K band value listed as such through all selection criteria.

\item \textbf{Scenario 2}: \textit{Objects listed with an upper limit in the J band}. First, we did not keep objects that present an upper limit in the J band and are located near the Galactic centre and plane ($|b|<2$ deg and $|l|<60$ deg). We considered that the search of RCB stars in this highly crowded and extinguished area is better suited to the VISTA/VVV \citep{2010NewA...15..433M} and Spitzer/GLIMPSE \citep{2009PASP..121..213C} datasets which have better spatial resolution. Then, outside that sky area, we decided to select objects that would pass all the seven main selection criteria listed in the previous section at the upper limit value given in the J band but also if they would be fainter by up to 1 mag. In practice, we decreased the J luminosity in 0.1 mag steps and tested all criteria each time. If an object was selected in all 11 steps (from 0 to 1 mag), it was validated for the next selection step. Furthermore, we requested that such objects present a unique possibility of association between the 2MASS and WISE catalogues, and that no blend was observed in WISE. On all six known RCB stars whose J band measurements present an upper limit, only three passed these criteria. The three rejected ones are: EROS2-LMC-RCB-6, EROS2-CG-RCB-5 and EROS2-CG-RCB-8. The candidate RCB star [RP2006] 1631, that is also listed with upper limit values in the J and H 2MASS bands, did not pass our selection criteria either.

\item \textbf{Scenario 3}: \textit{Objects listed with an upper limit only in the [22] band}. An object that would be fainter than the given [22] upper limit value would effectively have a bluer shell and would then move in the left side of the colour-colour diagram presented in Fig.~\ref{figcut_A}. By shifting the reddest selection limit of cut \#1 by 1 magnitude to the red:
\begin{equation}
$$[12]-[22] < 2.35\hspace{1mm}and\hspace{1mm}[4.6]-[12]\geqslant1.5\times([12]-[22]) -1.45) $$
\end{equation}
and applying the above criterion directly on the [22] upper limit values given, we have consequently widened the selection area and kept objects that could be up to one magnitude fainter than the [22] upper limit but still be potentially selected by the main selection criterion (see cut \#2). We also added a more stringent shell brightness lower limit by requesting a strict brightness threshold with $[12]< 8$ mag in selection cut \#4. Finally, no blend can be present in WISE and we added a stricter cut on the colour $[4.6]-[12]<3$ mag.
\end{itemize}		

In conclusion, we added 469 targets of interest (397 Galactic, 72 Magellanic) to our final catalogue (about 20\% of the total). They all present at least one upper limit value in the J, H, K, [3.4] or [22] bands. Most of them will be listed in the low priority group \#4 for spectroscopic follow-up (see Section~\ref{subsec_priority}).


\subsection{Results \label{sec_ToIResult}}

The summary of the selection criteria applied is stated in Table~\ref{tab.Selection}. Out of the sample of 101 Known RCB stars used as a reference, 14 were not selected by our analysis. It corresponds to a detection efficiency of $\sim$85\%. Overall, the selection suits well typical RCB stars with effective temperatures between $4000<T_\mathrm{eff}<8000$ K and surrounded by a thick circumstellar dust shell with a temperature between $400<T_{shell}<900$ K. 

Using RCB SED models, we found that RCB stars that possess one of these three characteristics: (1) a cold circumstellar shell (T$_{shell}<$400 K), (2) a very thin shell as their SED would appear similar to classical F or G stars, or (3) a second colder and thicker shell, like the one seen around MV Sgr \citep{2012A&A...539A..51T}, have a low detection efficiency. Furthermore, we are less efficient at detecting RCB stars that possess a warm shell whose 2MASS epoch coincides with a large decline in brightness or are already affected by a high extinction due to any combination of interstellar and circumstellar dust. The first reason is that RCB stars may not be detected in 2MASS J or H passbands, or both, and secondly, even if they are detected, RCB stars with warm shells would be so red that they may exit the defined selection area defined (see cut \#3 and Fig.~\ref{figcut_C}).

Due to the low resolution of 2MASS and WISE surveys ($>3$ arcsec) and the high interstellar extinction, our detection efficiency most certainly drops in the highly crowded central part of the Galactic bulge. We expect that it is strongly impacted in the sky area $-2<b<2$ deg and $-60<l<60$ deg where the interstellar dust extinction reaches $A_K>3$ mag. Higher resolution IR Surveys such as VISTA/VVV and Spitzer/GLIMPSE could be of great help to probe that crowded part of the sky.
	
Overall, out of 563 million objects catalogued by WISE, our final list of targets of interest, rich in RCB stars, contains 2356 objects, 2194 in the Galaxy and 162 in the Magellanic Clouds. Their distribution on the sky is presented in Fig.~\ref{fig_SpatialDistrib}.


\begin{figure*}
\centering
\includegraphics[scale=0.85]{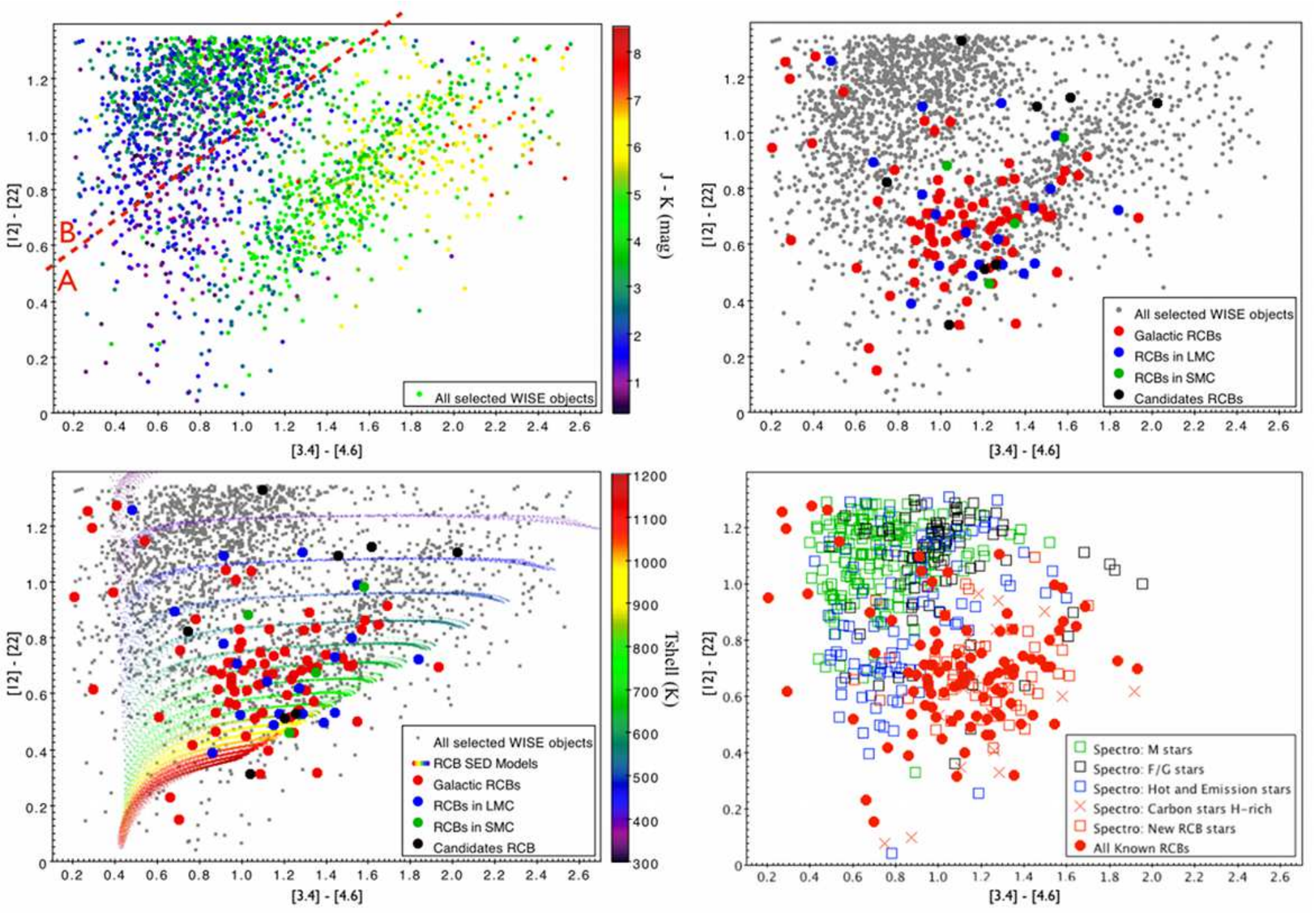}
\caption{Colour-colour $[3.4]-[4.6]$ versus $[12]-[22]$ diagrams. Top-left: distribution of all selected WISE targets of interest that have a valid measurement in all four WISE bands. The diagram is colour-coded with the $J-K$ colour index. Two groups of objects are emerging; we are separating them pragmatically into two zones, A and B (see text for more details). Top-right : the same sample of objects is represented in grey, while all known and candidate RCB stars are overlaid with large dots. Most RCB stars are located in zone A. Bottom-left: same diagram as the preceding except that the RCB SED models are added, colour coded with their shell temperatures. Bottom-right: here all known RCB stars are represented again by large red dots, while the large squares represent all the targets of interest that have already been followed-up spectroscopically. They are colour-coded respective to their spectral classifications. }
\label{fig_W1W2W3W4}
\end{figure*}


\subsection{Prioritisation for spectroscopic follow-up \label{subsec_priority}}

The second stage of our search is to follow-up spectroscopically each of the candidates selected. For that task, we classified them into five different groups of increasing priority. 

We considered as the lowest priority all targets that we found to have a genuine identification in the SIMBAD
database. We classified them in Group \#5. We carefully looked at the information reported for each object selected and marked the ones that either have already been reported with a spectral type, or have been classified as Mira/OH-IR stars from a radio survey or from optical photometric monitoring datasets with strong bibliographic support. We found also that our selection was contaminated by emission stars, RV Tauri stars and also some young stellar objects, T Tauri or pre-main sequence stars. Naturally, objects identified as carbon stars were not included in the low priority group as new observations with the same spectroscopic set-up as our on-going follow-up survey are necessary. We kept all objects from Group \#5 in case of mis-classification. RCB stars are sometimes confused with Mira type stars as they both present light curves with large changes in luminosity (for example, MSX-LMC-1795 was classified as Mira in SIMBAD). Group \#4 contains all objects that have been reported with a upper limit in at least one of the four WISE photometric bands or only in the 2MASS J band. We can expect that these objects are harder to follow-up as they could either be affected by a strong blending with neighbouring stars and/or have a strong extinction due to interstellar or circumstellar dust. At any given epoch, a fraction of RCB stars will be in a significant decline. 
Recently, the prototype R CrB returned to maximum light after a period of 9 years continuous decline, and V854 Cen was in a deep decline lasting at least 50 years before it was discovered \citep{1989MNRAS.238P...1K,1989MNRAS.240..689L}.

The categorisation of the three highest priority groups is explained with the colour-colour $[3.4]-[4.6]$ versus $[12]-[22]$ diagrams presented in Fig.~\ref{fig_W1W2W3W4}. In the top-left diagram, one can clearly recognise two distinct clumps of objects. We define two zones, A and B, to separate them. Interestingly, most of the known RCB stars are found in zone A (see top-right diagram). Using RCB SED models (bottom-left diagram), we understand that RCB stars located in zone B are the ones presenting a thin shell (therefore undergoing a low dust production phase) or having a thicker but colder shell than average ($T_{eff,shell}<500$ K), while, conversely, RCB stars with more typical warm thick shells are expected to be detected in zone A. Furthermore, looking at the distribution of objects classified as Mira or OH-IR stars in SIMBAD, or in our own on-going spectroscopic follow-up survey (see bottom-right diagram), we found that these stars occupy principally the top-left side of zone B. Therefore, we place all objects located in zone B in Group \#3. As it contains the highest number of targets, but is also highly contaminated by Mira type stars and RV Tauri stars, further specific selection within that sample would therefore be useful to not misuse telescope time. However, the targets of Group \#3 should not be disregarded as some less common RCB stars, that are useful to help to understand RCB's evolutionary path, can be discovered amongst them. Monitoring photometric surveys should be a very useful tool to detect and reject Mira type stars from this sample as their characteristic large periodic photometric oscillations are easily recognisable. RV Tauri stars are also expected to be present as some of them, the ones surrounded by dusty discs, show similar WISE colours \citep[see][]{2015MNRAS.453..133G}.

For the two highest priority groups, we separated targets in zone A based on their $J-K$ colour index as redder objects can be more difficult to follow-up due to high extinction, but also because these redder targets are contaminated by extreme AGB stars (see cut \#3 in Sect.~\ref{sec_mainana} and Fig.~\ref{figcut_B}). So, Group \#1 includes all objects located in zone A and bluer than $J-K<3.5$ mag, and Group \#2 includes all redder objects (see top-left diagram). Among the targets listed in priority Groups \#2 and \#4, are some intrinsically bright RCB stars, as they may have been observed in a high extinction phase during the 2MASS epoch and later returned to maximum light. They can also include uncommon highly-enshrouded RCB stars, such as MSX-SMC-014 and EROS2-SMC-RCB-4.

Group \#1 contains only 375 targets (329 in our Galaxy and 46 in the Magellanic Clouds). We consider that it should hold the highest percentage of bona-fide RCB stars. Indeed, of the 87 known RCB stars that passed the selection criteria, $\sim$70\% would be classified in Group \#1, while only $\sim$15\%, $\sim$12\%, $\sim$3\% would classified in Group \#2, \#3 and \#4 respectively. There are respectively 375, 463, 1005, 298 and 215 targets reported in Groups \#1 to \#5. An identification number is given to each target. This identifier was assigned depending on which location, Galactic or Magellanic, and which priority Group the target belongs to. As that corresponds to ten ensembles, each of them was given an interval of 1000 units between 1 and 10000, starting at 1 for the Galactic sample and 5001 for the Magellanic one. For example, the Galactic (Magellanic) Group \#1 targets are ranked between 1 (5001) and 1000 (6000), Group \#2 between 1001 (6001) and 2000 (7000), etc. 


\subsection{A new catalogue of candidate RCB stars \label{subsec_cat}}

We compiled a new catalogue of targets of interest in our search for RCB stars using the all-sky 2MASS and WISE surveys. This new catalogue supersedes the one created by \citet{2012A&A...539A..51T} using the WISE Preliminary data release. Further studies and spectroscopic follow-ups are now needed to classify each star. From the approximately 563 million objects catalogued by the WISE All-Sky survey, we have selected only 2356 targets (2194 in the Galaxy and 162 in the Magellanic Clouds) that present similar near- and mid-infrared colours and brightness to typical RCB stars. We used the 101 known RCB stars as a reference sample and determined a high detection efficiency of 85\%. Furthermore, we found that the HdC star, HD 175893, also passed all our photometric selection criteria. It is in fact not surprising as HD 175893 is the only HdC star known to possess a circumstellar dust similar to RCB stars \citep{2012A&A...539A..51T}. As we have a similar efficiency for selecting RCB stars in the Magellanic Cloud as in the Galactic sample, we conclude that our search should allow us to detect any RCB stars located within $\sim$ 50 kpc from the sun. 

We present a short version of the RCB enriched catalogue in Table~\ref{tab.short_version}. The entire catalogue will be available through the VizieR
catalogue service. 
Each target is listed with its identification number related to its priority group and its own WISE identification. Then in the following order, one will find their equatorial and Galactic coordinates, as well as all their four WISE and three 2MASS magnitudes with their associated 1-sigma errors. The last three columns of the catalogue give information listed by the SIMBAD (as of 2015-08-11) database using a 3 arcsec matching radius, for instance: name, object type and spectral classification. When no information was given, the number -99 is listed in replacement.

To simplify the coordination of the follow-up of each candidate, a dedicated web interface, http://www.rcb.iap.fr, has been created to be able to monitor the status of each star and give access to all relevant information: available light curves, spectra and/or observing charts. The authors invite everyone to report any information that would help them to identify new RCB stars. As of today, we have information that 313 of the candidates are Mira-type stars from their light curves or spectra. These targets do not need additional follow-up and their status is indicated on the website.

\begin{sidewaystable*}
\caption{First 10 rows of the published catalogue. \label{tab.short_version}}
\medskip
\begin{tabular}{cccccccccccc}
\hline
ID  & WISE ID & RA & Dec & l & b &  \multicolumn{2}{c}{WISE} & \multicolumn{2}{c}{WISE} & \multicolumn{2}{c}{WISE} \\
 & & (deg) & (deg) & (deg) & (deg) & $[3.4]$ & $\sigma_{[3.4]}$ & $[4.6]_{corr}$ & $\sigma_{[4.6]}$ &  $[12]$ & $\sigma_{[12]}$ \\
  & &   &   &   &   & mag & mag & mag & mag & mag & mag \\
\hline
\hline
1 & J000842.12+630033.6 & 2.1755086 & 63.0093535 & 118.1051790 & 0.5388797 & 7.875 & 0.023 & 7.088 & 0.020 & 5.521 & 0.014 \\
2 & J001910.61+520203.5 & 4.7942272 & 52.0343285 & 117.8955721 & -10.5215968 & 6.771 & 0.041 & 5.87827 & 0.026 & 4.603 & 0.015    \\
3 & J002921.18+644850.6 & 7.3382829 & 64.8140565 & 120.5838170 & 2.0431257 & 11.433 & 0.024 & 11.046 & 0.023 & 9.784 & 0.043    \\
4 & J003901.29+592805.0 & 9.7554102 & 59.4680639 & 121.3528413 & -3.3655336 & 11.025 & 0.024 & 10.476 & 0.021 & 7.428 & 0.017  \\
5 & J004628.27+585420.8 & 11.6177924 & 58.9057907 & 122.2891557 & -3.9596824 & 11.556 & 0.022 & 10.874 & 0.020 & 9.309 & 0.040   \\
6 & J004822.34+741757.4 & 12.0930919 & 74.2992817 & 122.7203357 & 11.4288814 & 6.679 & 0.040 & 5.17692 & 0.028 & 3.067 & 0.008  \\
7 & J004947.53+633810.0 & 12.4480736 & 63.6361242 & 122.7492110 & 0.7650405 & 9.247 & 0.023 & 8.343 & 0.021 & 6.871 & 0.018   \\
8 & J005128.08+645651.7 & 12.8670235 & 64.9477022 & 122.9351159 & 2.0760360 & 6.911 & 0.033 & 5.63923 & 0.029 & 3.876 & 0.015   \\
9 & J010500.10-054003.6 & 16.2504303 & -5.6676941 & 132.0909095 & -68.2979230 & 8.042 & 0.023 & 7.145 & 0.021 & 5.938 & 0.015 \\
10 & J013000.19+631044.5 & 22.5008221 & 63.1790352 & 127.2660314 & 0.6323574 & 10.320 & 0.024 & 9.821 & 0.020 & 8.673 & 0.030    \\
\hline
\multicolumn{12}{c}{}\\
\multicolumn{12}{c}{Columns continued}\\
\hline
\multicolumn{2}{c}{WISE} & \multicolumn{2}{c}{2MASS} &  \multicolumn{2}{c}{2MASS} &  \multicolumn{2}{c}{2MASS} & \multicolumn{3}{c}{SIMBAD} & \\
$[22]$  & $\sigma_{[22]}$ & J & $\sigma_{J}$ & H & $\sigma_{H}$ & K & $\sigma_{K}$ & Name & Object & Spectral &  \\
  mag & mag & mag & mag & mag & mag & mag & mag &  & type & class &  \\
\hline
\hline
4.622 & 0.024 & 10.285 & 0.018 & 9.603 & 0.017 & 9.059 & 0.020 & -99 & -99 & -99 &  \\
4.163 & 0.023 &  9.501 & 0.022 & 8.192 & 0.017 & 7.187 & 0.018 & V* V858 Cas & sr* & -99 &  \\
9.178 & 0.517 & 12.923 & 0.024 & 12.465 & 0.031 & 11.947 & 0.026 & -99 & -99 & -99 &  \\
7.318 & 0.097 & 13.939 & 0.024 & 12.983 & 0.021 & 12.104 & 0.021 & -99 & -99 & -99 &  \\
8.722 & 0.281 & 12.931 & 0.023 & 12.704 & 0.033 & 12.389 & 0.023 & -99 & -99 & -99 &  \\
2.200 & 0.014 & 10.286 & 0.023 & 8.990 & 0.027 & 7.642 & 0.024 & 2MASS J00482232+7417574 & IR & -99 &  \\
6.126 & 0.049 & 11.135 & 0.024 & 10.733 & 0.030 & 10.208 & 0.020 & -99 & -99 & -99 &  \\
3.367 & 0.022 & 10.936 & 0.027 & 10.066 & 0.030 & 9.177 & 0.019 & IRAS 00483+6440 & * & -99 &  \\
5.654 & 0.038 & 10.643 & 0.024 & 9.386 & 0.025 & 8.378 & 0.025 & [BEM91] 7 & C* & C &  \\
8.000 & 0.261 & 11.329 & 0.022 & 10.774 & 0.028 & 10.248 & 0.025 & -99 & -99 & -99 &  \\
\hline
\end{tabular}
\end{sidewaystable*}


\begin{table}
\caption{Priority classification of the WISE Targets of Interest followed-up spectroscopically.
(See section~\ref{subsec_priority} for a summary of the prioritisation process.) \label{tab.Priority}}
\medskip
\centering
\begin{tabular}{cccc}
\hline
\hline
Priority Id      & ToI Ids          &   \multicolumn{2}{c}{Number of ToI}  \\
for     & Galactic,           &  Galactic & Magellanic  \\
follow-up                  & Magellanic       &  \multicolumn{2}{c}{(with conclusive spectra)}     \\
     &   &  \multicolumn{2}{c}{(with clear Non-RCB variability)}  \\
                         &                      &  \multicolumn{2}{c}{New RCB stars / Candidates}  \\
              
\hline
\hline
     & 1-1000,      &   329  & 46  \\
1 & 5001-6000 &    (135)  & (10) \\
     &                   &   (6)  &  (19)  \\
     &                   &      31  / 13    &  4 / 0   \\
\hline
     & 1001-2000,   &  456  & 7  \\
2 & 6001-7000 &   (36)  & (1) \\
     &                   &   (12)  &  (5)  \\
     &                   &    6  / 0     &   1 / 0  \\
\hline
     & 2001-3000,     &  987  & 18  \\
3 & 7001-8000 &   (248)  & (8) \\
     &                   &   (143)  &  (1)  \\
     &                   &       1 / 0      &  0 / 0   \\
\hline
     & 3001-4000,       &  247  & 51  \\
4 & 8001-9000 &     (0)  & (0) \\
     &                   &   (7)  &  (12)  \\
     &                   &        0 / 0      &  0 / 0   \\
\hline
     & 4001-5000,        &  175  & 40  \\
5 & 9001-10000 &     (41)  & (9) \\
     &                   &   (18)  &  (11)  \\
     &                   &        0  / 1      &  0 / 0   \\
\hline
    & No Id          &     $-$  & $-$ \\
    &                    &     $(-)$  & $(-)$ \\
     &                   &     $(-)$  &  $(-)$  \\
     &                   &        2  / 1      &  0 / 0  \\
\hline
     &                    &         2194      &  162   \\
     &      Total      &        (460)      &  (28)   \\
     &                   &          (186)     &   (48)  \\
     &                    &       40  / 15      &  5 / 0  \\
\hline
\end{tabular}
\end{table}

\section{Spectroscopic data, light curves, and models \label{sec_data}}

\subsection{Spectroscopic data}
The spectroscopic follow-up of the ToI, listed above, was conducted with four telescopes. Table~\ref{tab.Instruments} lists the characteristics of the spectra obtained. Predominantly, we used the Wide Field Spectrograph (WiFeS) instrument \citep{2007Ap&SS.310..255D} attached to the 2.3 m telescope of the Australian National University at Siding Spring Observatory (SSO). WiFeS is an integral-field spectrograph permanently mounted at the Nasmyth A focus. It provides a $25\arcsec \times38\arcsec$~field of view with 0.5 \arcsec sampling along each of the twenty-five $38\arcsec \times1\arcsec$~slitlets. The visible wavelength interval is divided by a dichroic at around 600 nm, feeding two essentially similar spectrographs. The spectra have a two-pixel resolution of 2 $\AA$ and wide wavelength coverage, from 340 to 960 nm. We observed 415 targets with WiFeS during eight observational runs: 20-25 July 2011, 4-7 June 2012, 23-25 July 2012, 1-3 August 2012, 23-30 July 2013, 15-18 and 24 August 2013. We also observed 61 targets with the Goodman spectrograph
mounted on the 4.1 m Southern Astrophysical Research (SOAR) telescope located at Cerro Pach\'on, Chile on 6-8 July 2013. The 600 l/mm grating was used to achieve a 1.3 $\AA$ two-pixel resolution with a 435-702 nm wavelength coverage using the GG-385 blocking filter. All of the data were reduced using the spectral reduction package FIGARO. The telluric lines were removed using the IRAF task, TELLURIC, and standard stars observed during our runs. All spectra were flux calibrated using standard stars observed during the night. We also observed seven targets in automatic mode with the FRODOSpec
instrument mounted on the 2m Liverpool telescope at the Observatorio del Roque de los Muchachos. Using the low-resolution gratings, FRODOspec can obtain the spectra of northern targets between 390 and 940 nm with a two-pixel resolution of 4 $\AA$. Six Magellanic targets were observed with the AAOmega double-beam multi-fibre spectrograph \citep{2006SPIE.6269E..0GS} mounted on the 3.9 m Anglo Australian Telescope (AAT) at SSO as part of an observational campaign by \citet{2014MNRAS.439.2211K,2015MNRAS.454.1468K}.

\begin{table}
\caption{Characteristics of the spectra obtained \label{tab.Instruments}}
\medskip
\centering
\begin{tabular}{cccc}
\hline
Telescope      & Spectrograph          &  Wavelength &  Two-pixel \\
      &           &  range (nm) & resolution ($\AA$) \\
\hline
\hline
SSO/2.3m  & WiFes      &   340-960  & 2  \\
SOAR/4.1m & Goodman  &   435-702   & 1.3  \\
Liverpool/2m   & FRODOSpec  &  390-940  & 4  \\
AAT/3.9m  & AAOmega &  370-880  & 3  \\
\hline
\end{tabular}
\end{table}

\subsection{Light curves}
The photometric data accumulated by monitoring surveys are also of great importance for our search as RCB star light curves show very characteristic photometric brightness changes. In particular, RCB stars are well known to undergo unpredictable fast and deep declines in brightness due to newly formed dust clouds that obscure the photosphere (up to nine magnitudes in V band) in only two or three weeks. The photosphere usually reappears slowly after a few months when the dust grains disperse due to radiation pressure \citep[e.g.][]{2013AJ....146...23C}. However, in some cases, continual dust formation can keep the star in a deep decline for many years. If such fast and deep declines are observed for an RCB star candidate that also presents a carbon-rich spectrum, we use this information to strengthen our decision on its RCB nature. Furthermore, light curves allow us to remove some targets from the candidate list for spectroscopic follow-up, as they show periodic photometric variations typical of Mira and RV Tauri stars. This knowledge allows us to make more efficient use of telescope time.

Here is the list of monitoring surveys whose data were used for the present analysis: OGLE
\citep{2003AcA....53..291U}, CRTS
\citep{2009ApJ...696..870D}, ASAS
\citep{1997AcA....47..467P}, and the Bochum survey \citep{2015AN....336..590H}. The OGLE light curves of some new RCB stars presented in this article will be published in a dedicated article by Przmek et al., 2020.

\subsection{Atmospheric models}

Finally, we used a grid of hydrogen-deficient and carbon-rich MARCS
(Model Atmospheres in Radiative and Convective Scheme) atmospheric models for various $T_\mathrm{eff}$ (from 4000 to 7500 K), surface gravities ($logg =$ 0.5 or 1.0), and nitrogen abundances ($[N]$ from 7.0 to 9.4) \citep{1975A&A....42..407G, 1976A&AS...23...37B,2008PhST..133a4003P,2008A&A...486..951G}. We then created synthetic spectra using the Turbospectrum program \citep{1998A&A...330.1109A,2012ascl.soft05004P} and the Vienna Atomic Line Database (VALD) \citep{2015PhyS...90e4005R} to get information on atomic and molecular transition parameters.


\section{Spectroscopic analysis \label{sec_ana}}

First, we searched for our 2356 ToI in the databases of the monitoring surveys listed above. We found that light curves were available for 510 of them (394 Galactic and 116 Magellanic ToIs), and that 234 show typical photometric oscillations of large periodic variable stars such as Miras (226) and RV Tauri stars (8). We therefore did not observe these ToI spectroscopically to save telescope time, except for 35 ToI where the spectroscopic follow-up was done before the light curves became available.

We have obtained spectra of 488 targets so far.  A detailed summary of the number of spectra obtained for each priority group, as well as the number of objects presenting light curve with large periodic variability unseen in any known RCB stars, is given in Table~\ref{tab.Priority}. We immediately recognised and discarded 263 stars that present non-RCB spectra. Among those, 189 presented spectra of oxygen-rich M stars (Miras), with typical wide absorption features due to the TiO and VO molecules. The large majority of these stars belong to priority group \#3. In the Magellanic Clouds, the number of these targets is low because of the special selection criteria based on the K versus J-K H-R diagram, implemented early on to remove most Magellanic Mira-type objects (see cut \#5 in section~\ref{sec_mainana}). Another 74 were found to have spectra of hot, hydrogen-rich stars. All of these rejected stars are listed in our dedicated search website: http://rcb.iap.fr/trackingrcb/.

\begin{figure*}
\centering
\includegraphics[scale=0.58]{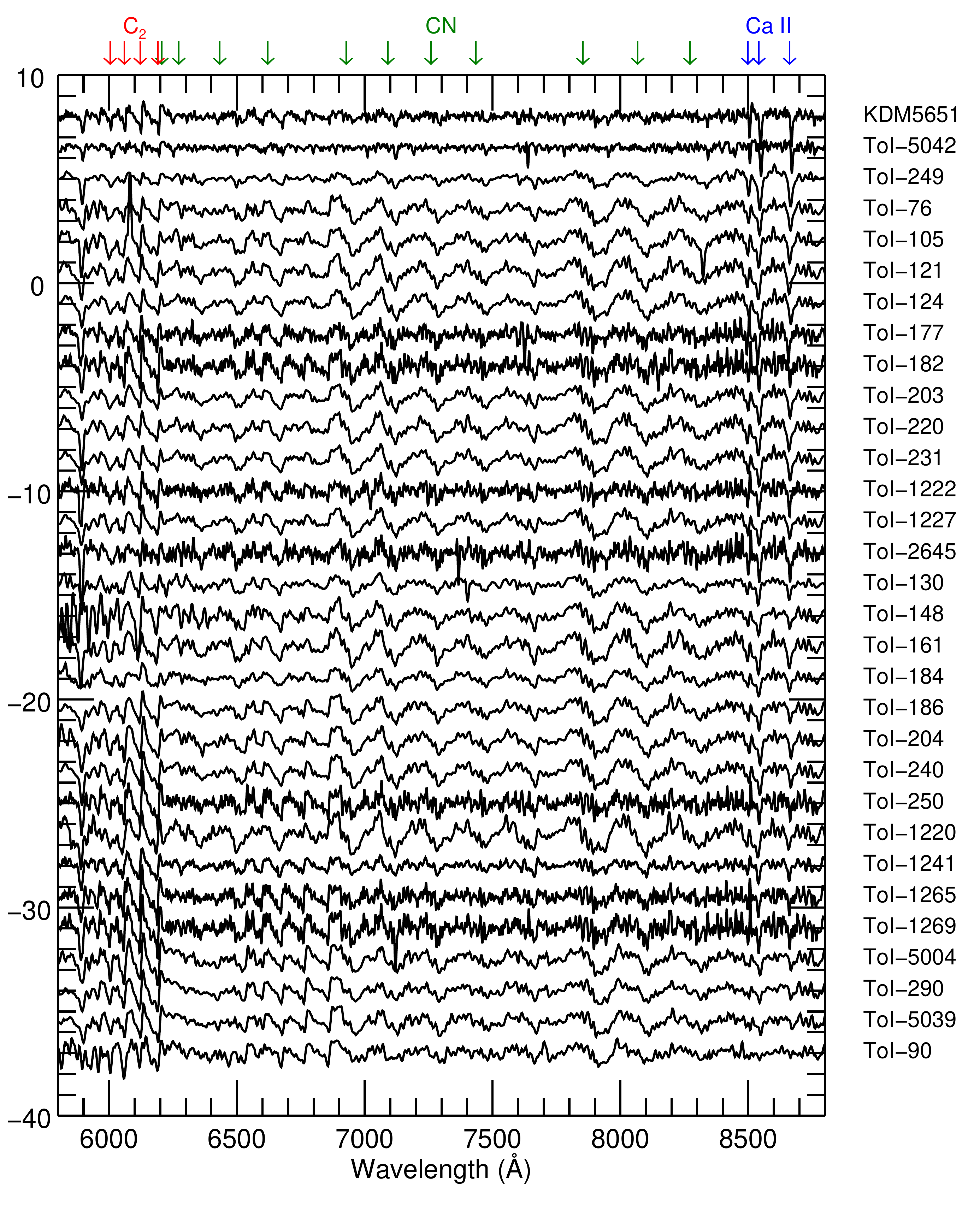}
\caption{Spectra, red region from 5800 to 8800 $\AA$, of KDM 5651 and the 30 newly discovered Galactic Cold RCB stars. The underlying blackbody curves were removed. We plot them in order of the Ca II IR triplet strength, from stronger to weaker lines (top to bottom), that is, in decreasing order of $T_\mathrm{eff}$. The spectra were smoothed (5 points were used) for a better presentation and comparison. The names of the corresponding stars are given on the right side. The ordinate is arbitrary.}
\label{fig_SpectroCoolRCB}
\end{figure*}

\begin{figure*}
\centering
\includegraphics[scale=1.0]{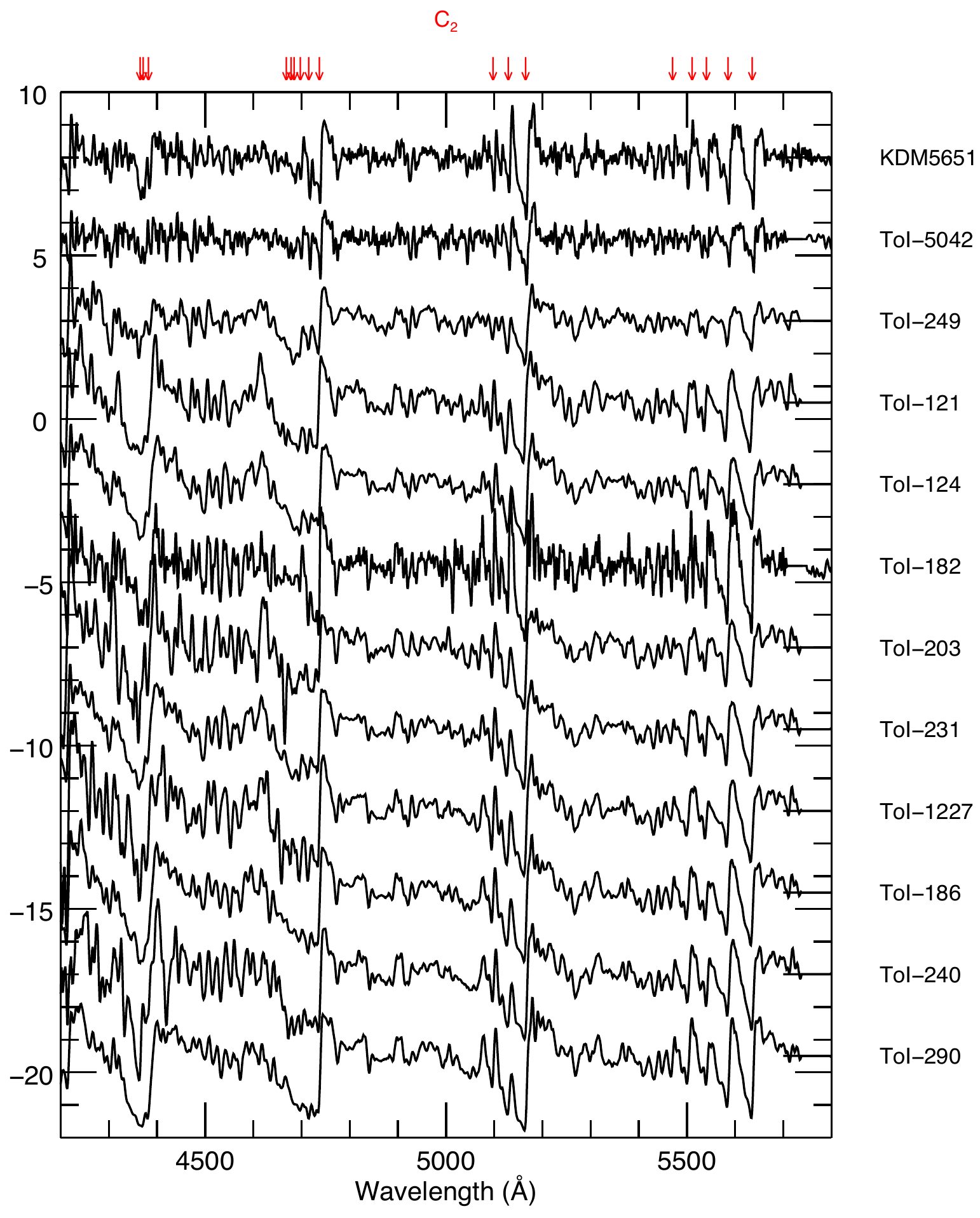}
\caption{Spectra, blue region from 4200 to 5800 $\AA$, of KDM 5651 and a selection of 11 newly discovered Galactic Cold RCB stars that present the best signal-to-noise ratio. The underlying blackbody curves were removed. The large C$_2$ band-heads features (2,0), (1,0), (0,0) and (0,1) are clearly visible. We plotted them in order of the Ca II IR triplet strength, from stronger to weaker lines (top to bottom), that is, in decreasing order of $T_\mathrm{eff}$. The spectra were smoothed (5 points were used) for a better presentation and comparison. The names of the corresponding stars are given on the right side.}
\label{fig_SpectroCoolRCBblue}
\end{figure*}

\begin{figure*}
\centering
\includegraphics[scale=1.0]{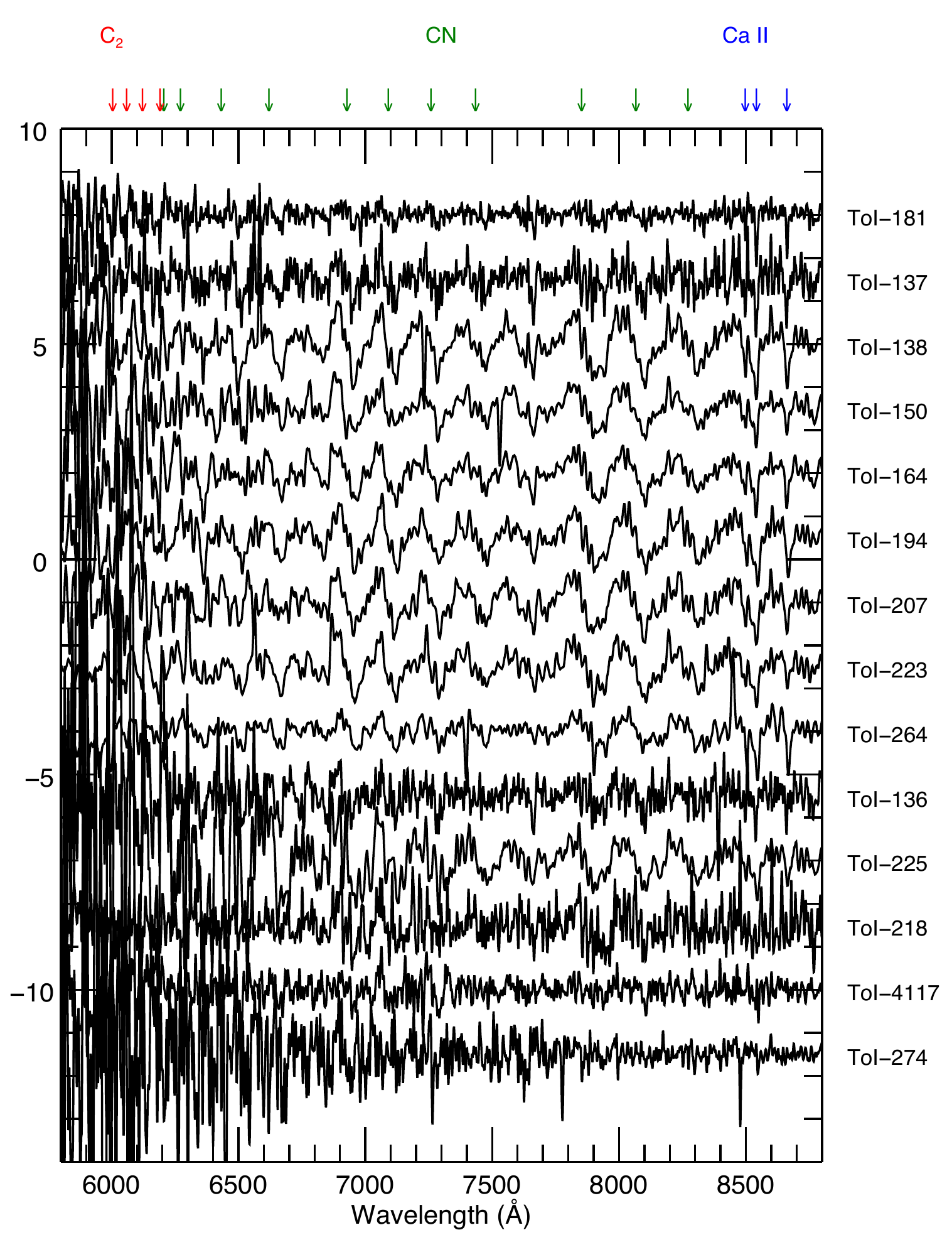}
\caption{Spectra, red region from 5800 to 8800 $\AA$, of the 14 new strong Cold RCB candidates found in this study. The underlying blackbodies curve were removed. The spectra were smoothed (5 points were used) for a better presentation and comparison. The names of the corresponding stars are given on the right side. They are strongly suspected to have been observed during dust obscuration events. We plotted them in order of their Ca II IR triplet strength, from stronger to weaker lines (top to bottom).}
\label{fig_SpectroCoolRCBCand}
\end{figure*}

To classify new RCB stars among the remaining spectra collected, we sorted the spectra into four groups, as RCB stars present a variety of optical spectra. Classical RCB stars are known to possess a wide range of $T_\mathrm{eff}$, from about 4000 to 8500 K, as they evolve through the H-R diagram. Indeed, below $\sim$6800 K, one can clearly observe absorption bands due to the C$_2$ and CN molecules (see Figs.~\ref{fig_SpectroCoolRCB}, ~\ref{fig_SpectroCoolRCBblue} and ~\ref{fig_SpectroCoolRCBCand}). These bands strengthen with lower $T_\mathrm{eff}$. Above the temperature threshold, none of these molecules exist and the spectra consist of atomic absorption lines (Fe I, C I, N I and O I, Ca II, in particular - see Figs.~\ref{fig_SpectroUXAnt} and ~\ref{fig_SpectroWarmRCB} for some examples). So, we classified our observed ToI into two distinct groups, cold and warm RCB stars (Sections ~\ref{subsec_targets_cold} and ~\ref{subsec_targets_warm}), based on the presence or absence of the molecular bands in their spectra. We also added two further ToI groups to search for RCB stars presenting spectra with emission lines. The first group has characteristic emission lines seen when an RCB star undergoes a decline (Section ~\ref{subsec_targets_dust}). The second group represents rare Hot RCB stars ($T_\mathrm{eff}>$15000 K) that present spectra with many strong emission lines (Section.~\ref{subsec_targets_em}).

We have defined and applied specific selection criteria for each of these four groups. These criteria are described in detail in the subsections below. A fifth and last group should also be mentioned, which is only seen when an RCB star is observed during a very deep minimum of a decline. In that particular scenario, only the cold and featureless spectrum of the circumstellar dust shell will be observed. We will discuss that particular case within the decline group (Section.~\ref{subsec_targets_dust}).

Finally, it is worth underscoring here that the vast majority of classical giant carbon stars were excluded during the IR selection process. Only extreme carbon-rich AGB stars are expected to be found within our ToI classified in priority groups \#2 and \#4. These groups were formed to contain ToI that are highly obscured due to thick circumstellar dust ($J-K>$3.5 mag) and therefore potentially more difficult to observe and identify.


All newly discovered RCB stars and all new candidates are listed in Table~\ref{tab.NewRCBcoord}. 

\begin{table*}[hbt!]
\caption{Status on RCB stars candidates listed by \citet{2003MNRAS.344..325M}  \label{tab.Morgan2003}}
\medskip
\centering
\begin{tabular}{cccl}
\hline
Names & Location & Classification & Comments \\
\hline
\hline
\multirow{3}{*}{\parbox{3.1cm}{\object{RAW 21, \newline EROS2-SMC-RCB-1}}} & \multirow{3}{*}{SMC}  &  \multirow{3}{*}{\parbox{2cm}{RCB \newline confirmed}} &  \multirow{3}{*}{\parbox{10.5cm}{A decline was observed in its EROS-2 light curve and no CH band was observed at 4300 $\AA$ \citep{2004A&A...424..245T}.}} \\ 
& &  & \\
& &  & \\
\hline
\multirow{5}{*}{\parbox{3.1cm}{ \object{KDM 2373, \newline EROS2-LMC-RCB-2}}}  & \multirow{5}{*}{LMC} & \multirow{5}{*}{\parbox{2cm}{RCB \newline confirmed}}& \multirow{5}{*}{\parbox{10.5cm}{A slow decline phase was observed by the MACHO survey as already reported by \citet{2003MNRAS.344..325M}. This decline and the following recovery were observed in its EROS2 light curve and no CH band was observed \citep{2009A&A...501..985T}.}} \\
& &  & \\
& &  & \\
& &  & \\
& &  & \\
\hline
 \multirow{12}{*}{\parbox{3.1cm}{\object{KDM 5651}}}  & \multirow{12}{*}{LMC} &  \multirow{12}{*}{\parbox{2cm}{RCB \newline confirmed}} &  \multirow{12}{*}{\parbox{10.5cm}{A series of small declines were observed by the three microlensing surveys, MACHO, EROS-2 and OGLE. A decline of $\sim$ 0.3 mag at JD$\sim$2450200 was detected by MACHO, then subsequently two similar declines of $\sim$ 0.3 mag were also monitored by EROS-2 at JD$\sim$2451800 and $\sim$2452100. The recovery stage of that last decline was observed by OGLE-III \citep[see][OGLE-LMC-RCB-20]{2009AcA....59..335S}. About $\sim$11.8 years later, a stronger decline of $\sim$1.7 mag was observed by the OGLE-IV RCOM survey\footnote{See OGLE-IV RCOM (OGLE Monitoring system of R Coronae Borealis type variable stars) website: http://ogle.astrouw.edu.pl/ogle4/rcom/kdm-5651a.html.}. We obtained a spectrum during our observational campaign and no CH band was observed, confirming a previous analysis made by \citet{1988ApJ...334..135H} of their candidate HC 119 spectrum. We observed absorption features due to C$_2$ and CN molecules typical of Cold RCB stars.}} \\
& &  & \\
& &  & \\
& &  & \\
& &  & \\
& &  & \\
& &  & \\
& &  & \\
& &  & \\
& &  & \\
& &  & \\
& &  & \\
\hline
\multirow{5}{*}{\parbox{3.1cm}{\object{KDM 2492}}}  & \multirow{5}{*}{LMC} & \multirow{5}{*}{\parbox{2cm}{RCB \newline confirmed}} & \multirow{5}{*}{\parbox{10.5cm}{This is the already identified Magellanic RCB star, HV 5637 \citep{2001ApJ...554..298A}. No decline was observed for more than 20 years between the MACHO, EROS-2 and OGLE-III microlensing surveys, but a $\sim$2.4 mag decline was monitored by the OGLE-IV RCOM monitoring system.}} \\
& &  & \\
& &  & \\
& &  & \\
& &  & \\
\hline
\multirow{3}{*}{\parbox{3.1cm}{\object{KDM 7101, \newline EROS2-LMC-RCB-5}}} &  \multirow{3}{*}{LMC} & \multirow{3}{*}{\parbox{2cm}{RCB \newline confirmed}} & \multirow{3}{*}{\parbox{10.5cm}{Two large and rapid declines were observed by the EROS-2 survey \citep{2009A&A...501..985T}. No CH band was observed. }}\\
 & &  & \\
& &  & \\
\hline
\multirow{6}{*}{\parbox{3.1cm}{\object{KDM 6546}}}  &  \multirow{6}{*}{Galactic} &  \multirow{6}{*}{\parbox{2cm}{Confirmed CH star}} & \multirow{6}{*}{\parbox{10.5cm}{We obtained a spectrum during our observational campaign and detected a strong CH band-head at $\sim$4300 $\AA$ (see Fig. ~\ref{fig_KDM6546spectrum}). It was already reported as a CH star by \citet{1988ApJ...334..135H}, HC 193, but as discussed by \citet{2003MNRAS.344..325M}, there was a possible issue due to a positional mismatch of more than 30\arcsec. }}\\
& &  & \\
& &  & \\
& &  & \\
& &  & \\
& &  & \\
\hline
\hline
\end{tabular}
\end{table*}

\begin{figure*}
\centering
\includegraphics[scale=0.58]{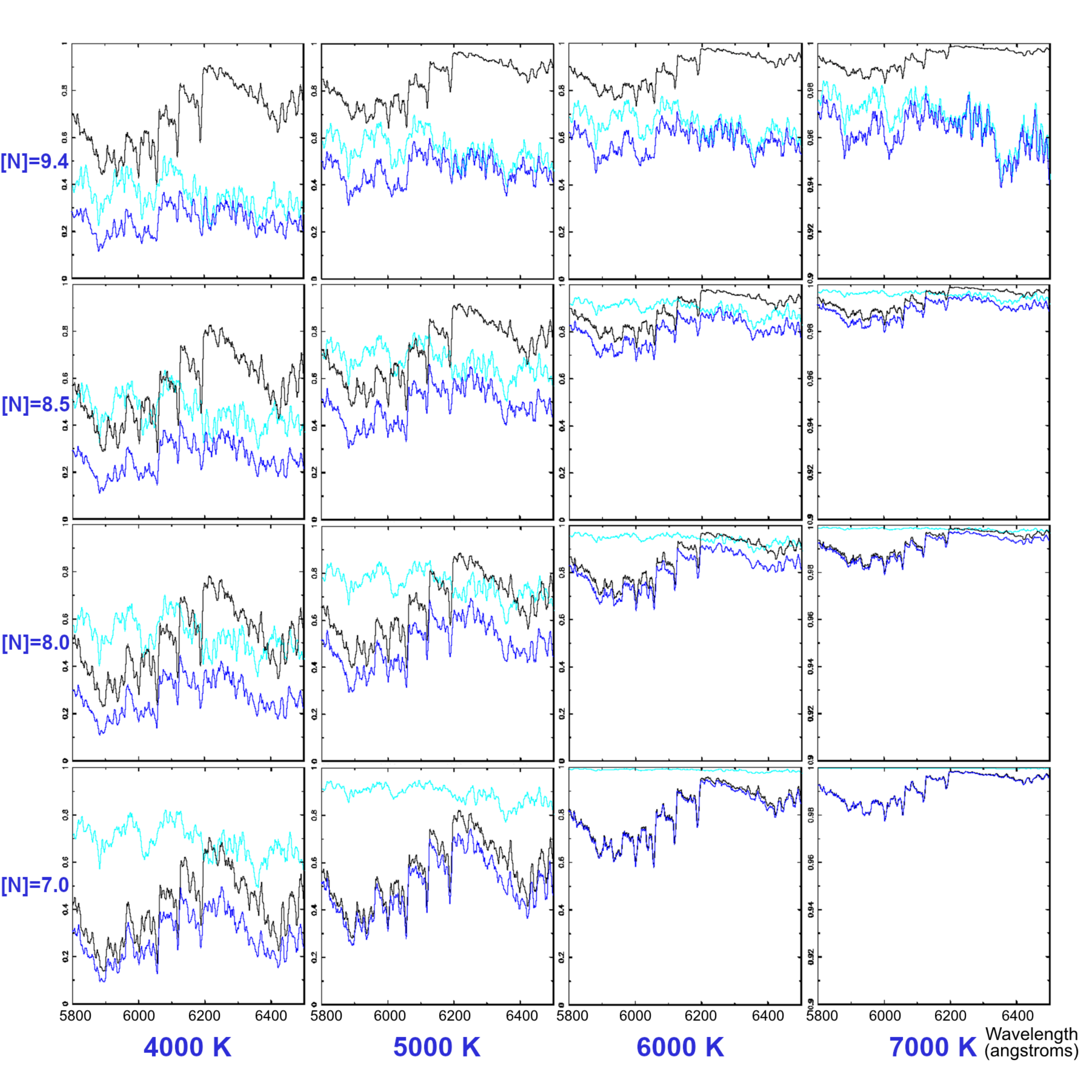}
\caption{Hydrogen-deficient synthetic spectra between 5800 and 6400 $\AA$ for four different nitrogen abundances (7.0 $\leqslant[$N$]\leqslant$ 9.4 dex) and four different $T_\mathrm{eff}$s (4000 to 7000 K). The spectra are normalised to unity. The intensity scales extend between 0 and 1 for all spectra except the ones with a $T_\mathrm{eff}$ of 7000 K whose intensity scales are between 0.9 and 1. Cyan represents the CN band-head features, black represents the C$_2$ bands, and blue is the sum of both. The abundances used for the models are those of a classical RCB star, i.e. [H]=7.5, [He]=11.5, [C]=9.0 and [O]=8.8, and the C/N ratio increased ($\sim$0.4, $\sim$3.0, 10, and 100) by decreasing the nitrogen abundances.}
\label{fig_CNC2bandheads}
\end{figure*}

\subsection{First group: Cold RCB stars \label{subsec_targets_cold}}

We focus here on a group of 72 ToI that show strong band-head features due to C$_2$ and CN, observable in the blue and red of the optical. Cold RCB stars have a $T_\mathrm{eff}$ ranging between $\sim$4000 and about 6800 K, and are the predominant group of RCB stars detected so far (they represent about 2/3 of the entire known RCB star sample). An excellent description of these features and other typical Cold RCB star absorption lines can be found in \citet{1983MNRAS.202P..31B} and \citet{2003MNRAS.344..325M}. They both underline that the main common characteristic observed in the spectra of Cold RCB stars, in addition to their hydrogen deficiency, is their very weak CN bands compared with classical carbon stars, and that as a result the C$_2$ bands between 6000 and 6200 $\AA$ are not degraded and thus are clearly visible. 

\subsubsection{Morgan et al. 2003 RCB stars selection}

In a catalogue of $\sim$8500 Magellanic carbon stars that were observed using the 2dF multi-object low-dispersion spectrograph, \citet{2003MNRAS.344..325M} have identified six stars presenting very weak CN features in their respective spectra. From further studies of these spectra as well as the near IR photometry, they suggested that five of them are strong Magellanic RCB candidates and the sixth, \object{KDM 6546}, presenting slightly stronger CN bands, should probably be a Galactic halo CH star located in front of the Large Magellanic Cloud. 
At the time, due to the lack of light curves and the narrow wavelength coverage of the spectra,
the nature of these six stars remained unconfirmed. An update on their classification status is given in Table~\ref{tab.organ2003}.

All five stars listed as strong Magellanic RCB star candidates by \citet{2003MNRAS.344..325M} are now confirmed RCB stars. It underscores that the spectroscopic analysis criteria described in \citet[Section 4 and references therein]{2003MNRAS.344..325M} can be considered a very reliable method for identifying new RCB stars. Here are the main criteria: firstly, very weak CN bands detected allowing four strong C$_2$ bands to be clearly distinguished between 6000-6200 $\AA$, secondly, no hydrogen Balmer lines detected, and thirdly, no $^{13}$C features observed.

Furthermore, the spectral analysis was supported by the study of each CN-weak star's near-IR luminosity and colour in comparison with other more classical carbon stars and already known RCB stars. \citet{2003MNRAS.344..325M} showed that the known RCB stars stand out compared to the classical carbon stars locus, but also more interestingly that the five candidates follow the same near-IR characteristics as RCB stars. 

All but one, \object{KDM 5651}, have passed our photometric selection criteria. KDM 5651 would have been selected if, as mentioned in Section~\ref{sec_mainana}, cut \#2, its $[22]$ WISE ALL-Sky measurement had not been strongly biased. Interestingly also, the Galactic CH star, KDM 6546, was rejected as it did not pass the first two selection criteria using near- and mid-IR colours.

\begin{figure}
\centering
\includegraphics[width=3.5in]{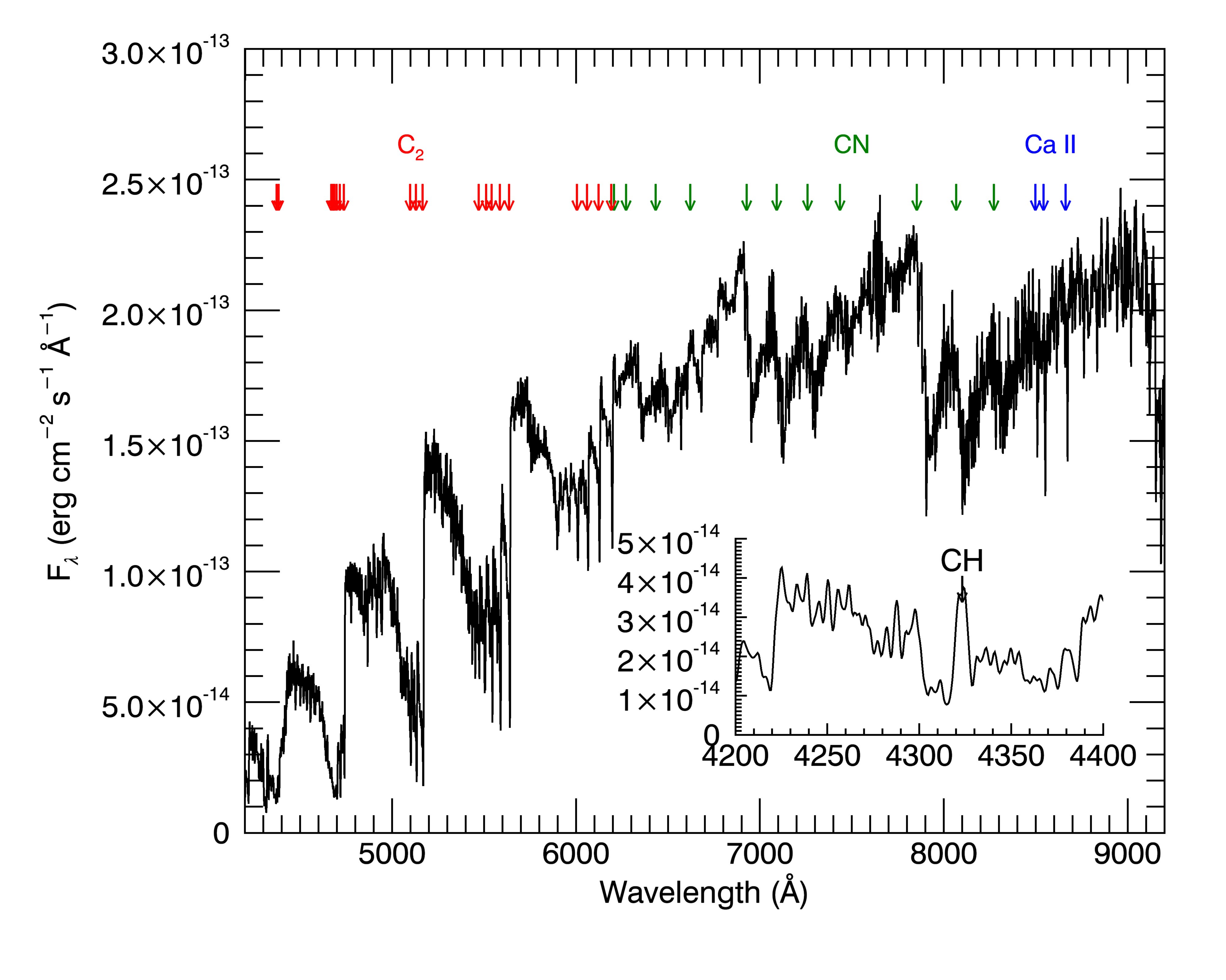}
\caption{Spectrum of KDM 6546 from 3400 to 9600 $\AA$ as observed on 2012-10-02 with the 2.3m/WiFeS spectrograph and R$\sim$3000 resolution. The CH band-head between 4280 and 4310 $\AA$ is clearly visible (see insert spectrum). We confirm the first speculation that KDM 6546 is indeed a Galactic halo CH star located in front of the Large Magellanic Cloud \citep{2003MNRAS.344..325M}. The four consecutive C$_2$ band-heads between 6000-6200 $\AA$ are visible also in the KDM 6546 spectrum as its C/N ratio is higher than classical carbon stars.}
\label{fig_KDM6546spectrum}
\end{figure}

\subsubsection{Our selection of Cold RCB stars \label{subsec_targets_coldselection}}


Dust reddening effects made the spectroscopic analysis of this cooler group of stars more difficult. In many cases, most of the signal detected was concentrated on the red side of the spectra, and some important features located in the blue ($\lambda < 5500$ $\AA$) was missing. In particular, these include the CH molecular band at $\sim$4300 $\AA$ (only 11 out of the 73 cold ToI have some signal that far into the blue), but also, the (1,0) $^{13}$C$^{12}$C absorption line at 4744 $\AA$ and the nearby (1,0) $^{12}$C$^{12}$C at 4737 $\AA$. Fortunately, the $^{13}$C abundance can also be estimated from the red side of the spectrum with the (1,3) and (0,2) $^{13}$C$^{12}$C bands, respectively at 6100 and 6168 $\AA$, which are near a series of $^{12}$C$^{12}$C bands, and also the $^{13}$CN band-head at $\sim$6260 $\AA$ with the nearby (4,0) $^{12}$CN band-head at 6210 $\AA$ if nitrogen abundance is high enough for the CN bands to be detectable. Most RCB star atmospheres are known to have a high C/N abundance ratio and a low $^{13}$C/$^{12}$C isotopic ratio, but this is not always the case \citep{2008MNRAS.384..477R,2012ApJ...747..102H}.

It has always been possible to detect the presence or absence of Balmer lines to test for hydrogen deficiency. RCB stars generally show weak or absent Balmer lines, but at least one RCB star, V854 Cen, shows significant hydrogen lines \citep{1989MNRAS.240..689L,1989MNRAS.238P...1K}. It is also now possible to detect the Ca II IR triplet located in the red ($\lambda\sim$8498, 8543 and 8662 $\AA$). Indeed, the intensity of these lines is, as shown by \citet{1971ApJ...167..521R}, a good indicator of carbon star temperature: the cooler the temperature, the weaker the lines. We have therefore empirically classified each of the 73 candidate RCB stars based on the Ca II IR triplet strength: none, weak, small, medium, and strong.

The CN bands observed in RCB stars are weaker than the ones observed in classical carbon stars. As discussed in the previous section, this allows the possibility of detecting four C$_2$ band-heads between 6000 and 6200 $\AA$ that are normally swamped by strong CN bands nearby. We studied synthetic spectra for a range of $T_\mathrm{eff}$ and C/N abundance ratios, and found that these C$_2$ bands indeed become detectable when the C/N ratio is higher than 3 and the $T_\mathrm{eff}$ ranges between 5000 and 6000 K (see Fig.~\ref{fig_CNC2bandheads}). The second C$_2$ band-head, around 6000-6060 $\AA$, fades first with a higher nitrogen abundance due to a strong CN band-head appearing nearby. At higher temperatures, $\sim$6000 K, the four C$_2$ band-heads can be seen even for lower C/N ratios as the CN bands vanish. On the other hand, at a low temperature like $\sim$4000 K, the C/N ratio needs to be higher than 100 to begin to observe these band-heads. For a large majority of Cold RCB stars, the observation of these C$_2$ features, rarely seen in classical carbon stars, are therefore common and can be used as a simple identification criterion. This criterion is not infallible, so for a very few RCB stars, like V4017 Sgr, there are no C$_2$ band-heads that are easily detectable in this wavelength range. We also note that after looking closely at the spectra of known RCB stars, we found that selecting stars based only on the extreme weakness of the CN bands, as did \citet{2003MNRAS.344..325M}, was too strict a criterion to detect all of the RCB stars.

Using already known RCB stars and synthetic spectra matching their $T_\mathrm{eff}$, we established a series of criteria to reveal new RCB stars. We describe these criteria below.

Among the 73 cold candidates, we recognised that 13 of them show H$_{\alpha}$ emission indicating that they are not hydrogen-deficient. They are most certainly carbon-rich AGB stars which show a cooler $T_\mathrm{eff}$ than RCB stars ($T_\mathrm{eff}< 3500$ K). This is confirmed by the strength of the Ca II IR triplet which is non-existent. These stars also present a very high IR excess and are located in the extreme top-right of the $J-H$ versus $H-K$ colour-colour diagram presented in \citet[Fig.6]{2012A&A...539A..51T}, where AGB carbon-rich stars are expected to be found. We rejected these stars. Another seven stars present strongly reddened spectra, but no H$_{\alpha}$ emission. These are also carbon stars due the presence of CN bands at wavelengths longer than 7900 $\AA$. They also show no Ca II IR triplet. So, these objects are most certainly carbon-rich AGB stars. We do not consider them further in our analysis.

Of the 60 remaining targets, 53 of them do not show any H$_{\alpha}$ emission or CH absorption. First, we focused on the 40 warmer ones, which are the ones showing Ca II IR triplets classified as small, medium, or strong. They are interesting ToI as RCB stars are known to be warmer than classical carbon stars. Among them, we recognise that 27 presented some C$_2$ features between 6000-6200 $\AA$ and no substantial $^{13}$C signal. We consider all these 27 stars as new bona fide RCB stars. Moreover, for eight of them, we even had enough signal in the blue spectrum to confirm the absence of the CH band which strengthens their identification. They are our "golden" Cold RCB sample as they match all of the following criteria: pass the near- and mid-IR photometric RCB selection, have a warm $T_\mathrm{eff}$ based on Ca II strength, no H$_{\alpha}$ emission or CH band detected, C$_2$ features between 6000-6200 $\AA$ observed and no substantial amount of $^{13}$C found. Finally, we found light curves showing characteristic RCB star declines for nine out of these 27 new bona fide RCB stars.


Second, we looked at the other group of 13 cooler candidates, presenting only weak or absent Ca II IR triplets. Their classification revealed to be more difficult as many of these stars show very noisy spectra at the interesting 6000-6200 $\AA$ wavelength range. However, first, we identified three ToI, numbers 90, 290, and 5039, presenting spectra with the same characteristics as listed above for the warmer RCB stars. Moreover, their CN bands were found to be very weak and typical RCB light curves were obtained for ToI 290 and 5039. We include all three of them in our list of new bona fide RCB stars. ToI 90 and 290 can also be included in our Cold "golden" sample as no absorption from CH molecules was detected.

Then out of the remaining ten cooler candidates, we kept seven of them: numbers 136, 138, 150, 164, 181, 207, and 225. They should remain as RCB candidates, as further analysis and observations will be needed before final confirmations can be made. These stars present very similar red carbon-rich spectra with the following characteristics: no detectable signal in the blue, low fluxes in the C$_2$ region of interest (6000-6200 $\AA$) and visible CN bands at wavelengths longer than 6900 $\AA$. It was also impossible to determine the abundance of $^{13}$C. Nevertheless, we found that the two strongest C$_2$ band-heads, (0,2) and (1,3), were visible in all cases and that CN bands around $\sim$6210 $\AA$ seem very weak or non-existent. They have weak Ca II IR triplet absorption and no H$_{\alpha}$ was detectable (except for ToI 136, which shows H$_{\alpha}$ and [O I] at 6300 $\AA$ in emission). Interestingly also, we note that all these ToI are part of the high priority group \#1, a group where no classical AGB carbon-rich stars are expected to be found.

Now we discuss the particular case of ToI 218, also belonging to the priority \#1 selection group. It also presents a very red spectrum and as for the previous seven stars, we could not determine its abundance of $^{13}$C or the presence or absence of CH, but again no H$_{\alpha}$ in absorption or emission was found indicating hydrogen-deficiency. The main difference is that ToI 218 shows no Ca II IR triplet at all and no sign of C$_2$ features but only wide CN features at wavelengths longer than 6900 $\AA$. Based on these characteristics, we would not have kept this star as a strong RCB candidate; however, there are light curves from the EROS-2 and OGLE surveys showing characteristic RCB-like declines. As discussed above, detecting the C$_2$ features around the 6000-6200 $\AA$ region is a common observation for Cold RCB stars but it is not mandatory as already seen with the known RCB, \object{V4017 Sgr}. These C$_2$ features are harder to detect at temperatures as low as 4000 K (see Fig.~\ref{fig_CNC2bandheads}) as it would need a higher C/N ratio. Then, ToI 218 could indeed be an RCB star with a low $T_\mathrm{eff}$, as indicated by the absence of the Ca II IR triplet. Another possible explanation for the absence of these Ca II absorption lines is that we could have observed ToI 218 in a decline phase and therefore emission in the Ca II lines fill in the absorption lines. More details on emission lines observed in Cold RCB spectra are discussed in section~\ref{subsec_targets_dust} below. We need to obtain another spectrum, perhaps at a brighter phase, to definitely confirm the status of ToI 218. We list it as a strong candidate.

The spectra of all new RCB stars are presented in Figs.~\ref{fig_SpectroCoolRCB} and~\ref{fig_SpectroCoolRCBblue} in decreasing order of the Ca II IR triplet strength (i.e. in decreasing order of $T_\mathrm{eff}$). One can see that the C$_2$ features in the interesting region between 6000 and 6200 $\AA$ are visible in all cases, but also that the next five C$_2$ band-heads located between 6500 and 6900 $\AA$ are getting stronger with the decreasing effective temperature. The strength of the CN band around $\sim$6210 $\AA$ is different for each new RCB star, confirming that selecting RCB stars based on the extreme weakness of the CN bands cannot be a strong selection criterion. For example, the CN band around $\sim$6210 $\AA$ is weak for ToI 290 and 5039, but much stronger in the case of ToI 1220 and 1241.

Finally, we note that ToI 5042, a member of our "golden" Cold sample, shows absorption bands due to C$_2$ and CN, as well as many absorption lines that are typical of Warm RCB stars. We discuss these lines below. ToI 5042 is the warmest of our 30 Cold new RCB stars (see Figs.~\ref{fig_SpectroCoolRCB} and ~\ref{fig_SpectroCoolRCBblue}).

\begin{table*}[hbt!]
\caption{Spectroscopic analysis of new Cold RCB stars \label{tab.NewRCBCold}}
\medskip
\centering
\begin{tabular}{ccccccccl}
\hline
\hline
WISE  &  WISE & 2MASS &Ca II IR   & $^{13}$C ?     & Strength C$_2$ & Hydrogen ?& Emission & Comments\\
ToI  &   [12] & K &  triplet  &  $\sim$4744,      & features        & H$_{\alpha}$, CH  & lines ? ($\AA$) &\\
        &   mag & mag &  strength &  $\sim$6168 $\AA$   & 600-620 nm         & &\\
\hline
\hline
\multicolumn{9}{c}{New Cold RCB stars}\\
\hline
76 & 3.55 & 8.32 & Medium & n/a, No & Strong & No, n/a &  & \\
90 & 5.04 & 8.82  & None & n/a, No & Strong & No, No &  & \\
105 & 4.41 & 9.46  & Medium & n/a, No & Small & No, n/a &  &  \\
121 & 4.10 & 8.54  & Medium & No, No & Strong & No, No & &  \\
124 & 4.96 & 9.20  & Medium & No, No & Strong & No, No &  &   \\
130 & 2.22 & 8.04  & Small & n/a, No & Small & No, n/a &  &   \\
148 & 5.07 & 9.26  & Small & n/a, No & Strong & No, n/a &  &   \\
161 & 4.68 & 8.84  & Small & n/a, No & Strong & No, n/a &  &  \\
177 & 2.87 & 7.86  & Medium & No, No & Strong & No, n/a &  &   \\
182 & 3.22 & 7.44 & Medium & No, No & Strong & No, No &  &     \\
184 & 3.64 & 8.50  & Small & n/a, No & Small & No, n/a &  &    \\
186 & 4.46 & 8.65  & Small & No, No & Strong & No, No &  &     \\
203 & 4.76 & 8.37  & Medium & No, No & Strong & No, No & &     \\
204 & 3.10 & 8.13  & Small & n/a, No & Strong & No, n/a &  &  \\
220 & 7.04 & 9.72  & Medium & No, No & Strong & No, n/a & &   \\
231 & 5.34 & 9.72  & Medium & No, No & Strong & No, No &  &    \\
240 & 3.54 & 7.48  & Small & No, No & Strong & No, n/a &  &   \\
249 & 3.66 & 7.56  & Strong & No, No & Strong & No, No &  &    \\
250 & 3.68 & 8.67  & Small & No, No & Strong & No, n/a &  &   \\
290 & 4.81 & 8.86  & Weak & No, No & Strong & No, No &  &    \\
1220 & 2.86 & 8.26  & Small & n/a, No & Strong & No, n/a &  &   \\
1222 & 2.95 & 9.38  & Medium & n/a, No & Strong & No, n/a &  &  \\
1227 & 4.53 & 9.91  & Medium & No, No & Strong & No, n/a &  &  \\
1241 & 3.04 & 8.65  & Small & n/a, No & Small & No, n/a &  &  \\
1265 & 2.84 & 9.30  & Small & No, No & Strong & No, n/a &  &   \\
1269 & 2.93 & 9.45  & Small & No, No & Strong & No, n/a &  &   \\
2645 & 5.07 & 7.95  & Medium & n/a, No & Weak & No, n/a &  &    \\
5004 & 6.72 & 12.36  & Small & n/a, No & Strong & No, n/a &  &    \\
5039 & 6.41 & 13.54  & Weak & No, No & Strong & No, n/a & [O III]$_{(4959+5007)}$,  & Emission lines observed  \\
				& &  &      	&    &            & &    H$_{\alpha,\beta,\delta,\gamma}$, [N II]$_{(6583)}$,  & only on the AAOmega \\
				& &  &        &    &            & &   [S II]$_{(6719+6730)}$   & spectrum   \\
5042 & 8.65 & 12.26  & Medium & No, No & Strong & No, No &  & Warm absorption lines\\
  &   & &  &   &   &  & & observed with some C$_2$ \\
  &   & &  &   &   &  & & features, but no CN \\
KDM 5651 & 8.96 & 12.75  & Strong & No, No & Strong & No, No &  &   \\
\hline
\hline
\multicolumn{9}{c}{New Cold RCB candidates that need further monitoring}\\
\hline
136 & 4.83 & 9.66  & Weak & n/a, n/a & Small & Em., n/a &  & \\
137 & 4.66 & 9.54  & Small & n/a, n/a & Weak & Em., n/a & [N II]$_{(6583)}$ & \\
138 & 4.76 & 9.67  & Small & n/a, n/a & Weak & No, n/a & [N II]$_{(6583)}$  &   \\
150 & 4.67 & 9.14  & Small & n/a, n/a & Weak & No, n/a &  &   \\
164 & 3.41 & 8.36  & Small & n/a, n/a & Small & No, n/a &  &   \\
181 & 2.99 & 8.73  & Medium & n/a, n/a & Weak & No, n/a &  &   \\
194 & 2.94 & 7.57  & Small & n/a, n/a & No & No, n/a & Na I D &  \\
207 & 3.08 & 7.93  & Small & n/a, n/a & n/a & No, n/a & & \\
218 & 6.25 & 11.07  & None & n/a, n/a & n/a &  No, n/a &   & \\
223 & 1.25 & 7.27  & Small & n/a, No & Strong & Em., n/a & Na I D & \\
225 & 3.25 & 7.95  & Weak & n/a, n/a & n/a & No, n/a &  & \\
264 & 5.88 & 10.90  & Small & n/a, n/a & Weak & No, n/a & Na I D  &   \\
274 & 1.03 & 6.33  & Small & n/a, n/a  & n/a   & Em., n/a &  & 2.3m/WiFeS featureless  \\
		&        &           &         &                &          &              &    &  spectrum  \\
4117 & 2.33 & 7.64 & None & n/a, n/a & Weak & No, n/a & Na I D   &  \\
\hline
\hline
\end{tabular}
\end{table*}

\begin{figure*}
\centering
\includegraphics[width=7in]{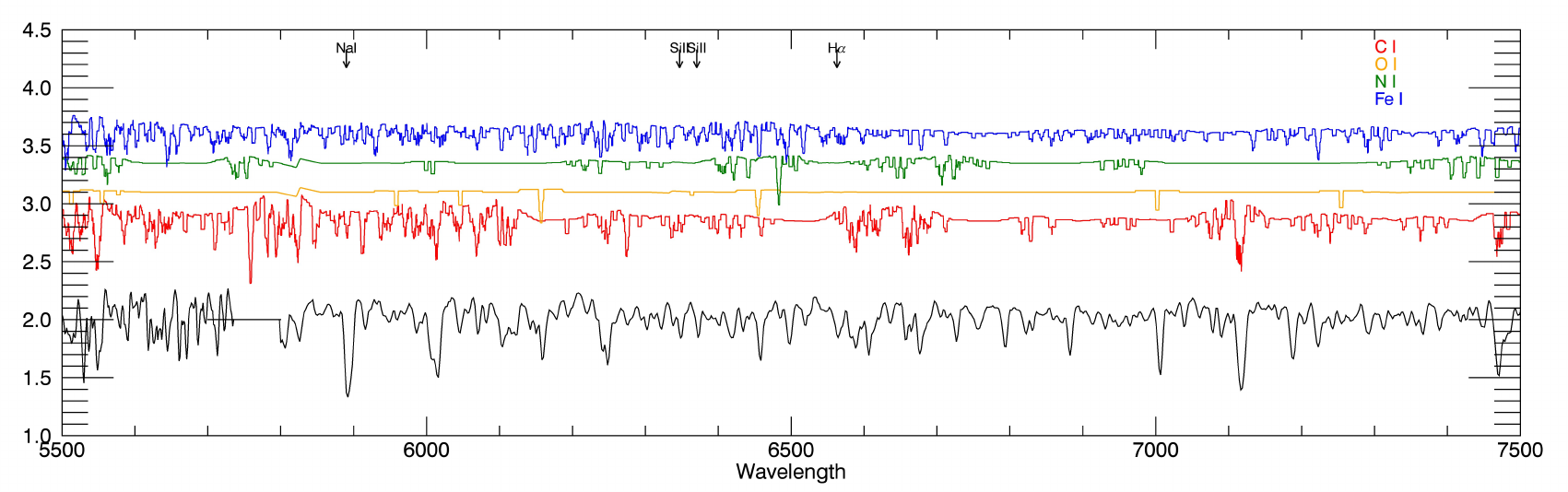}
\includegraphics[width=7in]{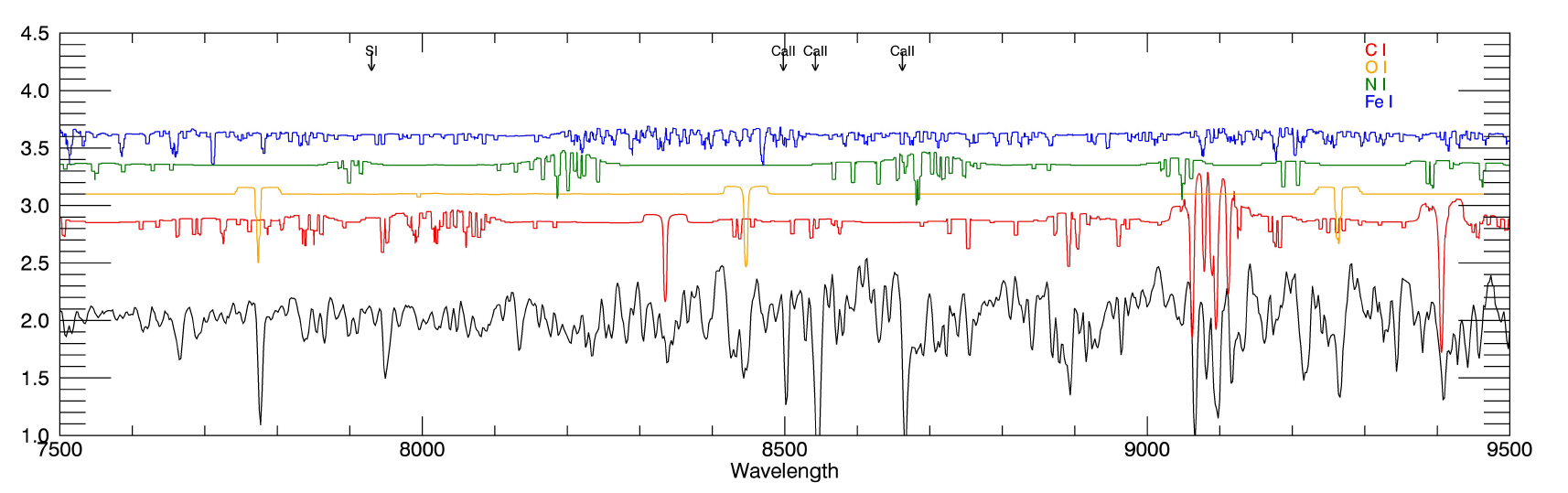}
\caption{The Warm RCB star UX Ant (black) plotted with model stellar atmosphere spectra of C I (red), O I (yellow), N I (green), and Fe I (blue).}
\label{fig_SpectroUXAnt}
\end{figure*}

\begin{figure*}
\centering
\includegraphics[width=6in]{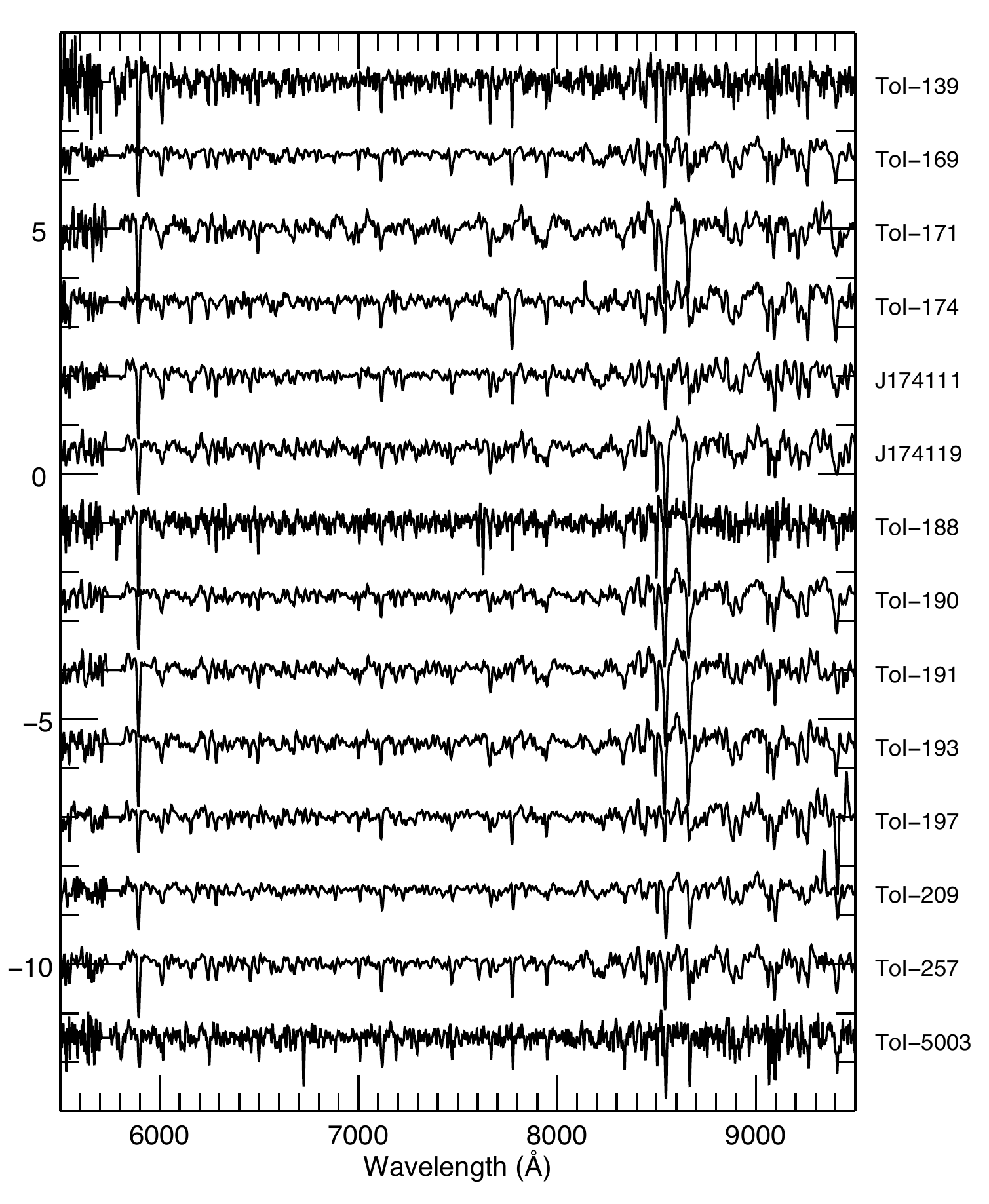}
\caption{Spectra of the newly discovered Warm RCB stars. From the top, the stars are ToI 139, ToI 169, ToI 171, ToI 174, WISE J174111.80-281955.3, WISE J174119.57-250621.2, ToI 188, ToI 190, ToI 191, ToI 193, ToI 197, ToI 209, ToI 257, and ToI 5003.}
\label{fig_SpectroWarmRCB}
\end{figure*}

\subsection{Second group: Warm RCB stars \label{subsec_targets_warm}}

Molecules do not exist in the atmospheres of RCB stars whose temperatures exceed $\sim$6800 K. Therefore, only atomic absorption lines of abundant elements, predominantly C I, N I, O I, but also Fe I, are seen in their spectra. In Fig.~\ref{fig_SpectroUXAnt}, we present the details of these atomic lines between 5500 and 9500 $\AA$ comparing synthetic spectra to the observed spectrum of the known Warm RCB star, \object{UX Ant}. Most of the lines observed at this resolution can be explained by the elements just mentioned, in particular, the strong C I absorption lines at 7112 $\AA$ and between 9050-9100 $\AA$, the O I line at 7774 $\AA$, and the Si II lines at 6345+6374 $\AA$. Other weaker C I, N I, and O I lines are mostly blended with Fe I lines, but the resulting features remain clearly identifiable.

Using these indicators, we searched the ToI spectra presenting no sign of Paschen lines and no or only weak H$_{\alpha}$ absorption. Thus, we identified 14 new Warm RCB stars. Their spectra are shown in Fig.~\ref{fig_SpectroWarmRCB}, they look remarkably similar. The main differences we found are in the strengths of the O I and Si II lines (Table~\ref{tab.NewRCBWarm}). ToI 5003 is the first Warm RCB star found in the Small Magellanic Cloud.

\begin{figure}
\centering
\includegraphics[scale=0.5]{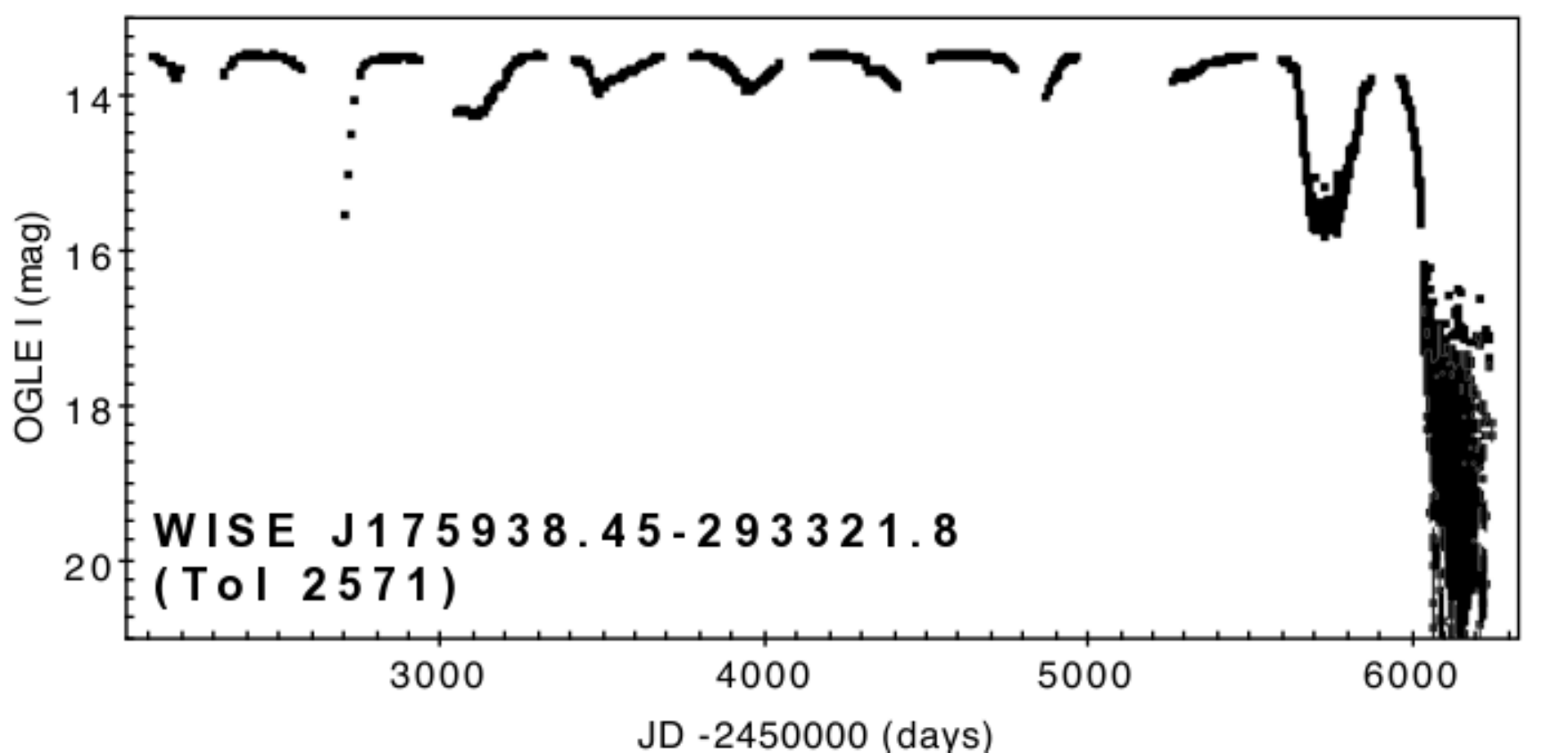}
\caption{OGLE light curve of WISE J175938.45-293321.8 (ToI 2571). This is an RVb type RV tauri star that is presenting some large photometric declines on top of periodic variations at maximum brightness.}
\label{fig_LC_RVtauri_decline}
\end{figure}

The principal difficulty in detecting new Warm RCB stars comes from similarities with RV Tauri stars. Indeed, these stars have similar photospheres, circumstellar shells, and $T_\mathrm{eff}$ to RCB stars, and therefore pass all the pre-selection criteria to be spectroscopically studied. Furthermore, they also show atomic absorption lines similar to Warm RCB stars, mainly the C I and Si II lines. However, RV Tauri star atmospheres are rich in hydrogen, and one can clearly detect the Paschen and H$_{\alpha}$ absorption lines, and sometimes CH. Interestingly, to emphasise even more the similarity observed between both classes of stars, we found that some RV Tauri star light curves may present very rapid and large declines due to dust obscuration events that are similar to RCB stars. The main difference being the longer and larger amplitude of RV Tauri star photometric pulsations at maximum brightness. As an example, we present in Fig.~\ref{fig_LC_RVtauri_decline} the light curve of WISE J175938.45-293321.8 (ToI 2571), an object whose brightness variation at maximum is similar to an RV Tauri star of RVb type with a period of about 500 days. On top of these variations, strong photometric declines can be seen, the last one being very large at JD$\sim$2455900 days. 

On the same subject, we found that the star, \object{OGLE-GC-RCB-2}, was wrongly reported as an RCB star in \citet{2011A&A...529A.118T}. Indeed, its identification as an RCB star was only based on its OGLE light curve as a sudden and large photometric decline was observed (see figure 5 in that article) but without any spectrum to support the claim. We later obtained a spectrum that shows all the features we have just indicated for an RV Tauri star. It stresses the need for spectroscopic follow-up to identify new RCB stars. The light curve is a useful indicator but it is not sufficient. 




ToI 1247 is a similar example. Its OGLE light curve shows multiple non-periodic declines very similar to RCB stars. It would have been classified as such after a classical dedicated search based only on its light curve, but the spectrum shows no classic Warm RCB star lines. It shows H$_{\alpha}$ but no Paschen lines. We do not have a classification for this star. We will observe it again in case there is confusion between a classical H-rich main sequence star and a real RCB star that could have undergone a decline at the time of our observation.

Two new Warm RCB stars are listed in Table~\ref{tab.NewRCBcoord} without any ToI identifier. They were followed up after extending our search just outside the colour cuts defined in Section~\ref{sec_mainana}. They are J174111.80-281955.3 and J174119.57-250621.2, located within 3 degrees of the Galactic centre. They did not pass the colour criteria on $J-H$ versus $H-K$ due to the very high reddening in that particular area of the sky.


With the detection of 14 new Warm RCB stars, we have reached a total of 45, representing nearly a third of the total number of RCB stars known (most of them being Cold). As an RCB star is supposedly evolving from a cold to a warm state before finishing its life as a heavy white dwarf, this ratio could indicate the relative time passed in both temperature regimes \citep{2002MNRAS.333..121S,2018AJ....156..148M,2019MNRAS.488..438L}.

\begin{table*}[hbt!]
\caption{Spectroscopic analysis of new Warm and Hot RCB stars \label{tab.NewRCBWarm}}
\medskip
\centering
\begin{tabular}{lcccccl}
\hline
\hline
WISE All-Sky & WISE & 2MASS & Ca II IR triplet & H$_{\alpha}$ ?    & Warm RCB  & Comments\\
designation  & [12] mag & K mag &  strength      &    & abs. lines?     &\\
\hline
\hline
\multicolumn{7}{c}{New Warm RCB stars}\\
\hline
139 & 5.43 & 9.60  & Small & No & Yes with Si II  & \\
169 & 5.54 & 9.66  & Weak & small H$_{\alpha}$ & Yes with Strong Si II  &  \\
171 & 4.47 & 9.61  & Strong & No & Yes with Si II &  \\
174 & 6.86 & 9.70  & Weak & small H$_{\alpha}$ & Yes with Strong Si II &  \\
J174111-28 & 5.27 & 9.40  & Small & No & Yes with Si II &  \\
J174119-25 & 6.00 & 9.89  & Medium & No & Yes with Si II  &  \\
188 & 3.99 & 8.30  & Strong & No & Yes with Si II& \\
190 & 5.14 & 9.58  & Strong & No & Yes &  \\
191 & 4.32 & 8.63  & Strong & No & Yes with Si II  &  \\
193 & 5.10 & 8.67  & Strong & No & Yes with Si II  &  \\
197 & 3.19 & 7.36  & Weak & No & Yes with Strong Si II &  Possible weak CH feature\\
209 & 2.75 & 7.88  & Medium & No & Yes with Si II &  \\
257 & 5.21 & 10.14  & Medium & small H$_{\alpha}$ & Yes with Strong Si II &  \\
5003 & 8.62 & 12.81  & Medium & No & Yes with Si II &  \\
\hline
\multicolumn{7}{c}{New Hot RCB star}\\
\hline
6005 & 5.55 & 13.42  & None & Emission & No & Many emission lines including   \\
          &     & &                           &                &       & He I$_{(5876+6678+7065)}$, O I$_{(7774,Pcyg.)}$,\\
          &    & &                           &                &       &  C I$_{(9050-9100)}$ and C II \\
\hline
\hline
\end{tabular}
\end{table*}


\subsection{Third group: possible new RCB stars undergoing a dust obscuration event \label{subsec_targets_dust}}

When an RCB star undergoes a dust obscuration event, multiple emission lines appear in the spectrum, which evolve depending on the decline phase. A summary of these lines was detailed by \citet{1996PASP..108..225C} and \citet{2004MNRAS.355..855K}. Early in the decline, many narrow emission lines blue-shifted from the stellar radial velocities \citep{1979A&A....80...61S,1990MNRAS.244..149C} are visible. A few weeks later and until the end of the decline phase, on top of a very reddened RCB star spectrum, one could see a few broad emission lines, the most common ones being Ca II H and K, and the Na I D lines. Only the Na I D line is detectable in the wavelength range of our optical spectra.


Detailed studies done at minimum light on three RCB stars, V854 Cen \citep{1993MNRAS.263L..27K}, UW Cen \citep{2004MNRAS.355..855K}, and R CrB \citep{2006MNRAS.370..941K} show that one can also expect to observe broad emission lines for the following elements: [N II] (5755, 6548, and 6583 $\AA$ in higher strength order), He I (5876 and 7065 $\AA$), Ca II IR triplet, and [Ca II] (7291 and 7323 $\AA$), but also, less commonly, [O I] (6300 and 6363 $\AA$), some weaker lines, [S II] (6717 and 6731 $\AA$) and K I (7664 and 7699 $\AA$) lines, and the C$_2$ Swan bands. In the particular case of the less hydrogen-deficient \object{V854 Cen} RCB star, H$_{\alpha}$ was also detected in emission \citep{Lawson_1992}. With that knowledge in hand, we searched our spectra for such features.

We searched for Na I D broad emission lines and found four members of the Cold group that show a red carbon spectrum with such emission. They are ToI 194, 223, 264, and 4117. For ToI 223, one can also observe weak H$_{\alpha}$ in emission as well as [O I] at 6300 $\AA$. No other broad emission lines were observed for the other three. ToI 223 could be a similar RCB star to \object{V854 Cen} which is less hydrogen-deficient than typical RCB stars. In its spectrum, one can also detect clear C$_2$ features between 6000-6200 $\AA$, some strong Ca II IR triplet absorption, but no detectable $^{13}$C lines. These characteristics support the case that it is a member of the RCB class. ToI 4117, also named IZ Sgr, is part of the lowest priority group, \#5, because of the classification given in the SIMBAD database as, surprisingly, a Mira M6 star \citep{1967IBVS..228....1H}, but it would have been part of the first priority group otherwise. There has clearly been a misclassification as we observe some weak CN features in our reddened spectra. Its light curve was also already selected by \citet{2013A&A...551A..77T} in their search for RCB stars in the ASAS database because two rapid and sudden photometric declines were observed. This target is therefore similar to ToI 218 discussed in section~\ref{subsec_targets_coldselection} and a spectrum near maximum brightness is needed for a full confirmation of its nature. The reddened spectra of ToI 194 and 264 show clear CN features observed above 6900 $\AA$, but no clear C$_2$ features in the interesting wavelength range as the signal level was weak and noisy. We consider all four targets as good Cold RCB candidates.

We also looked for spectra with broad emission lines from the other elements, mainly ToI that had [N II] (6583) in emission. This is the case of ToI 138 which we already classified as an RCB star in our group \#1 Cold star sample analysis. No other lines are, however, observed, Na I D being in absorption. We found another interesting target, ToI 137, having [N II] (6548 and 6583 $\AA$) in emission, but also He I (7065) and split emission lines of [O I] (6300) and of a weak H$_{\alpha}$ suggesting a symmetric structure. Its reddened spectrum also shows some CN features in absorption after 6900 $\AA$ and some weak Ca II IR triplet absorption. Again ToI 137 could be a less hydrogen-deficient RCB. We consider it also as a Cold RCB candidate. All other spectra presenting [N II] in emission also show strong H$_{\alpha}$ emission lines without an underlying red carbon spectrum, so we did not study them further.

Interestingly, we obtained a second spectrum of ToI 5039, a target that we already listed as a new cool Magellanic RCB star in the related section above, using the AAT/AAOmega spectrograph. The spectrum was observed on 2010-10-29 which corresponds to a moment of the deepest minimum as indicated by its OGLE-IV-RCOM light curve (I$\geq$20.8 mag). We observed some strong broad emission lines with the redshift expected for the LMC distance: [O III] (4959+5007 $\AA$), [N II] (6583 $\AA$), [S II] (6719+6730 $\AA$), and also strong emission in the Balmer series ($\alpha, \beta, \delta$, and $\gamma$). No obvious sign of hydrogen was observed in its spectrum taken closer to maximum brightness with the 2.3m/WiFeS a year earlier (on 2009-11-27). ToI 5039 could also be like the less hydrogen-deficient V854 Cen.

In the extreme cases of high obscuration, the RCB star photosphere disappears from the spectrum and only the featureless spectrum of the circumstellar shell remains visible \citep{2011ApJ...739...37G}. This is what was observed with ToI 274. Its spectrum (obtained the 2013-08-18 with the 2.3m/WiFeS spectrograph) is of a very cool featureless blackbody. Only a weak and broad Na I D emission line is detectable. A later observation obtained with the automatic Liverpool telescope revealed a reddened spectrum with some CN features above 6900 $\AA$, but without any C$_2$ band-heads. We also note a weak and broad H$_{\alpha}$ emission line. It was most certainly taken during a brighter phase. Further spectroscopic follow-up is needed on this particular ToI.

We consider ToI 274 as an RCB candidate as well as the other five ToI listed above. Their spectra show emission lines expected for RCB stars in decline but we need observations at maximum brightness to confirm their nature. ToI 223 and 4117 are the strongest candidates.

Recently \citet{2018A&A...610L...6O} reported an unidentified feature, located at 8692 $\AA$, observed on the spectra of few RCB stars. We looked for such broad emission in our mid-resolution spectra and found that only one of our Cold RCB star candidates, ToI 164, showed a feature at that wavelength. 

\begin{figure*}
\centering
\includegraphics[width=7in]{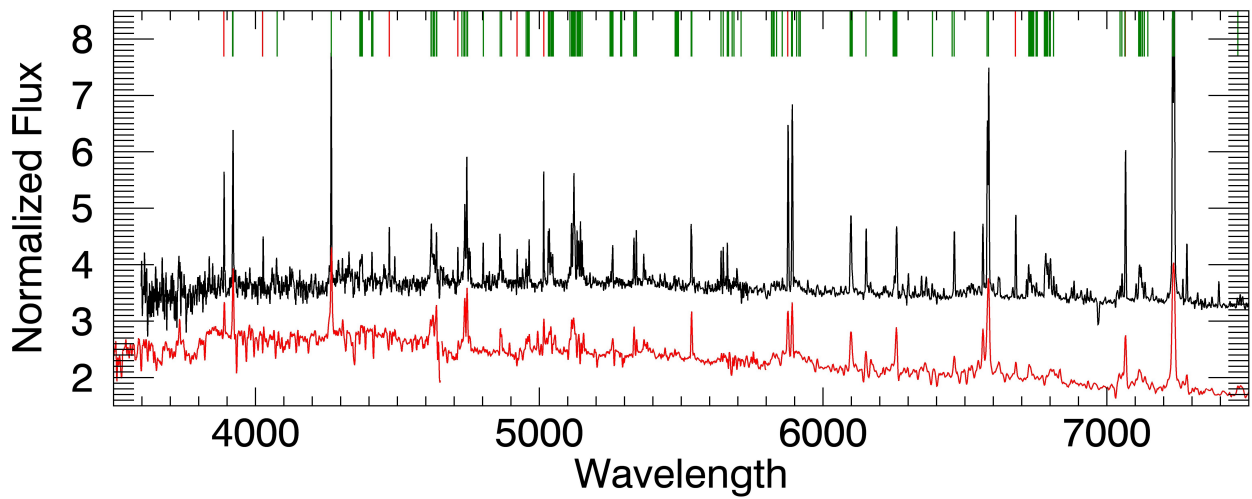}
\caption{Spectrum of the new LMC Hot RCB star (black), WISE J053745.70-635330.8 or ToI 6005, plotted with that of another Hot RCB star, V348 Sgr (red), for comparison \citep{2002AJ....123.3387D}. The emission lines are mostly C II (green lines ) and He I (red lines). The ordinate is arbitrary.}
\label{fig_SpectroHotRCB}
\end{figure*}

\subsection{Fourth group: Hot RCB stars \label{subsec_targets_em}}

Among the rare group of RCB stars, there exists an even rarer subgroup with only four members known so far: MV Sgr, V348 Sgr, DY Cen, and HV 2671 \citep{2002AJ....123.3387D}. They are the Hot RCB stars with $T_\mathrm{eff}$ between 15000 and 20000 K. In the context of the double degenerate merger scenario, they would represent the last phase of an RCB star evolution, where an RCB star evolves from Cold to Warm and then to Hot, while the atmosphere gets smaller and bluer. Such an evolution is supported by the long-term photometric analysis of known Hot RCB stars \citep{2002AJ....123.3387D,2016MNRAS.460.1233S}. The evolution to the blue is expected to continue and the Hot RCB stars would then become extreme helium stars \citep{2008ASPC..391....3J,2008ASPC..391...53J,2019MNRAS.488..438L}.

The spectra of Hot RCB stars show a hot blackbody continuum with many emission lines, in particular, C II and He I \citep{1994A&AS..103..445L,2002AJ....123.3387D}. Some low-resolution optical spectra of three Hot RCB stars are also presented by \citet[Fig.16]{2013A&A...551A..77T}. We found that one of our targets, ToI 6005, presents a similar spectrum. It is shown in Fig.~\ref{fig_SpectroHotRCB} with a spectrum of V348 Sgr for direct comparison. We consider it as a new Magellanic Hot RCB star. Along with many C II emission lines, are He I (5876, 6678 and 7065 $\AA$), O I (7774 $\AA$) and weak H$_{\alpha}$ as seen in \object{V348 Sgr}. A detailed description of these lines is given in \citet{1984ApJ...277..648D} who discussed the spectra of \object{V348 Sgr} observed at various epochs. Furthermore, the OGLE surveys have observed a light curve for ToI 6005 that clearly shows two large and rapid declines before a recovery to a brighter phase (the overall variation is $>$5 mag in I band).



\begin{figure}
\includegraphics[scale=0.33]{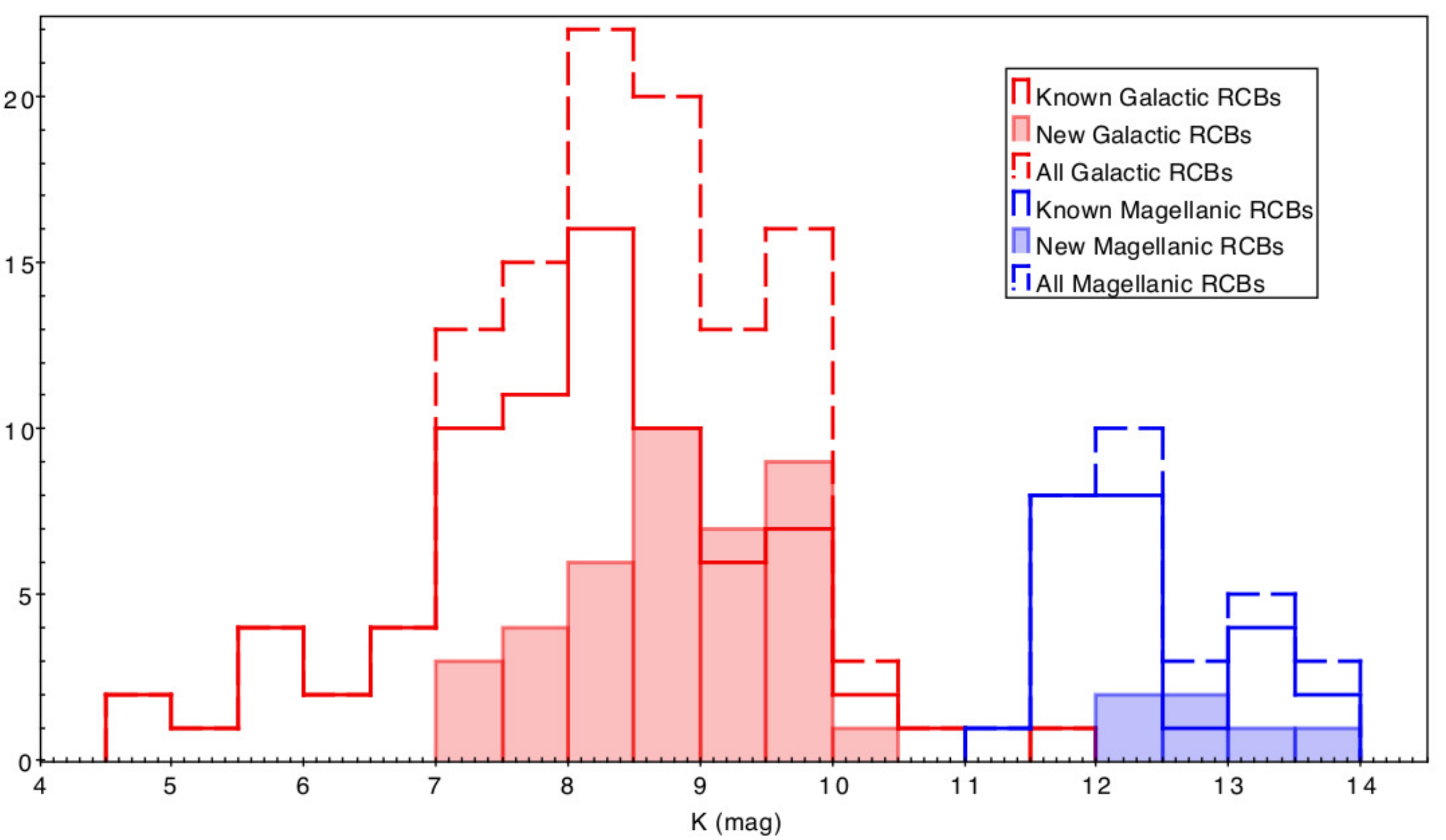}
\includegraphics[scale=0.33]{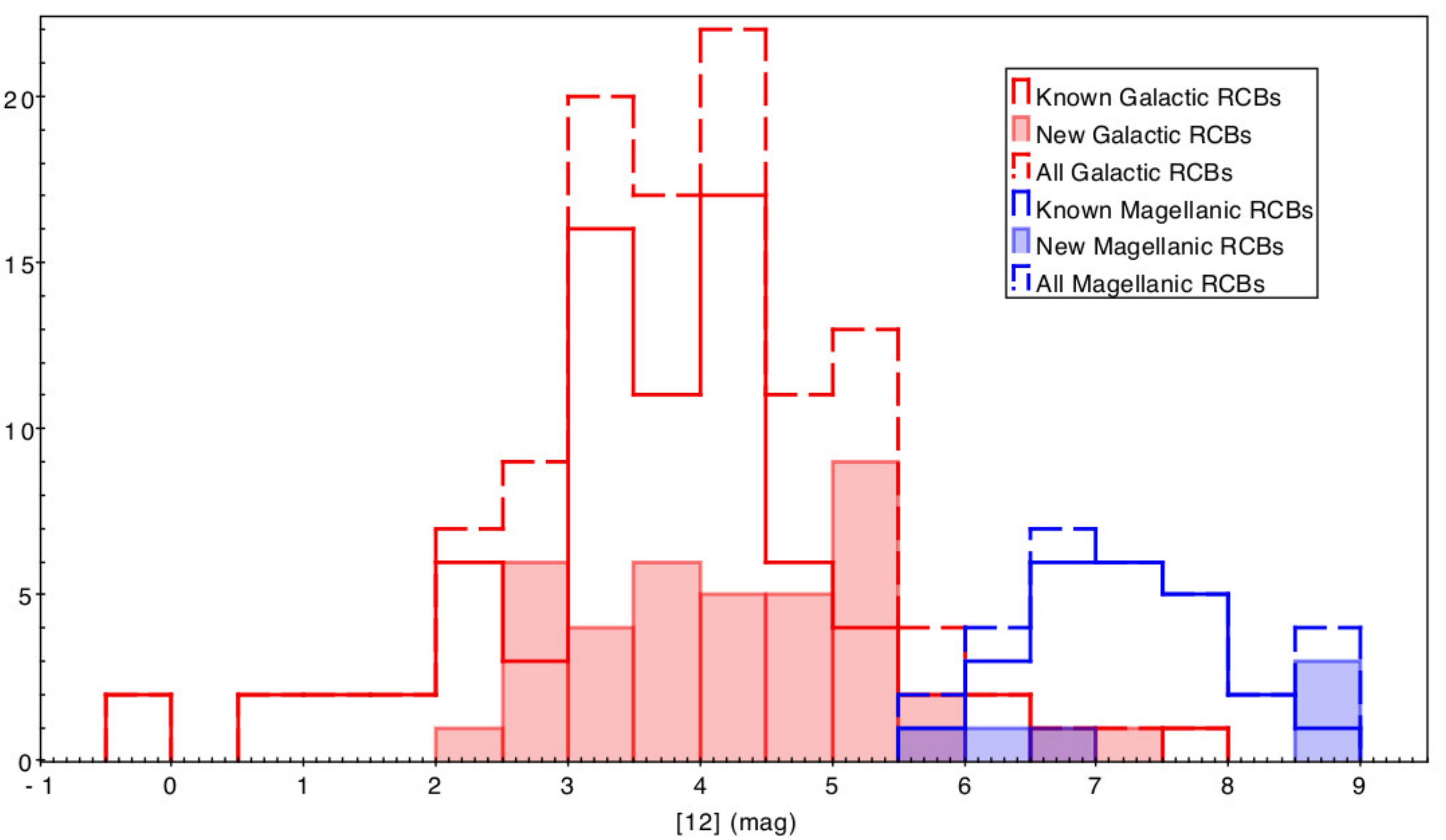}
\caption{Distributions of magnitude in the K (top) and [12] (bottom) bands for all previously known and newly discovered RCB stars.}
\label{fig_MagDistrib}
\end{figure}

\section{Status of previously proposed RCB candidates \label{sec_cand}}


Here, we discuss the status of strong RCB candidates accumulated during the years and for many of which we have update, and the status of candidates reported by four articles or research notes \citep{2009AcA....59..335S,2015A&A...575A...2L,2016IBVS.6190....1N,2019MNRAS.483.4470S} published lately.

\subsection{Strong candidates}

Ten strong candidate RCB stars are listed at the end of Table~\ref{tab.WISEa}, some because of photometric declines observed in their light curves but without any spectroscopic confirmation, others because they show interesting featureless mid-infrared spectra but have no other information yet to support their classification as RCB stars. 

Interestingly, four of them were selected in the present analysis, namely, GLIMPSE-RCB-Cand-1, GLIMPSE-RCB-Cand-2, MSX-LMC-1795 and EROS2-SMC-RCB-4. They are listed in our catalogue as priority Group \#1 targets with respectively the following Ids: 188, 204, 5039 and 5005. We succeeded in getting spectra for three of them and we can now confirm them as bona fide RCB stars. \object{GLIMPSE-RCB-Cand-1} (i.e. ToI 188) and \object{GLIMPSE-RCB-Cand-2} (i.e. ToI 204) were the first stars selected after a series of IR colour criteria using the 2MASS and Spitzer/GLIMPSE magnitudes \citep{2011A&A...529A.118T}, and for which a spectroscopic follow-up has shown interesting carbon features and very red continua, being located at less than 3\degr from the Galactic plane. There were no light curves available at the time.
We have classified these two stars as Warm and Cold RCB stars respectively, as they have satisfied all the criteria defined in each category. Furthermore, we now have a light curve available for ToI 188, as it was observed by the OGLE-IV survey. It shows a characteristic sudden 8 mag decline that occurred in 2015. The third known candidate that we have now confirmed is \object{MSX-LMC-1795} (i.e. ToI 5039) located in the Large Magellanic Cloud. It was first reported by \citet{2014MNRAS.439.1472M} as a potential RCB star due to its featureless mid-IR spectrum and the suspicion got stronger with its OGLE light curves. Indeed it showed a 4 mag variation during the OGLE III monitoring \citep[see object OGLE-LMC-RCB-21]{2009AcA....59..335S}, and six declines with maxima getting brighter each time during the last OGLE-IV phase\footnote{http://ogle.astrouw.edu.pl/ogle4/rcom/ogle-lmc-rcb-21.html}. We have now succeeded in getting a spectrum and it has been classified as a Cold RCB star.

We have not yet succeeded in obtaining a spectrum for the fourth remaining RCB candidate, \object{EROS2-SMC-RCB-4} (i.e. ToI 5005). This star is an interesting case as it has remained faint for the past 19 years. Indeed, after a short 4.5 mag decline that occurred in 1998 as reported by the EROS2 survey in \citet{2009A&A...501..985T}, the OGLE survey data have not reported any variations\footnote{RCOM OGLE III: http://ogle.astrouw.edu.pl/ogle3/rcom/eros2-smc-rcb-4.html and RCOM OGLE-IV: http://ogle.astrouw.edu.pl/ogle4/rcom/eros2-smc-rcb-4.html}. EROS2-SMC-RCB-4 has remained highly enshrouded, indicating a continuous production of dust in the line of sight. We also suspect that it did not reach its maximum brightness during the short recovery phase in 1998. If confirmed as an RCB star, EROS2-SMC-RCB-4 could be the first of a population of very cold, highly enshrouded RCB stars. Its brightness needs to be monitored to trigger a spectroscopic alert. It was reported by \citet{2015MNRAS.451.3504R} to present a mid-IR featureless spectrum which strongly supports its classification as an RCB star. Long duration decline phases is not uncommon in RCB stars. R CrB itself recovered recently from a 10 year decline, and V854 Cen spent several decades in a deep decline in the early 20th century \citep{1986IAUC.4245....2M}.

Of the six remaining RCB star candidates that did not pass the selection criteria defined in section~\ref{sec_IRcuts}, three of them, KDM 5651, OGLE-GC-RCB-Cand-1 and [RP2006] 1631, did not pass the criteria imposed on the shell colours (detailed explanation are given in section ~\ref{sec_IRcuts}, cut \#2). We obtained a spectrum of \object{KDM 5651} and \object{OGLE-GC-RCB-Cand-1}. We confirm KDM 5651 as a bona fide Cold RCB star. The spectrum of \object{OGLE-GC-RCB-Cand-1} is highly reddened, almost featureless, with only some weak Ca II IR triplet lines detected. It certainly corresponds to the bluest part of the dust shell as we observed this star during a decline phase. Indeed, at the time of the observation (epoch: 2010-07-16, I$\sim$15.43 mag), its OGLE-IV light curve shows that it was 1.2 mag fainter than its maximum magnitude. Most of the signal received was therefore from the circumstellar dust shell. We will wait for a brighter phase to observe it again. It is located only a few degrees from the Galactic centre at (l,b)$\sim$(0\fdg14, -1\fdg6), and therefore one expects a high interstellar extinction. From its light curve, presented in \citet{2011A&A...529A.118T}, and its position in the $J-H$ versus $H-K$ diagram near the giant carbon stars locus, OGLE-GC-RCB-Cand-1 could potentially be a DY Per type star. DY Per type stars are potentially the cooler counterparts of RCB stars (see \citet[Sect.3.4]{2013A&A...551A..77T} and \citet{2018ApJ...854..140B}). 
We still did not succeed in observing [RP2006] 1631 spectroscopically. A featureless mid-IR spectrum was presented by \citet{2011MNRAS.411.1597W}. That is typical of RCB stars circumstellar shell. 

The eighth RCB star candidate, EROS2-LMC-RCB-7, was found to have similar K brightness and $J-K$ colour to Magellanic extreme AGB stars (see Fig.~\ref{figcut_C}). Its light curve, presented in \citet[Fig.10 top-left]{2009A&A...501..985T}, shows only a weak variation typical of a highly enshrouded object. We think it is not an RCB star. 

The last two candidates, EROS2-LMC-RCB-8 and OGLE-GC-RCB-Cand-2, have shown similar IR photometric colour to the known RCB stars, EROS2-CG-RCB-12 and MACHO 308.38099.66 (see Fig.~\ref{figcut_B} right side). However, these four stars are located at the position where the AGB-star locus lies, outside the main selection area chosen for RCB stars. Their real nature is therefore suspicious, but a reddening effect during a photometric decline event could explain such position. That is the case for the two known RCB stars and OGLE-GC-RCB-Cand-2 whose 2MASS epochs were taken during a decline phase, but not for EROS2-LMC-RCB-8, that was observed during a bright phase. The light curves of EROS2-LMC-RCB-8, OGLE-GC-RCB-Cand-2 and EROS2-CG-RCB-12 are presented respectively in \citet[Fig.10]{2009A&A...501..985T}, \citet[Fig. 5]{2011A&A...529A.118T} and \citet[Fig.9]{2008A&A...481..673T}. Some sudden and large declines ($>$3 mag) were observed on top of large amplitude periodic oscillations ($\sim$1.5 mag) at maximum brightness for these three objects. These declines triggered their classification as possible RCB stars, however the large oscillations are not seen in any other RCB stars. The typical RCB star amplitude is between $\sim$0.1-0.3 mag \citep{1997MNRAS.285..266L}. Furthermore, the light curve of OGLE-GC-RCB-Cand-2 shows only a slow recovery phase, no fast photometric decline has ever been observed, and, in regards to EROS2-LMC-RCB-8, it is brighter in K than any other Magellanic RCB stars and is located just near the main distribution of AGB stars in the J versus $J-K$ diagram (see Fig.~\ref{figcut_C}). However, they all have been observed spectroscopically and are all cold carbon stars with no obvious hydrogen features. EROS2-CG-RCB-12 even shows the Ca II IR triplet in emission, as seen in RCB stars undergoing photometric declines. More investigations will be needed to understand if these stars are very cold RCB stars or more classical carbon Mira type stars undergoing dust production events similar to RCB stars. 

The light curve of MACHO-308.38099.66 (see \citet[Fig.1b]{2005AJ....130.2293Z}) does not show any large amplitude oscillations. Two decline episodes of two magnitudes in the visible were reported by the MACHO survey, while the star presented small $\sim$0.1 mag pulsations at maximum brightness. It was observed by 2MASS during the second decline phase and is located at Galactic coordinates $(+10.3,-2.9)$ where an interstellar extinction of A$_V\sim$3.0 mag \citep{2011ApJ...737..103S} is expected. MACHO-308.38099.66 is very likely an RCB surrounded by a hot ($>800$ K) circumstellar shell and is located at the AGB star locus because of reddening effects.



\begin{figure}
\centering
\includegraphics[width=3.5in]{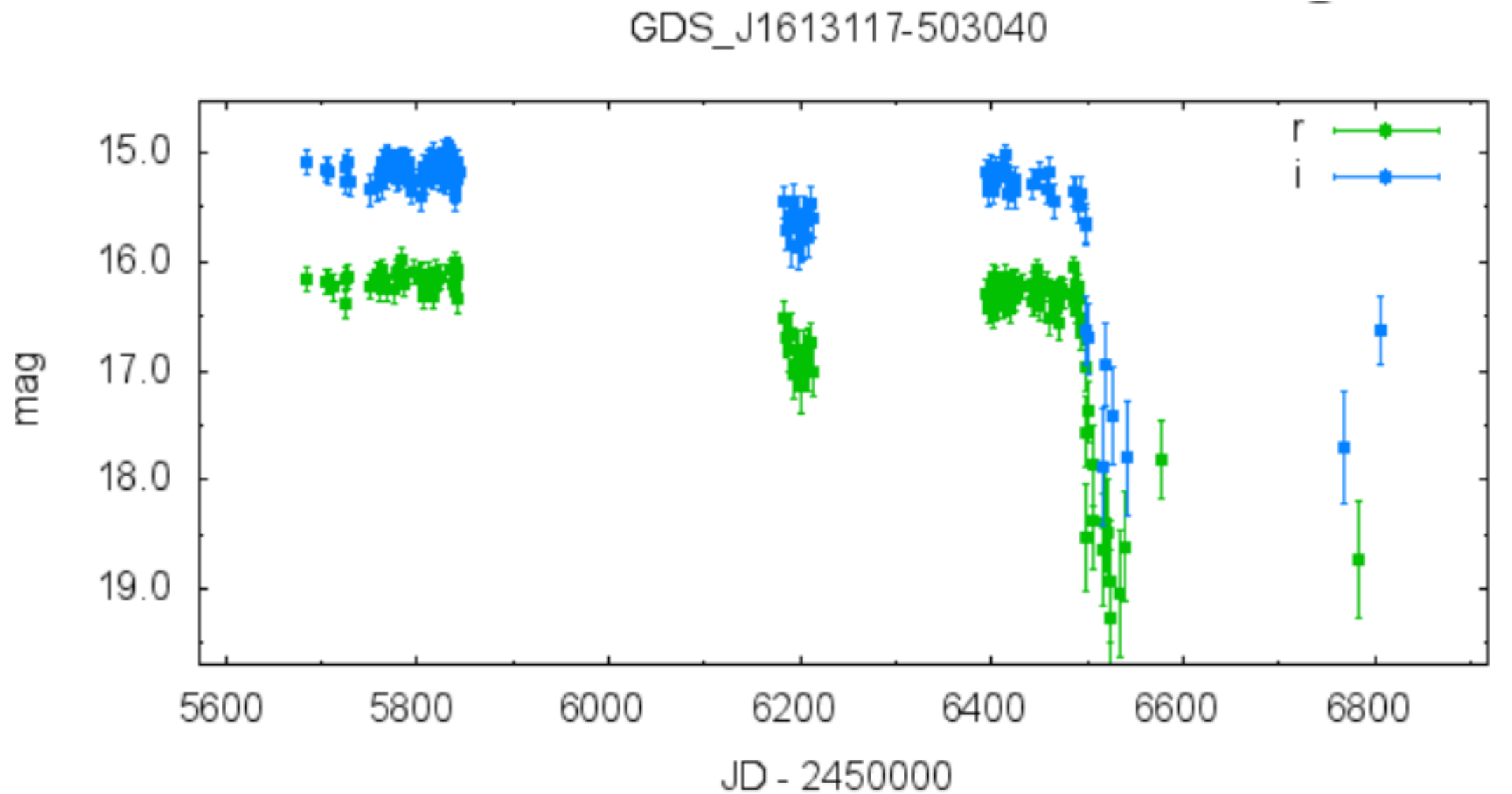}
 \caption{Light curve of GDS J1613117-503040 from the Bochum survey \citep{2015AN....336..590H}. A fast photometric decline of 3 mag is observable around JD$\sim$2456500 days in both the r and i bands. This star is surrounded by two distinct circumstellar dust shells and was therefore not selected by our IR photometric selection criteria. We consider it as an RCB candidate that needs a spectroscopic follow-up for confirmation.}
\label{fig_lc_bochum}
\end{figure}

Finally, we add as a new RCB star candidate, WISE J161311.79-503040.2, also known as GDS\_J1613117-503040. We list it as such because of the spectacular brightness declines of 3 mag observed in its Bochum survey light curve (see Fig.~\ref{fig_lc_bochum}). The SED of this star shows that it is surrounded by two thick circumstellar shells, as are the known RCB stars MV Sgr, DY Cen, and MACHO-11.8632.2507, and was therefore not selected by our criteria. We need a spectroscopic confirmation of its real nature. Similarly to OGLE-GC-RCB-Cand-1, one can expect a high extinction along the line of sight to this star as it is located near the Galactic plane at (l,b)$\sim$(332\fdg3, 0\fdg5). The presence of two dust shells around RCB stars seems to be common \citep{2018AJ....156..148M}.

\subsection{Candidates recently published in the literature}

\subsubsection{Candidates from \citet{2009AcA....59..335S}}

\citet{2009AcA....59..335S} went through their OGLE-III catalogue of LPV stars that show unusual IR characteristics and selected the ones that exhibit irregular fading episodes. They found six candidates towards the LMC, namely \object{OGLE-LMC-RCB-03}, -09, -15, -16, -20 and -21. In all cases except one, OGLE-LMC-RCB-21, the maximum variations observed were small (between 0.3 and 0.8 mag) and for the first four candidates, the fadings were not typical of RCB stars. None of these first four candidates are present in our list of targets of interest, because \object{OGLE-LMC-RCB-09} was not catalogued in the WISE AllSky or even ALLWISE databases, and because none of the other three had WISE colours that satisfied the second selection criterion (see cut \#2 in Sect.~\ref{sec_mainana}). We had followed them up spectroscopically nevertheless and found that \object{OGLE-LMC-RCB-03} and \object{OGLE-LMC-RCB-15} present spectra of F stars with typical Ca H and K absorption lines and balmer lines, while the other two, \object{OGLE-LMC-RCB-09} and \object{OGLE-LMC-RCB-16}, show spectra of Be stars with balmer, paschen en He I lines in emission. Concerning the last two candidates reported, OGLE-LMC-RCB-20 and -21, we already mentioned them in this present article as they are the now confirmed RCB stars, respectively KDM 5651 and WISE J054221.91-690259.3 (ToI 5039). 


\subsubsection{Candidates from \citet{2015A&A...575A...2L}}

\citet{2015A&A...575A...2L} used the IR selection criteria, that were developed by \citet{2012A&A...539A..51T} to retrieve a list of objects from the WISE ALL-Sky database. They cross-matched all of the selected objects with the catalogue of the Catalina survey, which has monitored the sky on both sides of the Galactic plane ($\mid$b$\mid>$15 deg) for over 8 years, to find those that have varied by more than 1 magnitude during that period. They found 26 interesting objects, five of them being already known RCB stars. In the second stage of their study, from the analysis of the Catalina light curves and the objects' SEDs, they reported that only three of the 21 remaining objects could be considered as RCB candidates. J194218.38-203247.6 shows seven large non-periodic photometric variations and a reported flat SED between 1 and 10 $\mu$m. This object was also selected by our analysis (ToI 290), and, after a spectroscopic follow-up, we classified it as a new RCB star (see Table~\ref{tab.NewRCBcoord}). J062741.84-681016.2 and J160701.19-143446.7 show similar long term increases in brightness (resp. 2 mag over $\sim$500 days, and 3 mag over $\sim$1500 days) typical of RCB stars during their phase of photometric recovery after a large decline. However, the authors are less optimistic with these last two objects. The former presents a single-component SED of a cold black-body and is therefore most certainly a N-type AGB star. The SED of the latter star has little optical photometry to pinpoint its overall shape. 
We have followed-up J160701.19-143446.7 spectroscopically as it was also selected in the present analysis, ToI 1189, and we found it to be a cool carbon-rich AGB star presenting H-alpha emission.

We cross-matched their additional 21 objects with our list of ToI and found eight objects are in common. That is not surprising, as the IR selection criteria used in both cases are quite similar. The criteria developed in the present study being more restrictive than those presented in \citet{2012A&A...539A..51T}. These objects are: J023405.26+373352.8 (ToI 1014), J063856.03+542940.4 (ToI 33), J070137.95-412816.5 (ToI 1042), J085725.83+172052.0 (ToI 1079), J160701.19-143446.7 (ToI 1189), J164327.28-141200.2 (ToI 1199), J194218.38-203247.6 (ToI 290) and J222704.54-165948.5 (ToI 323). Most of them belong to the priority \#2 group (i.e. ToI Id between 1000 and 1999), where the main background expected is very cold AGB stars ($J-K>$3.5 mag). Their Catalina light curves are also not convincing as many large periodic variation are seen. The only two stars we have already followed-up spectroscopically during our survey are ToI 290 and ToI 1189 (see our result in the previous paragraph).

\subsubsection{Candidates from \citet{2016IBVS.6190....1N}}

Some three RCB star candidates and 63 DY Per type candidates were listed by \citet{2016IBVS.6190....1N} based only on their OGLE light curves. Unfortunately, the IR $J-H$ versus $H-K$ that could have been simply used to differentiate between DY Per type stars and RCB stars, as in \citet{2003MNRAS.344..325M} and \citet{2009A&A...501..985T}, was not applied in that search. We cross-matched their list with the 2MASS and the WISE AllSky databases. We found that the first two RCB candidates listed, OGLE-SMC-LPV-01019 and OGLE-SMC-LPV-06216, should in fact be listed as good DY Per type candidates. Indeed, their IR characteristics are similar to those expected and their OGLE light curves are very similar to other DY Per type stars \citep{2001ApJ...554..298A,2004A&A...424..245T,2009A&A...501..985T}. 
Furthermore, they were both previously catalogued as carbon stars in the SMC surveys \citep{1995A&AS..113..539M,1993A&AS...97..603R}. They are named, respectively, [MH95] 219 and RAW 380. The reason to list the third and last object, OGLE-SMC-LPV-17611, as an RCB candidate, is only based on a visual similarity observed by the authors of its OGLE light curve with the one of the strong RCB candidates, MSX-SMC-014. However, we note that the light curve of MSX-SMC-014 shows some brightness increase and decline back to a faint background level (that is expected for star with a high rate of dust formation) while 
OGLE-SMC-LPV-17611 shows only two magnitudes of variability at a low median brightness level (I$\sim$20 mag). We checked the IR characteristics of OGLE-SMC-LPV-17611 and found that while it shows an excess in the $J-H$ versus $H-K$ diagram as expected for RCB stars, we would not have selected it in the mid-IR, as it did not pass the selection criteria \#2 (closely) and \#3 (more convincingly). That star need nevertheless a spectroscopic confirmation. Concerning the other 63 DY Per type candidates, we found that many of them do not have the IR characteristics expected for these stars. Two of them are listed in our catalogue of Targets of Interest for our search of RCB stars, namely, OGLE-SMC-LPV-16850 (ToI 6002) and OGLE-SMC-LPV-17267 (ToI 9006). We have not yet observed these stars spectroscopically. However, their OGLE light curves are not convincing as long period variabilities are observed. 

\subsubsection{Candidates from \citet{2019MNRAS.483.4470S}}

The authors published a list of 19 RCB and DY Per type candidates that were selected based on the photometric variability observed in the ASAS-SN survey \citep{2014AAS...22323603S},
and added another 16 weaker candidates. To do so, they started from a list of stars with similar IR characteristic than RCB stars. These objects are the 1602 targets listed by \citet{2012A&A...539A..51T} and 2006 new ones that they selected over the entire sky with the WISE ALLWISE database using the original idea of selecting objects after fitting their SED to each of the known RCB stars. They kept the ones with quasi-identical SED, extinction and distance corrected, to any of the known RCB stars they used.


Many of the candidates selected are in fact also listed in our catalogue of targets of interest, that is, 14 out of their 19 strong candidates, and a further eight out of their 16 weaker candidates. We list the ToI id for each of the candidates we have in common in Table~\ref{tab.Shields19}. We also give a classification based on our spectroscopic follow-up observations, if such data exist. That was the case for 13 objects out of the 22 candidates we have in common. In summary, six of them are also ToI that we have now confirmed as new RCB stars and two others are ToI that we have listed as strong RCB candidates (see Table~\ref{tab.NewRCBcoord}). We also report that five objects are not RCB stars, as two are Mira type objects, two others are potentially RV Tauri stars, and one shows an H-rich emission spectrum.   

\begin{table}
\caption{Status of candidates from \citet{2019MNRAS.483.4470S} in our study \label{tab.Shields19}}
\medskip
\centering
\begin{tabular}{lcccccl}
\hline
\hline
ASASSN-V & ToI & Status from our study \\
designation  & Id ? & \\
\hline
\multicolumn{3}{c}{Their Table 1 - "Candidates RCB stars"}\\
\hline
J053745.71-635330.9 &   6005 &   Confirmed RCB star   \\
J053213.93+340601.4 &    23 &      \\
J173819.81-203632.2 &    1227 &   Confirmed RCB star    \\
J173737.08-072828.2 &    1225  &      \\
J174257.20-362052.1 &    184 &     Confirmed RCB star    \\
J190309.89-302037.0 &    264 &     Strong RCB candidate     \\
J170737.02-314812.5 &      &      \\
J175031.71-233945.7 &    191 &    Confirmed RCB star    \\
J044531.02-683431.3 &       &        \\
J004822.94-734104.6 &      &      \\
\hline
\multicolumn{3}{c}{Their Table 1 - "Large amplitude candidates" } \\
\hline
J174317.53-182402.5 &    185 &      \\
J170552.81-163416.6 &    1213 &     \\
\hline
\multicolumn{3}{c}{Their Table 1 - "DY Per stars candidates"} \\ 
\hline
J191243.07+055313.1 &    274 &       Strong RCB candidate     \\
J175526.28-214214.1 &    2554 &      RV Tauri? Warm H-rich star \\
J202300.80+431111.5 &    302 &      \\
\hline
\multicolumn{3}{c}{Their Table 2 - "RCB Candidates discovered outside our search"}\\
\hline
J161156.22-575527.2 &    124 &     Confirmed RCB star      \\
J043259.32+415854.0 &      &      \\
J201504.29+462719.9 &      &      \\
\hline
\multicolumn{3}{c}{Their Table 2 - "DY Per candidates"}\\
\hline
J175700.51-213934.5 &    199 &   RV Tauri? Warm H-rich star    \\
\hline
\multicolumn{3}{c}{Their Table 3 - "Weak RCB Candidates"}\\
\hline
J195525.11+015601.6 &      &      \\
J075155.45-331057.2 &      &      \\
J185316.37-271352.7 &    2801 & M star     \\
J174445.73-362232.3 &    187 &      \\
J174328.51-375029.1 &    186 &     Confirmed RCB star     \\
J163750.78-644140.5 &    2317 & M star     \\
J174731.77-444501.4 &      &      \\
J173257.95-180435.6 &      &      \\
J211119.06+473847.7 &      &      \\
J054551.71+350300.0 &    2030 &      \\
J172216.67-281656.9 &    2409 &      \\
J174825.52-324240.5 &      &      \\
J160407.52-580250.6 &      &      \\
J181154.33-241827.3 &      &      \\
J054424.84-655814.2 &    5041 &      \\
J181214.33-252406.5 &    2642 &  Hot H-rich star    \\
\hline
\hline
\end{tabular}
\end{table}

\section{Results and discussion \label{sec_result}}

For the 488 WISE ToI that we have followed up spectroscopically, there is a success rate of nearly 10\% of bona fide new RCB stars. This rate is even higher when we only consider the priority group \#1 targets. In that subgroup, the success rate rises to $\sim$25\% with about 40\% of those targets followed up. The success rate for priority group \#2 is slightly lower, $\sim$19\%, while only $\sim$10\% of this group's targets were followed up. For group \#3, we found only one new RCB star among the 256 ToI (i.e. a quarter of all group \#3 ToI) observed. Half of our observational efforts were made on that group as it contains many bright targets. We found that most objects in this group are Mira and RV Tauri stars, but it should also contain some rare RCB stars that possess very thick circumstellar shells. Furthermore, we have light-curve information for a further 199 ToI (mostly from priority group \#3) that show clear photometric oscillations similar to Miras or RV Tauri stars. Overall, we have obtained a conclusive definition of the nature of 687 ToI, which is almost a third of all ToI listed. (It corresponds to $\sim$45\%, $\sim$12\%, $\sim$36\%, $\sim$6\%, and $\sim$37\% of the targets from the priority groups \#1 to \#5 respectively; See Table~\ref{tab.Priority}).

\subsection{Galactic distribution and magnitudes of new RCB stars}

As expected, the luminosities of the newly found RCB stars are on the fainter end compared to the known stars. This is illustrated in Fig.~\ref{fig_MagDistrib} showing the magnitude distribution in K and [12] bands for both Galactic and Magellanic RCB stars. However, it is worth noting that two Cold RCB candidates, ToI 223 and 274, have bright circumstellar dust shells with [12] $\sim$1.2 and $\sim$1.0 mag, respectively (see Tables~\ref{tab.NewRCBCold}). If confirmed as RCB stars, they would be respectively the fifth and sixth brightest RCB stars in the [12] band. Furthermore with an interstellar extinction of A$_V\sim$8 mag \citep{2011ApJ...737..103S} and an apparent K magnitude of 6.3 mag, ToI 274 could be one of the ten brightest Galactic RCB stars.

\begin{figure*}
\centering
\includegraphics[width=3.5in]{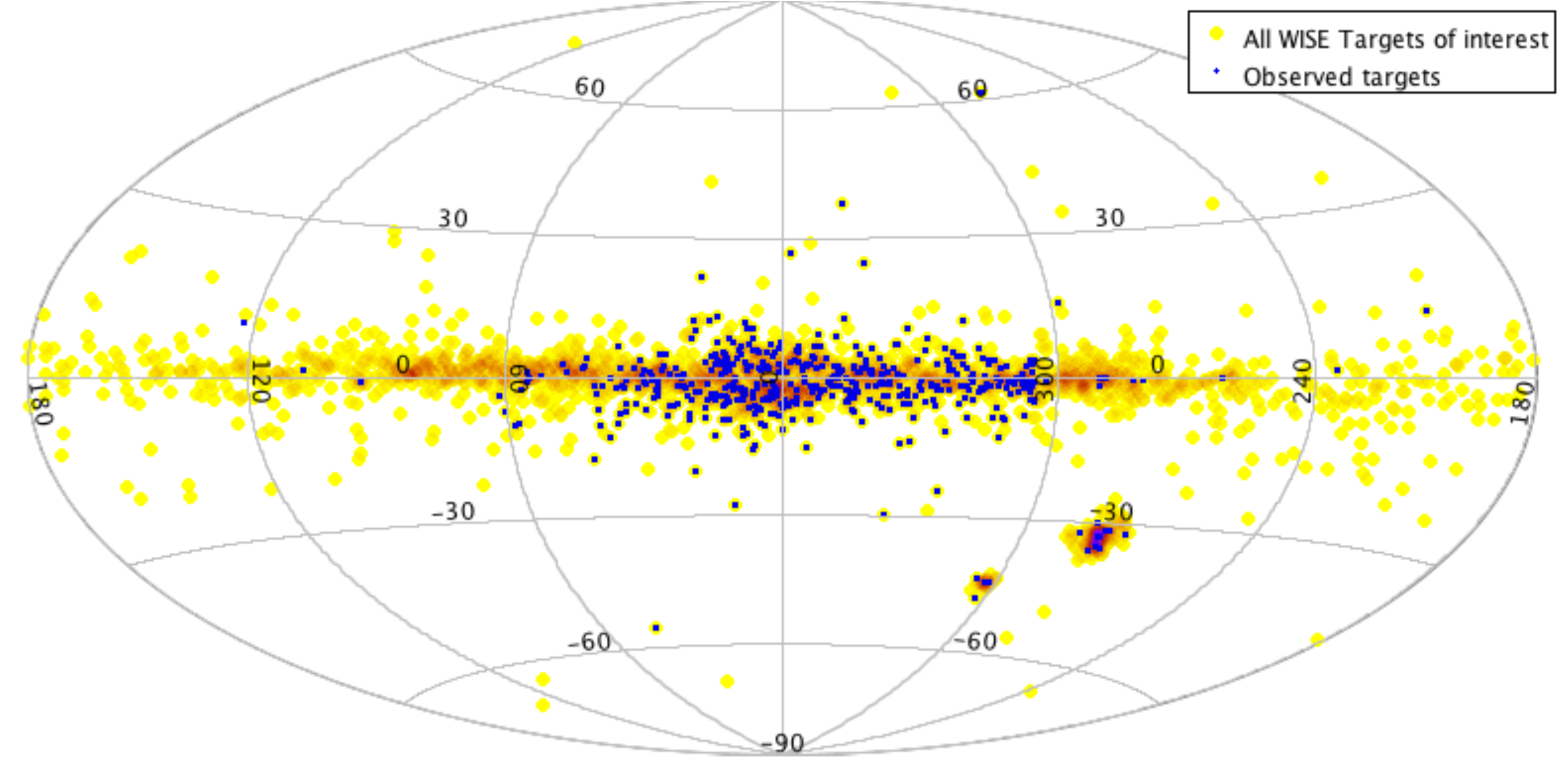}
\includegraphics[width=3.5in]{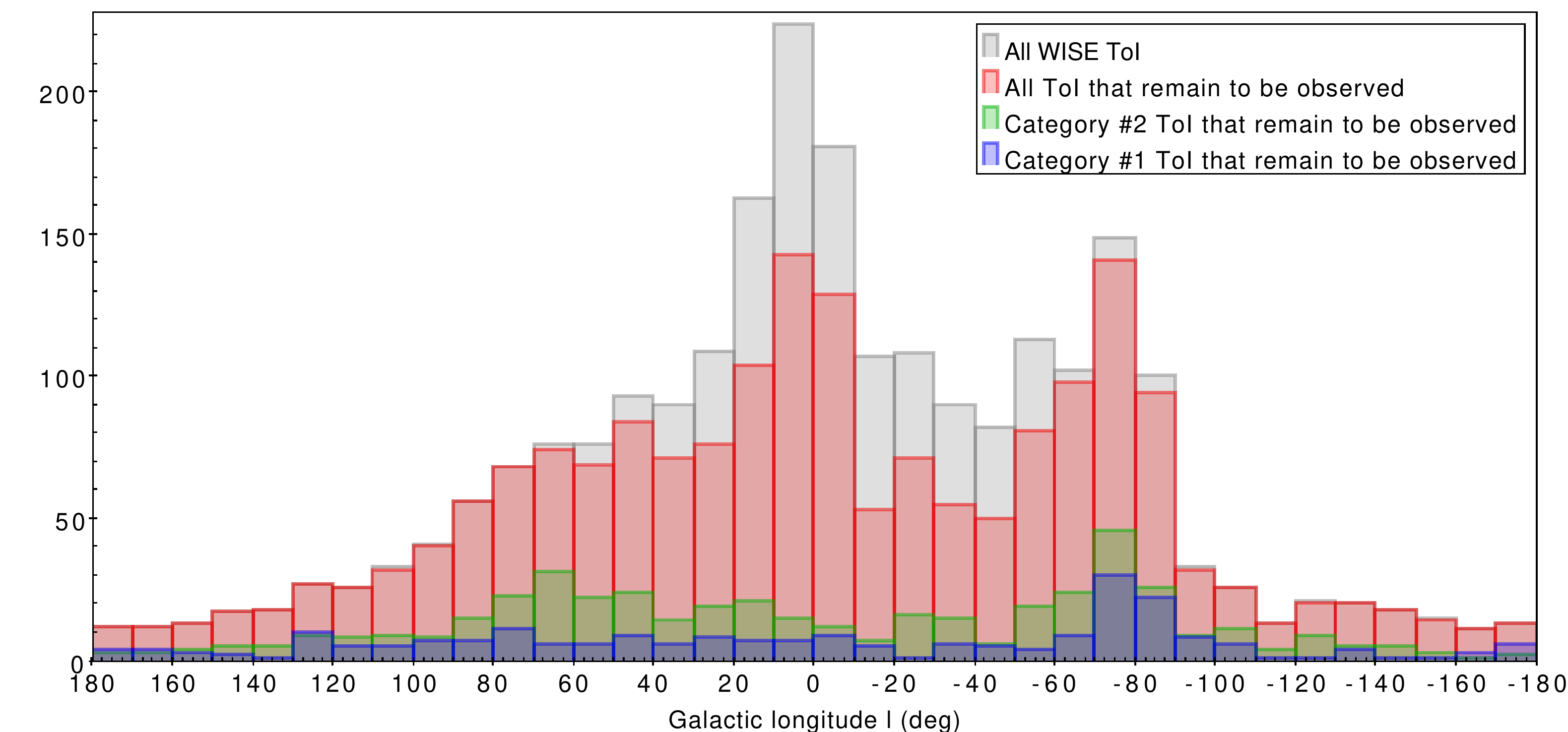}
\includegraphics[width=3.5in]{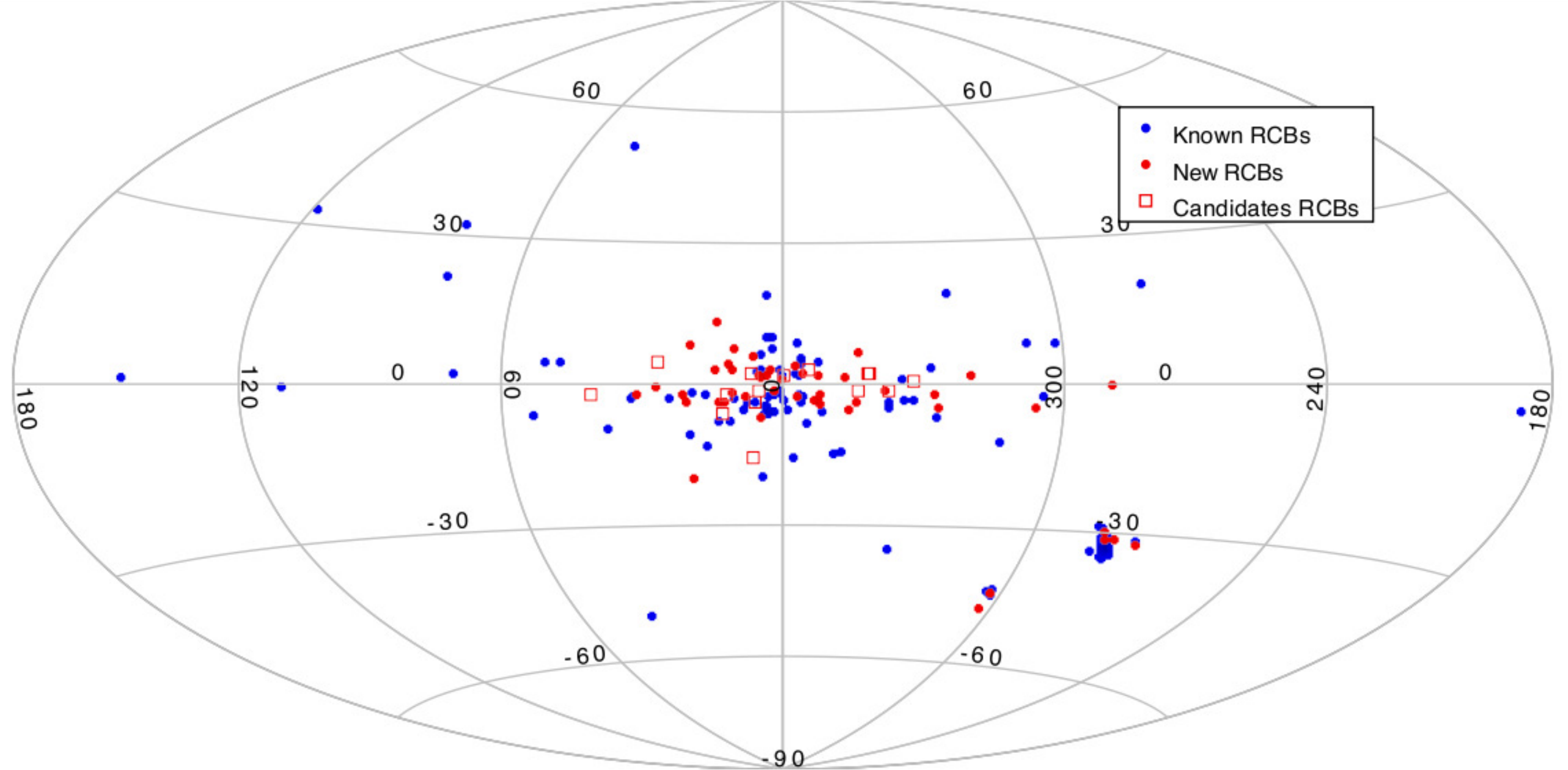}
\caption{Spatial distribution in Galactic coordinates of all selected targets of interest, overlaid with the ones already followed-up spectroscopically (top left). Distribution of all known and candidate RCB stars (bottom). Distribution in Galactic longitude of all ToI with the ones already observed and the remaining ones (top right).}
\label{fig_SpatialDistrib}
\end{figure*}

Figure~\ref{fig_SpatialDistrib} presents the spatial distribution of all ToI and of the 488 targets already observed spectroscopically. We found new RCB stars on both sides of the Galactic plane, and from a simple first distance estimate based on their K band luminosity, most of them are located beyond the Galactic centre. Our observational effort was concentrated mostly between the Galactic longitude range [-60\degr, +45\degr] as we used telescope facilities in the southern hemisphere. The distribution of the interesting ToI in the priority categories \#1 and \#2, that remain to be observed (see Fig.~\ref{fig_SpatialDistrib}, top-right), suggests that future observations will have to focus outside the Galactic Bulge and mostly along the Galactic disc. 

\subsection{How many Galactic RCB stars exist out there?}

Taking the results reported here into consideration, we can make an initial estimate of the total number of RCB stars in our Galaxy. A total of 77 Galactic RCB stars had been identified before this first dedicated all-sky search \citep{2013A&A...551A..77T}. This work identifies 40 new RCB stars, increasing the total to 117 Galactic RCB stars. So far only a few of the ToI listed in priority group \#4 have been followed up because most of them are faint highly enshrouded stars that require observations with 4m-class telescopes. So if we choose an optimistic rate of discovery of 20\% for this group, corresponding to about 50 new RCB stars, and similarly choose the favourable scenario that 20 new RCB stars can be discovered in both priority groups \#3 and \#5, we would reach a potential of about 220 new discoveries within the remaining list of ToI, assuming we add 50 and 77 potential new RCB stars from respectively the categories \#1 and \#2 groups, simply using the rate of discovery obtained so far. When corrected for the 85\% detection efficiency of the ToI photometric selection process, we estimate that in this favourable scenario there could be as many as about 380 RCB stars existing in the Galaxy (77+(220+40)/0.85).

This is a rough estimate, and, of course, our all-sky search has biases as it was optimized using the known RCB stars. We summarize briefly the selection biases as we know them. A more detailed discussion can be found in Section~\ref{sec_ToIResult}. Firstly, we do not know much about the highly enshrouded RCB stars, such as \object{EROS2-SMC-RCB-4} and \object{MSX-SMC-014}. They have remained faint most of the time during the past 20 years and have been found to be surrounded by a thick dust shell presenting featureless mid-IR spectra \citep{2005ApJ...631L.147K, 2015MNRAS.451.3504R}. They are members of priority group \#4, for which we assumed an optimistic discovery rate of 20\%. Secondly, our search does not target RCB stars that possess multiple bright dust shells (we know of three such RCB stars so far); and also RCB stars with either a very thin dust shell or no shell at all, like the HdC stars, will be missed. For the first two sets of stars, we have taken them into account via our detection efficiency, but their numbers could be underestimated. For the remaining group, the HdC stars, only five are known so far and considering that no dedicated deep all-sky searches were ever made to find more of them, we need to compare that number to the quantity of known bright Galactic RCB stars before the recent discoveries made using light curve surveys. We estimate there could exist up to one HdC star for every six RCB stars, which would thus correspond to a total of about 60 HdC stars in the Galaxy. Thirdly, our spectroscopic follow-up has mostly targeted ToI located within 60 degrees of the Galactic centre (see Fig.~\ref{fig_SpatialDistrib}) and the external disc region of the Galaxy was not covered. However, RCB stars seem to follow an old-disc spatial distribution and should be found in higher numbers in the central area of the Galaxy. This observational bias would support our optimistic conservative approach in the estimate of the total number of Galactic RCB stars as a lower rate of discovery is expected at higher Galactic longitude. Finally, we know that the detection efficiency of our all-sky ToI selection drops within a few degrees of the Galactic centre and along the Galactic plane at low Galactic latitude due to high interstellar extinction. Targeted surveys found many RCB stars located within the Galactic bulge, but still outside the high extinction sky areas \citep{2005AJ....130.2293Z,2008A&A...481..673T}. This could indicate that tens of new RCB stars potentially lie within the most highly crowded and reddened parts of the Galaxy.

Overall, we can confidently say from this first dedicated survey for RCB stars that it is unlikely that there are more than 500 HdC stars in the Galaxy, and that it is likely the population is somewhere between 300 and 500. These estimates are also consistent between the Galaxy and the Magellanic Clouds. Indeed, 30 RCB stars are now known in the Magellanic Clouds as we have added five in this study after the spectroscopic follow-up of 28 Magellanic ToI. With 86 Magellanic ToI remaining to be observed, mostly faint ones, we can estimate, in an optimistic scenario using a similar discovery rate, that about 20 Magellanic RCB stars are yet to be discovered. If the number of RCB stars scales with galaxy mass, the number we extrapolate in the Galaxy would then be $\sim$550 RCB stars (11$\times$(30+20)).

This result is in excellent agreement with theoretical estimates made from population synthesis via the double-degenerate channel. Indeed, with a He-CO white dwarfs merger birthrate ranging between $\sim10^{-3}$ and $\sim5\times10^{-3}$ per year \citep{2001A&A...365..491N,2009ApJ...699.2026R,2015ApJ...809..184K} and an RCB phase lifetime of about $10^5$ years, as predicted by theoretical evolution models \citep{2002MNRAS.333..121S,2019MNRAS.488..438L}, the expected theoretical number of Galactic RCB stars formed via the double-degenerate channel ranges between 100 and 500. 

As discussed in the Introduction, another scenario, the final helium shell flash, has also been suggested as a channel for the formation of RCB stars. If a fraction of Galactic RCB stars are indeed formed through this scenario, our estimate of the total number of Galactic RCB stars is in fact the sum of RCB stars formed by both channels. 
The overabundance of $^{18}$O and $^{19}$F, and the large $^{12}$C/$^{13}$C ratios seen in most RCB stars are a natural product of the WD merger channel, but are not expected from a final flash \citep{2012JAVSO..40..539C,2019MNRAS.488..438L}.
However, there are a small number of RCB stars which differ significantly from the majority characteristics. These include stars with particularly extreme values of Si/Fe and S/Fe, and those with measurable $^{13}$C, and Li, that may be more easily explained by the final-flash channel \citep{2000A&A...353..287A,2012JAVSO..40..539C}. However, Li production through the Cameron-Fowler mechanism \citep{Cameron:1971lr} may occur in the merger scenario \citep{2012A&A...542A.117L,2019MNRAS.488..438L}. Recent work also indicates that the large IR shells seen around RCB stars are not fossil planetary nebula shells which would point toward the final flash \citep{Montiel_2015,2018AJ....156..148M}. Overall, it is not unreasonable to envisage that the fraction of RCB stars formed by the final-flash scenario could be at level of only $\sim$10\%. In that case, the agreement between our estimate of the total number of Galactic RCB stars and the number predicted to exist via the double-degenerate channel is still holding. Of course, the fraction of RCB stars formed by either channel will need further observations to confirm this general impression.

\section{Summary \label{sec_summary}}

From the $\sim$563 million objects catalogued by the WISE All-Sky survey, we selected 2356 targets of interest that present similar near- and mid-IR colours and brightness to typical known RCB stars. We used the 101 known Galactic and Magellanic RCB stars as a reference sample and found that 85\% of them passed all our selection criteria. This emphasises the high detection efficiency of our selection criteria within $\sim$ 50 kpc from the sun. This list of 2356 ToI supersedes the one created using the WISE Preliminary data release \citep{2012A&A...539A..51T}. Further studies and spectroscopic follow-up are now needed to discover the true nature of each of them.

All 2356 targets of interest have been classified into five different groups to prioritise further spectroscopy follow-up. We have respectively 375, 463, 1005, 298 and 215 targets reported in Groups \#1 to \#5, in order of priority. Group \#1 is expected to result in the highest proportion of new RCB stars discovered. The majority of known RCB stars would have been reported in Group\#1, i.e typical RCB stars with a photospheric temperature between 4000$<T_{phot}<$8000 K and with a thick circumstellar shell of temperature between 500$<T_{shell}<$900 K. Group \#2 corresponds to similar RCB stars but with a higher $J-K>$3.5 mag colour index indicating high extinction. This Group is contaminated by highly enshrouded AGB stars, but can also reveal uncommon highly enshrouded RCB stars, such as MSX-SMC-014 and EROS2-SMC-RCB-4. 

Using RCB SED models, we found that our photometric selection has a low detection efficiency with RCB stars that possess one of these three characteristics: (1) a cold circumstellar shell (T$_{shell}<$400 K), (2) a very thin shell as their SEDs would appear similar to classical F or G stars, or (3) a second colder and thicker shell, like the one seen around MV Sgr \citep{2012A&A...539A..51T}. Furthermore, we are less efficient at detecting RCB stars whose 2MASS epoch coincides with a large decline in brightness. As reported in selection cut \#3, it particularly affects RCB stars that possess a warm shell or those with high extinction in any combination of interstellar and circumstellar dust. Finally, we note that we expect to be less sensitive in the sky area $-2<b<2$ deg and $-60<l<60$ deg, where the interstellar extinction is higher than $A_K>$3 mag. We report some bright targets in that region, but further work with datasets obtained with surveys of higher spatial resolution, such as the VISTA/VVV and Spitzer/GLIMPSE surveys, should help us to probe this crowded part of the sky more efficiently.

We have obtained spectra for nearly 500 ToI, and thanks to the light curves produced by OGLE, ASAS, and other surveys, we can give a definite classification to about a third of the 2356 ToI. All spectra, light curves, and charts accumulated for each of them are available online at URL: http://rcb.iap.fr/trackingrcb/. We encourage all observers to report their own observations and help to keep this database up to date.

Spectroscopic selection criteria were defined using known RCB star spectra and hydrogen-deficient stellar atmosphere models to reveal new RCB stars with a range of temperature and abundances. The special scenario of dust obscuration events was also considered. We found 45 new RCB stars and also confirmed the long-lasting candidate \object{KDM 5651} as an RCB star. Now, 117 Galactic and 30 Magellanic RCB stars are known. We also added a list of 14 strong Cold RCB star candidates for which further follow-up, particularly at a brighter phase, is needed. We strongly suspect them to be RCB stars as we have strong indications that their respective spectra were taken during a decline phase. These candidates are also mostly located near the Galactic plane, where the interstellar extinction is higher, making them more difficult to observe.

Light curve information is useful to identify other classes of variable stars, like Miras, but is not sufficient to identify an RCB star. Only the accumulation of evidence, and especially spectra taken in the bright phase, can confirm new RCB stars. For example, we have shown in the present work that the star \object{OGLE-GC-RCB-2} is not an RCB star. It was wrongly reported as such by \citet{2011A&A...529A.118T} based only on its light curve, but the spectrum once obtained shows it to be an RV Tauri star. Indeed RV Tauri type stars share the same photometric colours as RCB stars in the optical and IR, but can also display fast and large photometric declines.

This work is the most systematic survey of RCB stars over the whole sky ever undertaken. Considering that RCB stars account for about 85\% of all stars belonging to the larger class of HdC stars, we have estimated the total number of HdC stars located in the Milky Way to be no more than 500, with the most realistic range being between 300 and 500. This estimate is consistent with the total number of Magellanic RCB stars extrapolated to the mass of the Milky Way which corresponds to a total of about 550. Furthermore, this estimate matches well with theoretical predictions made from population synthesis. Indeed, between 100 and 500 RCB stars, formed from white-dwarf binary mergers are predicted to exist nowadays. As RCB stars could also be formed via the final helium shell flash mechanism, the challenge is now to measure the fraction of existing RCB stars formed via the double-degenerate and final-flash channels. 

Overall, we found 30 Cold ($T_\mathrm{eff}<6800$ K), 14 Warm ($T_\mathrm{eff}>6800$ K) and one Hot ($T_\mathrm{eff}> 15000$ K) RCB stars, reaching totals of 97, 45 and 5 respectively for known Cold, Warm, and Hot RCB stars. The ratios between these temperature regimes could tell us about the relative time spent during the RCB star evolution across the HR diagram while its atmosphere contracts and heats (see \citealt[figures 2 and 3]{2002MNRAS.333..121S}, and \citealt{2019MNRAS.488..438L}). 
This effect is expected to exist in both double-degenerate and final-flash formation scenarios. In the double-degenerate scenario, the situation is complex as there should exist a range of RCB stars masses, thus their $T_\mathrm{eff}$ and their evolution are highly governed by their respective original CO core mass and the envelope mass after accretion \citep{2002MNRAS.333..121S,2019MNRAS.488..438L}.

In the future, we will have to focus our observational efforts along the Galactic disc using northern telescopes. Also, to probe within the Galactic Centre area, and to examine fainter targets such as the ToI that compose priority group \#4 (i.e. to find and understand RCB stars that are highly enshrouded), we will have to use 4m class telescopes. Then, for these highly obscured RCB stars, the main difficulties will come from the lack of crucial information in the visible such as the $^{13}$C absorption lines, knowledge of the hydrogen abundance from the CH bands or the H$_{\alpha}$ line, but also the interesting C$_2$ band-heads between 6000 and 6200 $\AA$. Only the redder CN band-heads and the Ca II IR triplet will remain generally observable. The large dataset of Gaia will in the coming years be very useful to disentangle RCB stars from other variable stars. The infrared (0.9 to 2.0 $\mu$m) spectroscopic all-sky survey planned by the EUCLID mission \citep{2016SPIE.9904E..0TM} will also be a great help.




\begin{acknowledgements}

PT personally thank Tony Martin-Jones for his highly careful readings and comments. This research was conducted by the Australian Research Council Centre of Excellence for All-sky Astrophysics (CAASTRO), through project number CE110001020, and we acknowledge also financial support from "Programme National de Physique Stellaire" (PNPS) of CNRS/INSU, France. PT thanks the MARCS team in Uppsala (Sweden) for kindly providing a grid of hydrogen-deficient stellar models and the french embassy in Sweden for allowing a collaborative visit funded by the program TOR.
We also thanks the team located at Siding Spring Observatory that keeps the 2.3m telescope and its intruments is good shape, as well as the engineer, computer and technician teams located at Mount Stromlo Observatory that have facilitate the observations for the past 10 years. This research has made use of the SIMBAD database, operated at CDS, Strasbourg, France. This publication makes use of data products from the Wide-field Infrared Survey Explorer, which is a joint project of the University of California, Los Angeles, and the Jet Propulsion Laboratory/California Institute of Technology, funded by the National Aeronautics and Space Administration. This publication also makes use of data products from the Two Micron All Sky Survey, which is a joint project of the University of Massachusetts and the Infrared Processing and Analysis Centre, California Institute of Technology, funded by the National Aeronautics and Space Administration and the National Science Foundation. Finally, we heartily thank the OGLE team that have provided light curves for many of our candidates. The OGLE project has received funding from the National Science Centre, Poland, grant MAESTRO 2014/14/A/ST9/00121 to A.U.

\end{acknowledgements}

\bibliographystyle{aa}
\bibliography{WISE-Spectro-1}


 
\begin{appendix}
\onecolumn

\section{Previously known RCB stars and strong RCB stars candidates}

\begin{longtable}{|l|c|c|c|c|c|c|c|c|}
\hline
Name & \parbox{1.33cm}{\centering [3.4]} & \parbox{1.33cm}{\centering $\sigma_{[3.4]}$} & \parbox{1.33cm}{\centering [4.6]} & \parbox{1.33cm}{\centering $\sigma_{[4.6]}$} & \parbox{1.33cm}{\centering [12]} & \parbox{1.33cm}{\centering $\sigma_{[12]}$} & \parbox{1.33cm}{\centering [22]} & \parbox{1.33cm}{\centering $\sigma_{[22]}$} \\
\hline
  \endfirsthead 
\hline 
Name & \parbox{1.31cm}{\centering [3.4]} & \parbox{1.31cm}{\centering $\sigma_{[3.4]}$} & \parbox{1.31cm}{\centering [4.6]} & \parbox{1.31cm}{\centering $\sigma_{[4.6]}$} & \parbox{1.31cm}{\centering [12]} & \parbox{1.31cm}{\centering $\sigma_{[12]}$} & \parbox{1.31cm}{\centering [22]} & \parbox{1.31cm}{\centering $\sigma_{[22]}$} \\
\hline\hline
  \endhead 
\multicolumn{9}{c}{Galactic RCB stars}\\
\hline
\object{XX Cam} & 5.395 & 0.064 & 5.10$^a$ & 0.032 & 3.453 & 0.015 & 2.837 & 0.020 \\
\object{SU Tau} & 4.829 & 0.078 & 3.56$^a$ & 0.1 & 1.526 & 0.011 & 0.984 & 0.012 \\
\object{UX Ant} & 9.799 & 0.022 & 8.470 & 0.020 & 6.375 & 0.018 & 5.681 & 0.040 \\
\object{UW Cen} & 5.360 & 0.053 & 3.78$^a$ & 0.1 & 1.490 & 0.007 & 0.623 & 0.011 \\
\object{Y Mus} & 8.112 & 0.023 & 7.844 & 0.021 & 5.879 & 0.015 & 4.621 & 0.032 \\
\object{DY Cen} & 10.427 & 0.022 & 9.173 & 0.021 & 4.132 & 0.013 & 2.340 & 0.009 \\
\object{V854 Cen} & 3.187 & 0.144 & 2.49$^a$ & 0.1 & 0.533 & 0.007 & 0.381 & 0.008 \\
\object{Z Umi} & 5.994 & 0.051 & 5.02$^a$ & 0.037 & 3.455 & 0.012 & 2.894 & 0.016 \\
\object{S Aps} & 6.057 & 0.053 & 5.67$^a$ & 0.028 & 3.780 & 0.014 & 2.815 & 0.018 \\
\object{R CrB} & 3.455 & 0.140 & 2.37$^a$ & 0.1 & -0.498 & 0.020 & -0.813 & 0.008 \\
\object{RT Nor} & 5.838 & 0.052 & 4.48$^a$ & 0.048 & 2.901 & 0.011 & 2.583 & 0.018 \\
\object{RZ Nor} & 6.686 & 0.031 & 5.12$^a$  & 0.036 & 2.879 & 0.015 & 2.045 & 0.018 \\
\object{V517 Oph} & 4.933 & 0.069 & 3.71$^a$  & 0.1 & 2.096 & 0.010 & 1.577 & 0.014 \\
\object{V1783 Sgr} & 6.118 & 0.039 & 5.34$^a$  & 0.027 & 3.169 & 0.014 & 2.297 & 0.018 \\
\object{WX CrA} & 6.227 & 0.039 & 5.34$^a$  & 0.033 & 3.842 & 0.014 & 3.225 & 0.024 \\
\object{V739 Sgr} & 6.441 & 0.036 & 5.49$^a$  & 0.029 & 3.854 & 0.015 & 3.191 & 0.021 \\
\object{V3795 Sgr} & 7.122 & 0.027 & 5.60$^a$  & 0.024 & 3.081 & 0.013 & 2.375 & 0.017 \\
\object{VZ Sgr} & 7.392 & 0.026 & 6.458 & 0.019 & 4.962 & 0.015 & 4.249 & 0.027 \\
\object{RS Tel} & 7.068 & 0.030 & 5.79$^a$  & 0.028 & 3.705 & 0.014 & 3.021 & 0.021 \\
\object{GU Sgr} & 5.951 & 0.056 & 4.71$^a$  & 0.037 & 3.053 & 0.011 & 2.590 & 0.016 \\
\object{V348 Sgr} & 5.249 & 0.043 & 3.75$^a$  & 0.1 & 2.046 & 0.012 & 1.342 & 0.017 \\
\object{MV Sgr} & 8.124 & 0.022 & 7.401 & 0.020 & 4.725 & 0.014 & 2.193 & 0.016 \\
\object{FH Sct} & 6.062 & 0.043 & 5.11$^a$  & 0.037 & 3.625 & 0.015 & 2.977 & 0.022 \\
\object{V CrA} & 5.790 & 0.049 & 4.39$^a$  & 0.1 & 2.310 & 0.010 & 1.566 & 0.014 \\
\object{SV Sge} & 5.588 & 0.060 & 5.30$^a$  & 0.035 & 3.352 & 0.012 & 2.153 & 0.016 \\
\object{V1157 Sgr} & 6.288 & 0.043 & 4.94$^a$  & 0.045 & null & null & null & null \\
\object{RY Sgr} & 3.036 & 0.156 & 2.27$^a$  & 0.193 & -0.144 & 0.029 & -0.561 & 0.018 \\
\object{V482 Cyg} & 5.835 & 0.055 & 5.23$^a$  & 0.033 & 4.009 & 0.014 & 3.489 & 0.020 \\
\object{U Aqr} & 7.681 & 0.025 & 6.712 & 0.020 & 4.386 & 0.015 & 3.377 & 0.019 \\
\object{UV Cas} & 6.734 & 0.036 & 6.20$^a$  & 0.022 & 4.218 & 0.015 & 3.066 & 0.022 \\
\object{ES Aql} & 5.939 & 0.050 & 5.00$^a$  & 0.038 & 3.368 & 0.013 & 2.800 & 0.018 \\
\object{V2552 Oph} & 7.349 & 0.026 & 6.25$^a$  & 0.021 & 4.787 & 0.015 & 4.255 & 0.025 \\
\object{V4017 Sgr} & 7.332 & 0.028 & 6.35$^a$  & 0.020 & 4.691 & 0.010 & 3.925 & 0.020 \\
\object{V532 Oph} & 6.834 & 0.028 & 5.87$^a$  & 0.024 & 4.221 & 0.015 & 3.611 & 0.023 \\
\object{NSV11154} & 8.083 & 0.023 & 7.099 & 0.020 & 5.183 & 0.015 & 4.351 & 0.025 \\
\object{AO Her} & 5.436 & 0.065 & 4.10$^a$  & 0.1 & 2.371 & 0.007 & 1.797 & 0.009 \\
\object{ASAS-RCB-1} & 5.915 & 0.041 & 5.51$^a$  & 0.029 & 3.703 & 0.014 & 2.423 & 0.022 \\
\object{ASAS-RCB-2} & 4.412 & 0.084 & 3.39$^a$  & 0.1 & 1.755 & 0.012 & 1.140 & 0.011 \\
\object{ASAS-RCB-3} & 6.277 & 0.039 & 5.22$^a$  & 0.033 & 3.774 & 0.014 & 3.261 & 0.018 \\
\object{ASAS-RCB-4} & 6.669 & 0.040 & 5.65$^a$  & 0.031 & 4.172 & 0.014 & 3.723 & 0.020 \\
\object{ASAS-RCB-5} & 5.944 & 0.043 & 4.73$^a$  & 0.038 & 3.293 & 0.011 & 2.696 & 0.021 \\
\object{ASAS-RCB-6} & 9.380 & 0.024 & 8.337 & 0.020 & 5.964 & 0.015 & 4.922 & 0.026 \\
\object{ASAS-RCB-7} & 7.120 & 0.030 & 5.92$^a$  & 0.027 & 4.137 & 0.015 & 3.382 & 0.022 \\
\object{ASAS-RCB-8} & 9.481 & 0.023 & 9.277 & 0.021 & 7.669 & 0.028 & 6.718 & 0.163 \\
\object{ASAS-RCB-9} & 5.203 & 0.059 & 3.96$^a$  & 0.1 & 2.199 & 0.013 & 1.530 & 0.019 \\
\object{ASAS-RCB-10} & 6.933 & 0.030 & 5.81$^a$  & 0.027 & 4.329 & 0.016 & 3.931 & 0.023 \\
\object{ASAS-RCB-11} & 6.384 & 0.041 & 5.51$^a$  & 0.027 & 4.210 & 0.015 & 3.675 & 0.023 \\
\object{ASAS-RCB-12} & 5.573 & 0.053 & 4.71$^a$  & 0.040 & 3.153 & 0.013 & 2.469 & 0.019 \\
\object{ASAS-RCB-13} & 4.375 & 0.082 & 2.83$^a$  & 0.1 & 0.677 & 0.016 & 0.175 & 0.008 \\
\object{ASAS-RCB-14} & 6.158 & 0.047 & 4.93$^a$  & 0.038 & 3.166 & 0.013 & 2.511 & 0.020 \\
\object{ASAS-RCB-15} & 6.544 & 0.036 & 5.45$^a$  & 0.032 & 3.844 & 0.014 & 3.207 & 0.027 \\
\object{ASAS-RCB-16} & 6.794 & 0.035 & 5.92$^a$  & 0.023 & 4.445 & 0.015 & 3.979 & 0.022 \\
\object{ASAS-RCB-17} & 7.656 & 0.023 & 6.35$^a$  & 0.022 & 4.391 & 0.014 & 3.667 & 0.022 \\
\object{ASAS-RCB-18} & 6.512 & 0.036 & 5.51$^a$  & 0.030 & 3.843 & 0.015 & 3.056 & 0.018 \\
\object{ASAS-RCB-19} & 6.122 & 0.042 & 5.12$^a$  & 0.038 & 3.327 & 0.011 & 2.641 & 0.020 \\
\object{ASAS-RCB-20} & 5.648 & 0.054 & 4.33$^a$  & 0.1 & 2.203 & 0.008 & 1.309 & 0.014 \\
\object{ASAS-RCB-21} & 4.857 & 0.058 & 3.38$^a$  & 0.1 & 1.323 & 0.011 & 0.592 & 0.014 \\
\object{IRAS1813.5-2419} & 6.469 & 0.033 & 5.77$^a$  & 0.025 & 4.254 & 0.014 & 3.498 & 0.023 \\
\object{V391 Sct} & 8.069 & 0.023 & 7.119 & 0.021 & 5.421 & 0.016 & 4.787 & 0.031 \\
\object{MACHO 135.27132.51} & 8.312 & 0.023 & 7.022 & 0.021 & 5.067 & 0.014 & 4.236 & 0.025 \\
\object{MACHO 301.45783.9} & 8.515 & 0.024 & 7.167 & 0.020 & 5.382 & 0.014 & 4.544 & 0.025 \\
\object{MACHO 308.38099.66} & 7.639 & 0.022 & 6.509 & 0.020 & 4.850 & 0.014 & 4.016 & 0.024 \\
\object{MACHO 401.48170.2237} & 5.898 & 0.042 & 4.82$^a$  & 0.042 & 3.440 & 0.015 & 2.806 & 0.023 \\
\object{EROS2-CG-RCB-1} & 6.628 & 0.036 & 4.94$^a$  & 0.027 & 3.244 & 0.013 & 2.325 & 0.017 \\
\object{EROS2-CG-RCB-3} & 5.737 & 0.049 & 4.44$^a$  & 0.055 & 3.148 & 0.012 & 2.533 & 0.016 \\
\object{EROS2-CG-RCB-4} & 6.423 & 0.034 & 5.34$^a$  & 0.033 & 3.649 & 0.010 & 2.897 & 0.022 \\
\object{EROS2-CG-RCB-5} & 6.233 & 0.031 & 5.08$^a$  & 0.022 & 3.817 & 0.009 & 3.080 & 0.019 \\
\object{EROS2-CG-RCB-6} & 7.104 & 0.029 & 6.00$^a$  & 0.023 & 4.350 & 0.017 & 3.641 & 0.031 \\
\object{EROS2-CG-RCB-7} & 7.156 & 0.030 & 6.11$^a$  & 0.021 & 4.621 & 0.014 & 3.911 & 0.026 \\
\object{EROS2-CG-RCB-8} & 6.999 & 0.029 & 5.88$^a$  & 0.025 & 4.224 & 0.015 & 3.555 & 0.030 \\
\object{EROS2-CG-RCB-9} & 7.221 & 0.026 & 5.58$^a$  & 0.026 & 3.464 & 0.012 & 2.613 & 0.023 \\
\object{EROS2-CG-RCB-10} & 6.747 & 0.034 & 4.82$^a$  & 0.039 & 2.612 & 0.011 & 1.914 & 0.021 \\
\object{EROS2-CG-RCB-11} & 6.513 & 0.029 & 5.61$^a$  & 0.020 & 4.152 & 0.009 & 3.476 & 0.021 \\
\object{EROS2-CG-RCB-12} & 8.044 & 0.023 & 7.385 & 0.019 & 6.485 & 0.013 & 6.254 & 0.063 \\
\object{EROS2-CG-RCB-13} & 6.815 & 0.032 & 5.67$^a$  & 0.026 & 4.017 & 0.014 & 3.345 & 0.023 \\
\object{EROS2-CG-RCB-14} & 6.404 & 0.031 & 5.05$^a$  & 0.032 & 3.337 & 0.010 & 2.640 & 0.015 \\
\object{OGLE-GC-RCB-1} & 6.683 & 0.035 & 5.72$^a$  & 0.024 & 4.061 & 0.014 & 3.347 & 0.020 \\
\hline
\multicolumn{9}{c}{Magellanic RCB stars}\\
\hline
\object{HV 5637} & 11.780 & 0.024 & 11.101 & 0.020 & 8.319 & 0.017 & 7.422 & 0.055 \\
\object{W Men} & 9.998 & 0.022 & 9.138 & 0.020 & 7.966 & 0.016 & 7.577 & 0.078 \\
\object{HV 12842} & 10.568 & 0.023 & 9.131 & 0.020 & 7.036 & 0.015 & 6.303 & 0.029 \\
\object{MACHO-11.8632.2507} & 10.043 & 0.023 & 8.761 & 0.020 & 6.403 & 0.014 & 4.178 & 0.019 \\
\object{MACHO-81.8394.1358} & 10.807 & 0.023 & 9.658 & 0.020 & 7.891 & 0.019 & 7.400 & 0.067 \\
\object{MACHO-6.6575.13} & 9.898 & 0.021 & 8.355 & 0.019 & 5.922 & 0.012 & 4.926 & 0.019 \\
\object{MACHO-6.6696.60} & 10.427 & 0.023 & 9.141 & 0.020 & 6.815 & 0.016 & 5.707 & 0.025 \\
\object{MACHO-12.10803.56} & 10.269 & 0.023 & 9.356 & 0.020 & 7.738 & 0.015 & 6.954 & 0.034 \\
\object{MACHO-16.5641.22} & 10.030 & 0.023 & 9.042 & 0.021 & 7.304 & 0.016 & 6.779 & 0.049 \\
\object{MACHO-18.3325.148} & 10.475 & 0.023 & 9.205 & 0.020 & 7.090 & 0.015 & 6.467 & 0.058 \\
\object{MACHO-79.5743.15} & 9.993 & 0.023 & 8.606 & 0.022 & 6.785 & 0.016 & 6.285 & 0.042 \\
\object{MACHO-80.6956.207} & 10.253 & 0.024 & 9.136 & 0.020 & 7.481 & 0.019 & 6.834 & 0.039 \\
\object{MACHO-80.7559.28} & 9.695 & 0.022 & 8.786 & 0.020 & 6.959 & 0.015 & 5.859 & 0.030 \\
\object{EROS2-LMC-RCB-1} & 10.499 & 0.022 & 9.211 & 0.019 & 7.453 & 0.018 & 6.924 & 0.061 \\
\object{EROS2-LMC-RCB-2} & 10.918 & 0.024 & 9.741 & 0.020 & 7.462 & 0.019 & 6.933 & 0.081 \\
\object{EROS2-LMC-RCB-3} & 11.574 & 0.023 & 10.601 & 0.021 & 7.803 & 0.020 & 7.091 & 0.100 \\
\object{EROS2-LMC-RCB-4} & 9.981 & 0.023 & 8.541 & 0.020 & 6.671 & 0.015 & 6.137 & 0.031 \\
\object{EROS2-LMC-RCB-5} & 11.842 & 0.023 & 11.362 & 0.021 & 8.658 & 0.018 & 7.395 & 0.062 \\
\object{EROS2-LMC-RCB-6} & 10.591 & 0.023 & 9.073 & 0.021 & 6.885 & 0.015 & 6.085 & 0.034 \\
\object{ASASJ050232-7218.9} & 11.116 & 0.023 & 9.279 & 0.019 & 6.473 & 0.015 & 5.747 & 0.031 \\
\object{EROS2-SMC-RCB-1} & 11.000 & 0.024 & 9.775 & 0.021 & 7.957 & 0.018 & 7.494 & 0.098 \\
\object{EROS2-SMC-RCB-2} & 11.148 & 0.023 & 10.119 & 0.021 & 8.133 & 0.018 & 7.246 & 0.083 \\
\object{EROS2-SMC-RCB-3} & 9.807 & 0.023 & 8.458 & 0.021 & 6.552 & 0.015 & 5.874 & 0.034 \\
\object{MSX-SMC-014} & 10.269 & 0.024 & 8.690 & 0.020 & 6.365 & 0.014 & 5.378 & 0.028 \\
\hline
\multicolumn{9}{c}{Strong RCB stars candidates\hspace{0.5cm} (look at discussions and at possible update in their status in Section~\ref{sec_cand})} \\
\hline
\object{OGLE-GC-RCB-Cand-1}$^{r1}$ & 7.011 & 0.030 & 6.07$^a$  & 0.023 & 4.057 & 0.013 & 2.405 & 0.018 \\
\object{OGLE-GC-RCB-Cand-2}$^{r1}$ & 7.792 & 0.025 & 7.048 & 0.022 & 5.463 & 0.015 & 4.635 & 0.033 \\
\object{GLIMPSE-RCB-Cand-1}$^{r1}$ & 6.822 & 0.029 & 5.57$^a$  & 0.025 & 3.992 & 0.013 & 3.463 & 0.018 \\
\object{GLIMPSE-RCB-Cand-2}$^{r1}$ & 5.689 & 0.059 & 4.48$^a$  & 0.051 & 3.098 & 0.012 & 2.583 & 0.019 \\
\object{MSX-LMC-1795}$^{r2}$ & 10.03 &  0.023 & 8.577 &  0.02 &  6.414 & 0.016 & 5.317 & 0.031 \\
\object{[RP2006] 1631}$^{r3}$ & 10.738  &  0.023 & 9.644 &  0.020 & 7.707 & 0.017 & 6.371 & 0.045 \\
\object{KDM 5651}$^{r4}$  & 11.419 & 0.023 & 10.389 & 0.021 & 8.960 & 0.027 & 9.690 & 0.392 \\
\object{EROS2-LMC-RCB-7}$^{r5}$ & 9.577 & 0.022 & 7.965 & 0.020 & 5.504 & 0.014 & 4.373 & 0.021 \\
\object{EROS2-LMC-RCB-8}$^{r5}$ & 9.737 & 0.023 & 8.701 & 0.020 & 7.290 & 0.015 & 6.976 & 0.055 \\
\object{EROS2-SMC-RCB-4}$^{r5}$  & 11.538 & 0.023 & 9.519 & 0.019 & 6.697 & 0.014 & 5.585 & 0.028 \\
\hline
\multicolumn{9}{l}{$a$: [4.6] original magnitude was corrected for photometric bias observed at high saturation level (see Sect.~\ref{sec_ana}) } \\
\multicolumn{9}{l}{r1: listed in \citet{2011A&A...529A.118T}}\\
\multicolumn{9}{l}{r2: featureless mid-IR spectrum presented by \citet{2014MNRAS.439.1472M} and listed in \citet{2009AcA....59..335S} as OGLE LMC-RCB-21} \\
\multicolumn{9}{l}{r3: featureless mid-IR spectrum presented by \citet{2011MNRAS.411.1597W}} \\
\multicolumn{9}{l}{r4: listed in \citet{2003MNRAS.344..325M} as a strong RCB candidate.}\\
\multicolumn{9}{l}{r5: listed in \citet{2009A&A...501..985T} due to their interesting light curves}\\
\caption{WISE All-Sky magnitudes and errors for all previously known RCB stars and strong RCB candidates \label{tab.WISEa}}
\end{longtable}

\section{New Galactic and Magellanic RCB stars discovered and the new strong RCB stars candidates}

\begin{longtable}{|l|c|r|r|c|c|l|}

\hline
WISE All-Sky & WISE-ToI & \multicolumn{2}{c}{Galactic coordinates} &  Temperature  & Spectroscopic  & Light Curve, survey and   \\
designation  &    Id    & l (deg)   & b (deg)     &  group  & instrument    & largest variations observed   \\
\hline
 \endfirsthead 
\hline
WISE All-Sky & WISE-ToI & \multicolumn{2}{c}{Galactic coordinates} &  Temperature  & Spectroscopic  & Light Curve, survey and   \\
designation  &    Id    & l (deg)   & b (deg)     &  group  & instrument    & largest variations observed   \\
\hline
  \endhead 
\multicolumn{7}{c}{}\\  
\multicolumn{7}{c}{New Galactic RCB stars}\\
\hline
J110008.77-600303.6 & 76 &  289.45754 & -0.12330 & Cold & 	2.3m/WiFeS &  \\
J132354.47-673720.8 & 90 & 306.02499 & -4.94583 & Cold &  2.3m/WiFeS & \\
J150104.50-563325.1 & 105 & 320.12182 & 1.92522 & Cold & 	2.3m/WiFeS & Bochum: 1.0 mag variation\\
J160205.48-552741.6 & 121 & 327.72638 & -2.05059 & Cold & 	2.3m/WiFeS & Bochum: 0.4 mag variation\\
J161156.23-575527.1 & 124 & 327.07187 & -4.78229 & Cold & 	2.3m/WiFeS & ASAS-SN: 2.5 mag drop$^{\#}$ \\
J163450.35-380218.5 & 130 & 343.92434 & 6.46016 & Cold & 	2.3m/WiFeS &  \\
J164704.67-470817.8 & 139 & 338.56953 & -1.23860 & Warm & 	2.3m/WiFeS & Bochum: 3.0 mag variation \\ 
J170343.87-385126.6 & 148 & 346.92365 & 1.59123 & Cold & 	2.3m/WiFeS &  \\
J171815.36-341339.9 & 161 & 352.38354 & 1.97860 & Cold & 	2.3m/WiFeS &  \\
J171908.50-435044.6 & 1220 & 344.60551 & -3.68706 & Cold & 	2.3m/WiFeS &  \\
J172447.52-290418.6 & 169 & 357.42448 & 3.75908 & Warm & 	2.3m/WiFeS &  \\ 
J172553.80-312421.1 & 171 & 355.62021 & 2.25628 & Warm & 	2.3m/WiFeS &  \\ 
J172951.80-101715.9$^{r1}$ & 174 & 14.05807 & 12.92716 & Warm &  	2.3m/WiFeS & CRTS: no variation \\
J173202.75-432906.1 & 1222 & 346.20875 & -5.42525 & Cold & 	2.3m/WiFeS &  \\
J173553.02-364104.3 & 177 & 352.34341 & -2.37032 & Cold & 	2.3m/WiFeS &  \\
J173819.81-203632.1 & 1227 & 6.22662 & 5.78226 & Cold & 	2.3m/WiFeS &  \\
J174111.80-281955.3 & No Id & 0.00465 & 1.14323 & Warm & 	2.3m/WiFeS &  \\ 
J174119.57-250621.2 & No Id & 2.76109 & 2.82268 & Warm &  	2.3m/WiFeS &  \\
J174138.87-161546.4 & 182 & 10.37103 & 7.38144 & Cold & 	2.3m/WiFeS & ASAS: 0.9 mag (oscillations?) \\
J174257.19-362052.1 & 184 & 353.38413 & -3.39531 & Cold & 	2.3m/WiFeS & OGLE: 8.0 mag (4 drops) \\
J174328.50-375029.0 & 186 & 352.16205 & -4.26413 & Cold & 	2.3m/WiFeS & OGLE: 4.0 mag (2 drops) \\
J174645.90-250314.1$^{r2}$ & 188 & 3.44972 & 1.80007 & Warm &  	2.3m/WiFeS & OGLE: 8.0 mag (2 drops) \\
J174851.29-330617.0 & 190 & 356.79067 & -2.75043 & Warm &  	2.3m/WiFeS &  \\
J175031.70-233945.7 & 191 & 5.08443 & 1.78184 & Warm &  	2.3m/WiFeS & OGLE: 1.2 mag (1 small drop) \\ 
J175107.12-242357.3 & 193 & 4.52017 & 1.28935 & Warm &  	2.3m/WiFeS & OGLE: no variation \\
J175521.75-281131.2 & 1241 & 1.73524 & -1.45559 & Cold & 	SOAR/Goodman & OGLE: 8.0 mag (5 drops) \\
J175558.51-164744.3 & 197 & 11.66355 & 4.15326 & Warm & 	2.3m/WiFeS &  \\ 
J175749.76-075314.9 & 203 & 19.68570 & 8.14974 & Cold & 	2.3m/WiFeS & ASAS-SN: >1.8 mag$^{M}$  \\
	                                &	      &	               &               	&	             &	                     & (1 drop)\\
J175749.98-182522.8$^{r3}$ & 204 & 10.47189 & 2.95757 & Cold & 	2.3m/WiFeS & \\
J180550.49-151301.7 & 209 & 14.21051 & 2.86870 & Warm &  	2.3m/WiFeS & ASAS: 1.0 mag variations \\
J181252.50-233304.4 & 2645 & 7.71430 & -2.61157 & Cold & 	2.3m/WiFeS & OGLE: 1.6 mag (2 drops) \\
	                                &	      &	               &               	&	             &	                     & ASAS-SN: >1.5 mag$^{M}$ \\
	                                &	      &	               &               	&	             &	                     & (1 drop)\\
J181538.25-203845.7 & 220 & 10.57510 & -1.78842 & Cold & 	2.3m/WiFeS & Pan-STARRS: >2.0 mag$^{M}$  	\\
J182334.24-282957.1 & 1265 & 4.45107 & -7.04796 & Cold & 	2.3m/WiFeS &  \\
J182723.38-200830.1 & 1269 & 12.31294 & -3.98522 & Cold & 	2.3m/WiFeS &  \\
J182943.83-190246.2 & 231 & 13.54224 & -3.97017 & Cold & 	2.3m/WiFeS & Bochum: 3.0 mag (1 drop) \\
J183649.54-113420.7 & 240 & 20.97746 & -2.05850 & Cold & 	2.3m/WiFeS & Bochum: 5.0 mag (1 drop) \\
J184158.40-054819.2 & 249 & 26.68830 & -0.54812 & Cold & 	2.3m/WiFeS & Bochum: 0.3 mag \\
	                                &	      &	               &               	&	             &	                     & (1 small drop)\\
J184246.26-125414.7 & 250 & 20.45347 & -3.95830 & Cold & 	2.3m/WiFeS &  \\
J185525.52-025145.7 & 257 & 30.83712 & -2.19132 & Warm &  	2.3m/WiFeS & Bochum: 0.2 mag and  \\
	                                &	      &	               &               	&	             &	                     & ASAS-SN: 1.6 mag$^{\#\#}$ (1 drop)\\
J194218.38-203247.5 & 290 & 19.48694 & -20.08972 & Cold & 	2.3m/WiFeS & CRTS: 8.0 mag (7 drops) \\
\hline
\multicolumn{7}{c}{}\\
\multicolumn{7}{c}{New Magellanic RCB stars}\\
\hline
J005010.67-694357.7$^{r4}$:  & 5003 & 303.09311 & -47.39506 & Warm &  2.3m/WiFeS & OGLE: 6.0 mag (2 drops) \\ 
J005113.58-731036.3$^{r5}$ & 5004 & 302.95316 & -43.95139 & Cold & 	2.3m/WiFeS & OGLE: 7.0 mag (2 drops)\\
J053745.70-635330.8 & 6005 & 273.35887 & -32.22225 & Hot &  	2.3m/WiFeS & OGLE: 5.0 mag (3 drops) \\
J054123.49-705801.8$^{\star\star}$ & No Id &  281.60571 & -31.22409 & Cold & 2.3m/WiFeS &  \\ 
J054221.91-690259.3$^{r6}$:  & 5039 & 279.36499 & -31.35237 & Cold & 	2.3m/WiFeS & OGLE: 6.0 mag (9 drops)  \\
										&			&					&					&					&	 AAT/AAOmega &   \\
J055643.56-715510.7$^{r7}$ & 5042 & 282.57199 & -29.91933 & Cold/Warm & 	2.3m/WiFeS & OGLE: no variation \\
\hline
\multicolumn{7}{c}{}\\
\multicolumn{7}{c}{}\\
\multicolumn{7}{c}{}\\
\multicolumn{7}{c}{}\\
\multicolumn{7}{c}{New Galactic RCB candidates}\\
\hline
J161311.79-503040.2$^{\star}$ & No Id & 332.29137 &  0.49038 & n/a & not observed & Bochum: 3.0 mag (1 drop) \\
J164424.53-481205.1 & 136 & 337.46936 & -1.58860 & Cold &  2.3m/WiFeS &  \\
J164433.19-423032.2 & 137 & 341.79607 & 2.11155 & Cold &  2.3m/WiFeS &  \\
J164440.88-421522.3 & 138 & 342.00320 & 2.25855 & Cold & 	2.3m/WiFeS &  \\
J170738.27-431019.7 & 150 & 343.92942 & -1.60193 & Cold & 	2.3m/WiFeS &  \\
J172044.89-315031.7 & 164 & 354.64091 & 2.91539 & Cold & 	2.3m/WiFeS &  \\
J173837.00-281734.5 & 181 & 359.73602 & 1.64518 & Cold & 	2.3m/WiFeS & Pan-STARRS: 5.9 mag$^{M}$ \\
  &   &   &  &   & 	 &  (1 drop)  \\
J175136.80-220630.6 & 194 & 6.54934 & 2.36043 & Cold & 		2.3m/WiFeS &  \\
J180313.12-251330.1 & 207 & 5.18282 & -1.50019 & Cold &   2.3m/WiFeS & \\
J181316.97-253135.1 & 218 & 6.02007 & -3.63537 & Cold &   2.3m/WiFeS &  OGLE: 4.2 mag (1 drop) \\
  &   &   &   &    &  & EROS2: 2.5 mag (1 drop) \\ 
J182010.96-193453.4 & 223 & 12.01864 & -2.22483 & Cold &  2.3m/WiFeS &  \\
J182235.25-033213.2 & 225 & 26.47598 & 4.78381 & Cold &   2.3m/WiFeS & \\
J183631.25-205915.1$^{r8}$ & 4117 & 12.53382 & -6.27324 & Cold & 	2.3m/WiFeS & ASAS: 1.5 mag (2 drops)  \\
J190309.89-302037.1 & 264 & 6.44446 & -15.64277 & Cold & 	2.3m/WiFeS &  \\
J191243.06+055313.1 & 274 & 40.59924 & -2.02375 & Cold & 	2.3m/WiFeS &  \\
\hline
\multicolumn{7}{c}{}\\  
\multicolumn{7}{l}{$^{\#}$: Reported as a likely RCB star based on a brightness drop observed by the ASAS-SN survey \citep{2017ATel11017....1J} and } \\
\multicolumn{7}{l}{being flagged in \citet{2012A&A...539A..51T} due to its IR colours characteristic.} \\
\multicolumn{7}{l}{$^{\#\#}$: A decline of $\sim$1.6 mag at JD$\sim$2458000 days was observed by the monitoring survey ASAS-SN \citep{2018MNRAS.477.3145J} } \\
\multicolumn{7}{l}{$^{\star}$: named GDS\_J1613117-503040 in the Bochum survey.} \\ 
\multicolumn{7}{l}{$^{\star\star}$: also known as KDM 5651 : previously known candidate RCB \citep{2003MNRAS.344..325M} } \\ 
\multicolumn{7}{l}{as reported by Stefan H{\"u}mmerich (light curve available on AAVSO variable star website)} \\  
\multicolumn{7}{l}{$^{r1}$: also known as the variable star AC Ser } \\ 
\multicolumn{7}{l}{$^{r2}$: listed as candidate RCB \object{GLIMPSE-RCB-Cand-1} in \citet{2011A&A...529A.118T} } \\ 
\multicolumn{7}{l}{$^{r3}$: listed as candidate RCB \object{GLIMPSE-RCB-Cand-2} in \citet{2011A&A...529A.118T} } \\  
\multicolumn{7}{l}{$^{r4}$: also known as \object{[MH95] 580}, and GAIA 16aau, a GAIA Alert ATEL http://www.astronomerstelegram.org/?read=8681 } \\
\multicolumn{7}{l}{$^{r5}$: also known as \object{RAW 658} in the SMC carbon stars catalogue from \citet{1993A&AS...97..603R}} \\
\multicolumn{7}{l}{$^{r6}$: also known as \object{MSX-LMC-1795} \citep{2014MNRAS.439.1472M} and \object{OGLE-LMC-RCB-21} \citep{2009AcA....59..335S}} \\ 
\multicolumn{7}{l}{$^{r7}$: also known as \object{HV 12862}  and 	\object{KDM 6829}} \\ 
\multicolumn{7}{l}{$^{r8}$: also known as the variable star IZ Sgr. It was surprisingly classified as a Mira type star (M6 spectral type)  by \citet{1967IBVS..228....1H} after } \\
\multicolumn{7}{l}{a spectroscopic follow-up. } \\ 
\multicolumn{7}{l}{$^{M}$: variations observed by Gabriel Murawski (private communication, 2019), light curves available on AAVSO variable star website. }\\ 
\caption{Summary on spectroscopic temperature group and light curves variation found for all newly confirmed Galactic and Magellanic RCB stars and of the new strong RCB stars candidates \label{tab.NewRCBcoord}}
\end{longtable}


\twocolumn

\end{appendix}

\end{document}